\documentclass[twocolumn,showpacs,preprintnumbers,amsmath,amssymb,pre]{revtex4}

\usepackage{graphicx}
\usepackage{dcolumn}
\usepackage{bm}
\usepackage{epsfig} 

\newcommand{\dd}{{\rm d}}
\renewcommand{\pl}{\partial}
\newcommand{\ii}{{\rm i}}
\newcommand{\inta}{\int_{-i\infty}^{+i\infty}} 
\newcommand{\beq}{\begin{equation}} 
\newcommand{\eeq}{\end{equation}} 
\newcommand{\beqa}{\begin{eqnarray}} 
\newcommand{\eeqa}{\end{eqnarray}} 
\newcommand{\bea}{\begin{array}} 
\newcommand{\ea}{\end{array}} 
\newcommand{\lag}{\langle} 
\newcommand{\rag}{\rangle}
\newcommand{\bu}{{\bf u}}
\newcommand{\bx}{{\bf x}}
\newcommand{\bk}{{\bf k}}
\newcommand{\bq}{{\bf q}}
\newcommand{\thetat}{{\tilde{\theta}}}
\newcommand{\but}{{\tilde{\bf u}}}
\newcommand{\bxref}{{\bf x}_0}
\newcommand{\deltat}{{\tilde{\delta}}}

\newcommand{\sigr}{\sigma_{\delta_{Lr}}}
\newcommand{\sigq}{\sigma_{\delta_{Lq}}}
\newcommand{\cP}{{\cal P}}
\newcommand{\cD}{{\cal D}}

\newcommand{\bxh}{\hat{\bf x}}
\newcommand{\cF}{{\cal F}}
\newcommand{\cG}{{\cal G}}
\newcommand{\cS}{{\cal S}}
\newcommand{\ctheta}{\Theta}

\newcommand{\sigpsir}{\sigma_{\psi_{0r}}}
\newcommand{\bCpsi}{\overline{C}_{\psi_{0r}}}
\newcommand{\cL}{{\cal L}}
\newcommand{\Ai}{{\rm Ai}}
\newcommand{\law}{\stackrel{\rm law}{=}}
\newcommand{\cPb}{\overline{\cal P}}

\begin{document}


\title{Quasi-linear regime and rare-event tails of decaying Burgers turbulence}

\author{P.~Valageas}
\affiliation{Institut de Physique Th\'eorique, CEA Saclay, 91191 Gif-sur-Yvette, France}

\date{\today}

\begin{abstract}
We study the decaying Burgers dynamics in $d$ dimensions for random Gaussian
initial conditions. We focus on power-law initial energy spectra, such that
the system shows a self-similar evolution. This is the case of interest
for the ``adhesion model'' in cosmology and a standard framework for
``decaying Burgers turbulence''. We briefly describe how the system can be
studied through perturbative expansions at early time or large scale 
(quasi-linear regime). Next, we develop a saddle-point method, based on
spherical instantons, that allows to obtain the asymptotic probability
distributions $\cP(\eta_r)$ and $\cP(\ctheta_r)$, of the density and
velocity increment over spherical cells, reached in the quasi-linear regime.
Finally, we show how this approach can be extended to take into account
the formation of shocks and we derive the rare-event tails of these probability
distributions, at any finite time and scale. This also gives the high-mass
tail of the mass function of point-like singularities (shocks in the
one dimensional case).
\end{abstract}

\pacs{Valid PACS appear here}

\maketitle

\section{Introduction}
\label{Intro}

The Burgers equation \cite{Burgersbook,Kida1979,Bec2007}, which describes 
the evolution
of a compressible pressureless fluid, with a non-zero viscosity, was first
introduced as a simplified model of fluid turbulence, as it shares the same
hydrodynamical (advective) nonlinearity and several conservation laws with the
Navier-Stokes equation. It also displays strong intermittency, associated
with anomalous scaling exponents for the velocity structure functions,
but this arises from the formation of shocks (i.e. singular structures in the
inviscid limit $\nu\rightarrow 0^+$) where energy is dissipated, whereas
the structures that appear in Navier-Stokes turbulence seem to be more varied
and less singular (because of pressure effects) \cite{Frisch1995}.
Nevertheless, due to its greater simplicity -  it can actually be explicitly
integrated through the Hopf-Cole transformation \cite{Hopf1950,Cole1951} -
the Burgers dynamics retains much interest for hydrodynamical studies,
particularly as a useful benchmark for approximation schemes \cite{Fournier1983}.
On the other hand, the Burgers equation also appears in many physical problems,
such as the propagation of nonlinear acoustic waves in
non-dispersive media \cite{Gurbatov1991}, the study of disordered systems and
pinned manifolds \cite{LeDoussal2008}, or the formation of large-scale
structures in cosmology \cite{Gurbatove1989,Vergassola1994},
see \cite{Bec2007} for a recent review.
In the cosmological context, where one considers the inviscid limit without
external forcing, it is known as the ``adhesion model'' and it provides a
good description of the large-scale filamentary structure of the cosmic 
web \cite{Melott1994}.
Then, one is interested in the statistical properties of the dynamics, 
as described by the density and velocity fields, starting with a random Gaussian
initial velocity \cite{Kida1979,Gurbatov1997} and a uniform density.
These initial conditions are the signature of quantum fluctuations generated
in the primordial Universe and agree with the small Gaussian fluctuations
observed on the cosmic microwave background. In the hydrodynamical context,
this setup corresponds to ``decaying Burgers turbulence'' \cite{Gurbatov1997}.

This problem has led to many studies, focusing on power-law initial energy
spectra (fractional Brownian motion) in one dimension, $E_0(k) \propto k^n$,
especially for the two peculiar cases of white-noise initial velocity ($n=0$)
\cite{Burgersbook,Kida1979,She1992,Frachebourg2000} or Brownian motion initial
velocity ($n=-2$) \cite{She1992,Sinai1992,Bertoin1998,Valageas2008}. 
Indeed, in these two cases the initial velocity field is built from a white-noise 
stochastic field (either directly or through one integration), which gives rise
to Markovian processes and allows to derive many explicit analytical results.
For more general $n$, it is not possible to obtain full explicit solutions,
but several properties of the dynamics are already known
\cite{Gurbatov1991,Gurbatov1997}. In particular,
for $-3<n<1$, the system shows a self-similar evolution as shocks merge
to form increasingly massive objects separated by a typical length, $L(t)$
- the integral scale of turbulence - that grows as $L(t) \sim t^{2/(n+3)}$,
while the shock mass function scales as $\ln[n(>m)] \sim -m^{n+3}$ at large
masses \cite{She1992,Molchan1997,Gurbatov1997,Noullez2005}.
In spite of these common scalings, the range $-3<n<1$ can be further split
into two classes, as shocks are dense for $-3<n<-1$ but isolated for $-1<n<1$ 
\cite{She1992}.

In this article, we consider the decaying Burgers dynamics in $d$
dimensions, for random Gaussian initial conditions and power-law initial energy
spectra such that the system displays a self-similar evolution. This is
in particular the case of interest in the cosmological context, which
shows a hierarchical evolution as increasingly large scales turn nonlinear
as time goes on. Applying to the Burgers dynamics methods that have been used
to study the collisionless gravitational dynamics encountered in cosmology,
we present a saddle-point approximation (instanton technique) that allows
to derive some properties of the velocity and density fields in two regimes,
i) the quasi-linear regime associated with early times or large scales,
and ii) the rare-event tails of the velocity and density distributions
at any time or scale.

This article is organized as follows. We first introduce in 
section~\ref{Burgers-dynamics} the equations of motion and the initial conditions
that define our system and we recall the geometrical interpretation of the
Hopf-Cole solution of the dynamics. We also define the overdensity, $\eta_r$,
and the velocity divergence (i.e. spherical velocity increment), $\ctheta_r$,
within spherical cells of radius $r$, that are the two quantities that we study in
this paper. Then, we briefly describe in section~\ref{Perturbative-expansion}
how the dynamics can be studied through perturbative expansions, that hold
at early times or large scales, and we make the connection with the Zeldovich
dynamics that is equivalent from a perturbative point of view.
Next, we present in section~\ref{Quasi-linear} a saddle-point approximation
that allows to derive the asymptotic probability distributions $\cP(\eta_r)$
and $\cP(\ctheta_r)$ reached in the quasi-linear limit (i.e. at early times
or large scales). Then, we show in section~\ref{Asymptotic-tails} how
to modify this approach to take into account shocks, and we derive the 
rare-event tails of these probability distributions, at any fixed time and scale.
This also yields the high-mass tail of the mass function of point-like
objects (shocks in the one dimensional case).
Finally, we conclude in section~\ref{conclusion}.

\section{Burgers dynamics}
\label{Burgers-dynamics}

\subsection{Equations of motion and initial conditions}
\label{eq-motion}

We consider the $d$-dimensional Burgers equation in the inviscid limit
(with $d\geq 1$),
\beq
\pl_t \bu + (\bu.\nabla)\bu = \nu \Delta \bu , \hspace{1cm} \nu\rightarrow 0^+ ,
\label{Burgers}
\eeq
for the velocity field $\bu(\bx,t)$, and the evolution of the density field 
$\rho(\bx,t)$ generated by this dynamics, starting from a uniform density $\rho_0$
at the initial time $t=0$. The latter obeys the usual continuity equation
\beq
\pl_t \rho + \nabla.(\rho\bu) = 0 \;\;\; \mbox{and} \;\;\; \rho(\bx,0)=\rho_0 .
\label{continuity}
\eeq
Then, since there is no external forcing in 
Eqs.(\ref{Burgers})-(\ref{continuity}),
the stochasticity arises from the random initial velocity $\bu_0(\bx)$, which we 
take to be Gaussian and isotropic, whence $\lag\bu\rag=0$ by symmetry. Moreover, 
as is well-known \cite{Bec2007}, if the initial velocity is potential,
$\bu_0=-\nabla\psi_0$, 
it remains so forever, so that the velocity field is fully defined by its 
potential $\psi(\bx,t)$, or by its divergence $\theta(\bx,t)$, through
\beq
\bu=-\nabla\psi, \;\;\;\; \theta = -\nabla.\bu = \Delta \psi .
\label{thetadef}
\eeq
Normalizing Fourier transforms as
\beq
\theta(\bx) = \int\dd\bk \; e^{\ii\bk.\bx} \; \thetat(\bk) ,
\label{Fourier}
\eeq
the initial divergence $\theta_0$ is taken as Gaussian, homogeneous and 
isotropic, so that it is fully described by its power spectrum $P_{\theta_0}(k)$ 
with
\beq
\lag\thetat_0\rag=0 , \;\;\; \lag\thetat_0(\bk_1)\thetat_0(\bk_2)\rag = 
\delta_D(\bk_1+\bk_2) P_{\theta_0}(k_1) ,
\label{Ptheta0def}
\eeq
where we note $\delta_D$ the Dirac distribution.
In this article we focus on the power-law initial power spectra,
\beq
P_{\theta_0}(k) \propto k^{n+3-d} \;\;\; \mbox{with} \;\; -3<n<1 .
\label{ndef}
\eeq
Thus, the initial conditions obey the scaling laws
\beqa
\lambda>0 : \;\;\; \thetat_0(\lambda^{-1}\bk) & \law & \lambda^{d-\frac{n+3}{2}}
\; \thetat_0(\bk) , \label{scalingthetat0} \\
\theta_0(\lambda\bx) & \law & \lambda^{-\frac{n+3}{2}} \; \theta_0(\bx) ,
\label{scalingtheta0}
\eeqa
where ``$\law$'' means that both sides have the same statistical properties.
This means that there is no preferred scale in the system and the Burgers dynamics
will generate a self-similar evolution for $-3<n<1$, as seen in 
section~\ref{self-similarity}. This is why we only consider the range $-3<n<1$
in this article. For the initial velocity and potential this yields for any
$\lambda>0$,
\beq
\bu_0(\lambda\bx) \law \lambda^{-\frac{n+1}{2}} \; \bu_0(\bx) , \;\;\;
\psi_0(\lambda\bx) \law \lambda^{\frac{1-n}{2}} \; \psi_0(\bx) .
\label{scalingu0psi0}
\eeq
Since we have $\but(\bk,t)=\ii(\bk/k^2)\thetat(\bk,t)$, the initial energy
spectrum is a power law,
\beq
\lag\but_0(\bk_1).\but_0(\bk_2)\rag = \delta_D(\bk_1+\bk_2) E_0(k_1) ,
\label{E0def}
\eeq
with
\beq
E_0(k) = k^{-2} P_{\theta_0}(k) \propto k^{n+1-d} .
\label{E0n}
\eeq
The initial velocity correlation at distance $x$ reads as
\beqa
\lefteqn{\hspace{-0.7cm} \lag \bu_0(\bx_1).\bu_0(\bx_2)\rag 
= \int\dd\bk \, e^{\ii\bk.\bx} E_0(k)} \nonumber \\
&& \hspace{-0.7cm} = (2\pi)^{\frac{d}{2}} \int_0^{\infty}\dd k \,k^{d-1}
\frac{J_{\frac{d}{2}-1}(kx)}{(kx)^{\frac{d}{2}-1}} E_0(k) \propto x^{-n-1} ,
\label{u0xu00}
\eeqa
where $\bx=\bx_2-\bx_1$ and $J_{\frac{d}{2}-1}(kx)$ is the Bessel function of
the first kind of order $d/2-1$, whereas the initial one-point variance is
\beq
\lag |\bu_0|^2\rag = \int\dd\bk \, E_0(k) = \frac{2\pi^{\frac{d}{2}}}{\Gamma(d/2)}
\int_0^{\infty} \dd k \, k^{d-1} E_0(k) .
\label{u0u0}
\eeq
Thus, for $-1<n<1$ the initial velocity correlation decreases at large distance
as the power law (\ref{u0xu00}), in agreement with the scaling 
(\ref{scalingu0psi0}), while the one-point variance at $x=0$, Eq.(\ref{u0u0}),
diverges because of the contribution from high wavenumbers. 
Then, the initial velocity field is singular (e.g., a white noise for $d=1$
and $n=0$) but this ultraviolet divergence is regularized as soon as $t>0$
by the infinitesimal viscosity \cite{Burgersbook}.
For $-3<n<-1$ the integral (\ref{u0u0}) shows an infrared divergence.
In this case, the initial velocity field is no longer homogeneous and only has
homogeneous increments (but the divergence $\theta_0$ is still homogeneous)
\cite{Frisch1995}.
Then, to build the initial velocity from its divergence one must
choose a reference point, such as the origin $\bx_0=0$, with $\bu_0(\bxref)=0$,
and define the initial velocity in real space as
\beq
\bu_0(\bx) = \int\dd\bk \left( e^{\ii\bk.\bx}-e^{\ii\bk.\bxref} \right) 
\but_0(\bk) , \;\; \mbox{for} \;\; -3<n<-1 .
\label{uxukn}
\eeq
Then, Equation (\ref{u0xu00}) no longer applies but the initial second-order 
structure function, $\lag |\bu_0(\bx)-\bu_0(\bxref)|^2\rag$, 
grows as $x^{-n-1}$.
Note that because of the nonlinear advective term in the Burgers equation 
(\ref{Burgers}), the increments of the velocity field are no longer 
homogeneous for $t>0$, which also means that the divergence $\theta(\bx,t)$
is no longer homogeneous either.
However, at large distance from the reference point (i.e. taking the
limit $|\bxref|\rightarrow \infty$ or $|\bx|\rightarrow \infty$), 
we can expect to recover
an homogeneous system (in terms of velocity increments and matter distribution),
see \cite{Frisch2005} for more detailed discussions.
This can be shown explicitly for the case $d=1$ and $n=-2$,
where the initial velocity field is a Brownian motion 
\cite{Bertoin1998,Valageas2008}. On the other hand, we may add a low-$k$ cutoff
$\Lambda$ to the initial power spectrum and restrict ourselves to finite times
and scales where the influence of the infrared cutoff is expected to vanish
for equal-time statistics.

\subsection{Density contrast and linear mode}
\label{Linear-theory}

In order to follow the evolution of the matter distribution we define the density
contrast, $\delta(\bx,t)$, by
\beq
\delta(\bx,t) = \frac{\rho(\bx,t)-\rho_0}{\rho_0} .
\label{deltadef}
\eeq
Then, if we linearize the equations of motion (\ref{Burgers})-(\ref{continuity})
we obtain the solution
\beq
\thetat_L(\bk,t) = \thetat_0(\bk) e^{-\nu k^2 t} , \;\;
\deltat_L(\bk,t) = \thetat_0(\bk) \frac{1-e^{-\nu k^2 t}}{\nu k^2} ,
\label{linearnu}
\eeq
where the subscript $L$ stands for the ``linear'' mode. In the inviscid limit,
$\nu\rightarrow 0^+$, this yields
\beq
\nu \rightarrow 0^+ : \;\; \thetat_L(\bk,t) = \thetat_0(\bk) , \;\;
\deltat_L(\bk,t) = t \, \thetat_0(\bk)  ,
\label{linear}
\eeq
which could also be obtained by setting $\nu=0$ in Eq.(\ref{Burgers}).
Then, when we study the system at a finite time $t>0$, we can as well define the
initial conditions by the linear density field $\delta_L(\bx,t)$, which is
Gaussian, homogeneous and isotropic, with a power spectrum
\beq
-3 <n <1 : \;\;\; P_{\delta_L}(k,t) = t^2 \, P_{\theta_0}(k) 
\propto t^2 \, k^{n+3-d} ,
\label{PdeltaL}
\eeq
and an equal-time two-point correlation
\beqa
\lefteqn{\hspace{-0.6cm} C_{\delta_L}(\bx_1,\bx_2) 
= \lag \delta_L(\bx_1,t) \delta_L(\bx_2,t)\rag } \nonumber \\
&& \hspace{-0.8cm} = (2\pi)^{\frac{d}{2}} \int_0^{\infty}\dd k \, k^{d-1}
\, \frac{J_{\frac{d}{2}-1}(kx)}{(kx)^{\frac{d}{2}-1}} 
P_{\delta_L}(k) \propto t^2 x^{-n-3} ,
\label{CdeltaL}
\eeqa
where $x=|\bx_2-\bx_1|$.
Note that for any $n>-3$ the initial density field is homogeneous, even though
the initial velocity only shows homogeneous increments when $-3<n<-1$.

Here we may add a few comments on the initial conditions that are relevant to the
cosmological context. Let us first briefly recall how the Burgers equation
(\ref{Burgers}) arises in this case.
In the standard cold dark matter scenario \cite{Peebles1982},
about $83\%$ of the matter content of the Universe is in the form of a cold dark
matter component, whereas ordinary baryonic matter only forms the remaining
$17\%$ (in addition, there is a dark energy component, which is consistent with a
cosmological constant in the Einstein equations, which makes about $72\%$ of
the energy content of the Universe, while the previous two matter components only
form the remaining $28\%$), see \cite{Komatsu2009}. The cold dark matter has
a negligible velocity dispersion (whence the label ``cold'') and it has only very
weak non-gravitational interactions (whence the label ``dark'', as it has only
been ``seen'' through its gravitational effects so far). Then, it is well
described as a pressure-less fluid coupled to its own gravity (here we focus on
the late Universe, after the end of the radiation-dominated era, about
$5\times 10^4$ years after the Big Bang, and on scales smaller than the Hubble
scale, where the Newtonian approximation is valid). Therefore, the growth of
matter density fluctuations is governed by the pressure-less Euler equation and
the continuity equation, coupled to the Poisson equation, in an expanding
background \cite{Peebles1980}. Since the gravitational force derives from the
scalar gravitational potential, it does not generate any vorticity, and any
primordial vorticity is diluted by the expansion of the Universe (this only
holds in the linear regime, as shell-crossings can generate vorticity in a
non-perturbative fashion). Then, using a rescaling of time and velocity field,
that brings out the deviations from the mean Hubble flow (and also absorbs the
effect of the uniform cosmological constant), and making the approximation that
the velocity and gravitational potentials remain equal (this is exact in the
linear regime and in one dimension, $d=1$, before shell-crossing), one obtains
the Zeldovich equation \cite{Zeldovich1970}. This corresponds to the Burgers
equation (\ref{Burgers}) with $\nu=0$. Then, one adds an infinitesimal viscosity,
$\nu\rightarrow 0^+$, to prevent shell-crossing
\cite{Gurbatove1989,Vergassola1994}. This induces a sticking of particles within
shocks, that is intended to mimic the trapping within gravitational potential
wells \cite{Melott1994}.

Next, in the cosmological context, the present matter density fluctuations are
assumed to arise from the growth of tiny quantum fluctuations generated during
an inflationary stage in the early Universe. Moreover, these Gaussian initial
fluctuations almost have a Harrison-Zeldovich power spectrum, that corresponds
to $n=1$ in Eq.(\ref{PdeltaL}) above (observations give $n \simeq 0.96$
\cite{Komatsu2009}). The case $n=1$ is also called ``scale-invariant'', as it
gives a gravitational potential power spectrum of the form
$P_{\psi_0}(k) \propto k^{n-1-d}=k^{-d}$, so that all wavenumbers contribute
with the same weight and the two-point correlation is formally scale-invariant,
$C_{\psi_0}(\bx) \propto \int \dd k k^{d-1} P_{\psi_0}(k) W(k x)$ is independent
of $x$ (where $W(k x)$ is some filtering function on scale $x$). Within the
inflationary scenario, this property arises from the fact that the only relevant
scale is the Hubble scale, that remains roughly constant during this stage (this
can also be understood from the fact that during an exponential expansion there
is no genuine origin of time, i.e. the de Sitter spacetime is invariant under
time translations, so that wavelengths generated at different times share the same
properties). Then, since these fluctuations have remained small until recent times
they have evolved through linear theory until the matter-dominated era and the
Newtonian regime. Therefore, they have remained Gaussian and different wavenumbers
have evolved independently (the linearized equations of motion are diagonal
in Fourier space)
until a redshift $z \sim 10^3$. However, the primordial spectrum with $n \simeq 1$
has been modified in-between, during the radiation-dominated era. Indeed, during
this stage, density fluctuations on scales larger than the Horizon keep growing
whereas they oscillate on small scales, due to the pressure associated with the
coupling to the radiation component of the Universe (photons). This implies
that fluctuations $\deltat_L(\bk)$ are multiplied by a transfer function $T(k)$
that decays as $k^{-2}$ at high wavenumbers. Then, the ``initial'' density power
spectrum $P_{\delta_0}(k)$ used to study the formation of large-scale structures
in the late Universe is the primordial one, with $n\simeq 1$, multiplied by
$T(k)^2$. This yields a curved cold dark matter power spectrum, with a local
slope $n$ that runs from $1$ at low $k$ to $-3$ at high $k$. Thus, today at
$z=0$ we have $n \simeq -2$ on galactic scales and $n \simeq -1$ slightly above
cluster scales \cite{Peebles1980,Bernardeau2002}.
This corresponds to the range studied in this article.
Then, power-law power spectra with $-3<n<1$ can model the dynamics on the ranges
of interest for specific purposes. On the other hand, the simplifications
associated with power-law power spectra, such as the scaling laws
(\ref{scale_psi})-(\ref{scale_q}) and the self-similar evolution
(\ref{selfsimilar}) seen below, can be used to check the accuracy of numerical
algorithms and to shed light on the dynamics \cite{Colombi1996}.

\subsection{Spherically symmetric statistics}
\label{Spherical-statistics}

\begin{table*}
\begin{center}
\begin{tabular}{c|c||c|c|c}
\, $n$ \, & \, $d$ \, & \, $C_{u_{0r}}(r_1,r_2)$ \, & $C_{\psi_{0r}}(r_1,r_2)$ & $C_{\delta_{Lr}}(r_1,r_2)$ \rule[-0.25cm]{0cm}{0.7cm} \\ \hline\hline
0 & 1 & $\delta_D(r_1-r_2)$ & $r_1$ & \,$t^2\delta_D(r_1-r_2)/(r_1r_2)$ \rule[-0.2cm]{0cm}{0.6cm} \\ \hline
0 & 3 & $r_1/r_2^2$ & $r_1(3r_2-r_1)/(2r_2)$ & $9t^2/r_2^3$ \rule[-0.25cm]{0cm}{0.6cm} \\ \hline
-1 & 2 & $r_1/r_2$ & $r_1^2[1+\ln(r_2/r_1)]/2$ & $4t^2/r_2^2$ \rule[-0.25cm]{0cm}{0.6cm} \\ \hline
-2 & 1 & $r_1$ & $r_1^2(3r_2-r_1)/6$ & $t^2/r_2$ \rule[-0.25cm]{0cm}{0.6cm} \\ \hline
-2 & 3 & \,$r_1-r_1^3/(5r_2^2)$\, & \,$r_1^2(r_1^2-5r_1r_2+10r_2^2)/(20r_2)$ & $9t^2[1/r_2-r_1^2/(5r_2^3)]$ \!\rule[-0.25cm]{0cm}{0.7cm} \\ \hline
$n$ & \,$d\rightarrow\infty$\, & \,$r_1r_2(r_1^2+r_2^2)^{-(n+3)/2}/d^2$\, & \,$[(r_1^2+r_2^2)^{(1-n)/2}-r_1^{1-n}-r_2^{1-n}]/(d^2(n^2-1))$\, & $t^2 (r_1^2+r_2^2)^{-(n+3)/2}$ \rule[-0.25cm]{0cm}{0.7cm} \\
\end{tabular}
\end{center}
\caption{The initial velocity and potential correlations $C_{u_{0r}}(r_1,r_2)$
and $C_{\psi_{0r}}(r_1,r_2)$ of the spherical component of the initial 
conditions, for some values of $n$ and $d$ where Eqs.(\ref{Cdr1dr2_3})
and (\ref{Cpsi0r}) simplify (with a dimensional normalization factor set to
unity). The last column shows the covariance of the linear density contrast
$\delta_{Lr}$ within radius $r$ at time $t$. 
Here we assumed $r_1<r_2$, except for the velocity and density correlations in 
the case $\{n=0,d=1\}$, and formulae for $r_1>r_2$ are obtained by exchanging
$r_1$ and $r_2$. The correlations are singular at $r_1=r_2$, except in the
limit $d\rightarrow\infty$ shown in the last line (where the velocity and 
potential correlations also become vanishingly small as compared with the density
correlation).}
\label{Table_correl}
\end{table*}

In this article, in order to take advantage of the statistical isotropy of the
system, we focus on two spherically symmetric quantities, the overdensity,
$\eta_r$, and the mean divergence, $\ctheta_r$, within spherical cells of radius
$r$, which we define as
\beq
\eta_r = \frac{m(<r)}{\rho_0 V} = \frac{\rho_r}{\rho_0} = 1+\delta_r \;\;\; 
\mbox{with} \;\;\; \delta_r = \int_V \frac{\dd\bx}{V} \, \delta(\bx) ,
\label{deltardef}
\eeq
and
\beq
\ctheta_r= t \int_V\frac{\dd\bx}{V} \, \theta(\bx) = -\frac{t}{V} \int_S
\dd \bx \; \bu(\bx).\bxh ,
\label{cthetadef}
\eeq
where we used Eq.(\ref{thetadef}). Here $V$ and $S$ are the volume and the surface
of the $(d\!-\!1)-$sphere of radius $r$, $\bxh$ the unit radial vector, and we 
multiplied the divergence $\theta$ by time $t$ in the definition (\ref{cthetadef})
to have a dimensionless quantity $\ctheta_r$.
The moments $\lag\ctheta_r^p\rag$ can be understood as dimensionless
spherical velocity structure functions, the usual longitudinal velocity 
structure functions being defined as $\lag[(\bu(\bx)-\bu(0)).\bxh]^p\rag$
for a given direction $\bxh$ and length $|\bx|$, while in (\ref{cthetadef}) 
we integrate over all directions. In one dimension, $d=1$, up to a sign, 
$\ctheta_r$ is simply the dimensionless velocity increment over the distance
$2r$,
\beq
d=1 : \;\; \ctheta_r= - \frac{t}{2r} [ u(r)-u(-r) ] .
\label{theta1d}
\eeq
In arbitrary dimension, $-\ctheta_r$ is the dimensionless velocity increment
over distance $2r$ averaged over all directions about a given point.
We investigate in this article the probability
distributions $\cP(\eta_r)$ and $\cP(\ctheta_r)$ in the quasi-linear regime
(i.e. at large scales or early times), and their tails in any regime.
The system being homogeneous we can focus on the cell that is centered on
the origin, and this gives in Fourier space
\beq
\delta_r = \int\dd\bk \, \deltat(\bk) W(kr) , \;\;\;\; \mbox{with}
\label{deltarFkr}
\eeq
\beq 
W(kr) = \int_V \frac{\dd\bx}{V} \, e^{\ii\bk.\bx} = 
2^{\frac{d}{2}} \, \Gamma\left(\frac{d}{2}+1\right) \,
\frac{J_{\frac{d}{2}}(kr)}{(kr)^{\frac{d}{2}}} .
\label{Fkrdef}
\eeq

In the linear regime we obtain from Eq.(\ref{linear})
\beq
\delta_{Lr} = \ctheta_{Lr} = -\frac{t}{V} \int_S \dd \bx \; \bu_0(\bx).\bxh .
\label{cthetadeltalinear}
\eeq
For the initial conditions (\ref{PdeltaL}) the linear density contrast 
$\delta_{Lr}$ is Gaussian, of mean zero, $\lag\delta_{Lr}\rag=0$, and covariance
$C_{\delta_{Lr}}(r_1,r_2)$ with
\beqa
\hspace{-0.6cm} C_{\delta_{Lr}}(r_1,r_2) & = & 
\lag \delta_{Lr_1}\delta_{Lr_2}\rag \label{Cdr1dr2def} \\
&& \hspace{-1.9cm} = \frac{2\pi^{\frac{d}{2}}}{\Gamma[\frac{d}{2}]}
\int_0^{\infty} \dd k \, k^{d-1} P_{\delta_L}(k) W(kr_1) W(kr_2) 
\label{Cdr1dr2_3} \\
&& \hspace{-1.9cm} \propto \frac{t^2}{(r_1\!+\!r_2)^{n+3}} \; 
_2F_1\!\left[\frac{n\!+\!3}{2},\frac{d\!+\!1}{2};d\!+\!1;
\frac{4r_1r_2}{(r_1\!+\!r_2)^2}\!\right] 
\label{Cdr1dr2}
\eeqa
where the last relation (\ref{Cdr1dr2}) only holds for $-3<n<d-1$
(if $n\geq d-1$ the integral (\ref{Cdr1dr2_3}) diverges at high $k$ and the
correlation $C_{\delta_{Lr}}(r_1,r_2)$ is a distribution, such as a Dirac
distribution for $\{n=0,d=1\}$).

Then, the linear variance reads as
\beq
n<d-2 : \;\;\; \sigr^2 = \lag \delta_{Lr}^2\rag = C_{\delta_{Lr}}(r,r) 
\propto t^2 \, r^{-n-3} .
\label{sig2def}
\eeq
Indeed, we can note that the integral (\ref{Cdr1dr2_3}) converges at 
$k\rightarrow 0$ for any $n>-3$ but only converges at $k\rightarrow \infty$ for 
$n<d-2$, when $r_1=r_2$. Therefore, in dimensions $d<3$, the variance 
$\sigr^2$ shows an ultraviolet divergence for a power index in the range 
$d-2\leq n<1$ (we only consider the range $-3<n<1$ in this article).
Of course, as soon as $t>0$, the nonlinear evolution associated with the Burgers
dynamics (shocks) makes the nonlinear variance $\lag\delta_r^2\rag$ finite.
In such a case one could also study the density field smoothed by a Gaussian 
window, $\propto e^{-|\bx|^2/(2r^2)}$, instead of the spherical top-hat 
(\ref{deltardef}), to obtain a finite linear variance $\sigr^2$, but we shall
not investigate this alternative in this paper.

We also introduce the spherical component of the initial radial velocity, 
$u_{0r}$, which reads from Eq.(\ref{cthetadeltalinear}) as
\beq
u_{0r} = - \frac{r}{t\,d} \, \delta_{Lr} ,
\label{u0rdef}
\eeq
since $V=(r/d)S$. This is the mean initial radial velocity at radius $r$.
As with $\eta_r$ and $\ctheta_r$, for spherical components we note the dependent
coordinate $r$ as an index, to distinguish from the $d$-dimensional field
$\bu_0(\bx)$ (but contrary to $\eta_r$ and $\ctheta_r$, $u_r$ is the mean
radial velocity at radius $r$, rather than within the volume $V$).
It will also be useful to consider the spherical component of the initial
velocity potential, which we define from Eq.(\ref{thetadef}) as
\beq
\psi_{0r} = - \int_0^r \dd r' \, u_{0r'} = \frac{1}{t\,d} \int_0^r \dd r' \, r'
\delta_{Lr'} ,
\label{psi0rdef}
\eeq
that is, we choose to normalize the initial potential by $\psi_0(0)=0$.
Then, the initial radial velocity and potential two-point correlations are
\beq
C_{u_{0r}}(r_1,r_2) = \frac{\pl^2}{\pl r_1 \pl r_2} C_{\psi_{0r}}(r_1,r_2) 
= \frac{r_1 r_2}{t^2 d^2} C_{\delta_{Lr}}(r_1,r_2) ,
\label{Cu0r}
\eeq
which can be obtained from Eqs.(\ref{Cdr1dr2def})-(\ref{Cdr1dr2}), and
\beqa
C_{\psi_{0r}}(r_1,r_2) & = & 
\frac{2^{d+1}\pi^{\frac{d}{2}}\,\Gamma(\frac{d}{2}+1)^2}{\Gamma(d/2)\;d^2}
\int_0^{\infty} \dd k \, k^{d-5} P_{\theta_0}(k) \nonumber \\
&& \hspace{-2.5cm} \times \left( 
\frac{J_{\frac{d}{2}-1}(kr_1)}{(kr_1)^{\frac{d}{2}-1}} -
\frac{2^{1-\frac{d}{2}}}{\Gamma[\frac{d}{2}]} \right) 
\left( \frac{J_{\frac{d}{2}-1}(kr_2)}{(kr_2)^{\frac{d}{2}-1}} -
\frac{2^{1-\frac{d}{2}}}{\Gamma[\frac{d}{2}]} \right) ,
\label{Cpsi0r}
\eeqa
with a variance
\beq
-3<n<1 : \;\;\; \sigpsir^2=C_{\psi_{0r}}(r,r) \propto r^{1-n} .
\label{sigpsirdef}
\eeq
Note that $\sigpsir^2$ is finite and well-defined over the whole range
$-3<n<1$.
We give in Table~\ref{Table_correl} the initial radial velocity and potential 
correlations $C_{u_{0r}}(r_1,r_2)$ and $C_{\psi_{0r}}(r_1,r_2)$ for
a few low integer values of $n$ and $d$ where they take a simple form,
as well as the limit $d\rightarrow\infty$.
We also show the covariance $C_{\delta_{Lr}}(r_1,r_2)$ of the linear density
contrast at time $t$ within radius $r$, from Eq.(\ref{Cdr1dr2_3}).
We can check that they satisfy Eq.(\ref{Cu0r}).
The formulae are written for $r_1<r_2$ (except for the velocity and density
correlations in the case $\{n=0,d=1\}$) with a dimensional normalization factor
set to unity. For these power-law initial power spectra, the normalizations
used in Table~\ref{Table_correl} can always be achieved by a rescaling of spatial
coordinates.

\subsection{Hopf-Cole solution and self-similarity}
\label{self-similarity}

As is well known, the nonlinear Burgers equation (\ref{Burgers}) can be solved
through the Hopf-Cole transformation \cite{Hopf1950,Cole1951}, by making the
change of variable $\psi(\bx,t)=2\nu\ln\Xi(\bx,t)$. This yields the linear heat
equation for $\Xi(\bx,t)$, which leads to the solution
\beq
\psi(\bx,t) = 2\nu\ln\int\frac{\dd\bq}{(4\pi\nu t)^{d/2}} \, 
\exp\left[ \frac{\psi_0(\bq)}{2\nu}-\frac{|\bx-\bq|^2}{4\nu t}\right] .
\label{Hopf1}
\eeq
Then, in the inviscid limit $\nu\rightarrow 0^+$, a steepest-descent method
gives \cite{Burgersbook,Bec2007}
\beq
\psi(\bx,t) = \max_{\bq}\left[\psi_0(\bq)-\frac{|\bx-\bq|^2}{2t}\right] .
\label{psixpsi0q}
\eeq
If there is no shock, the maximum in (\ref{psixpsi0q}) is reached at a unique
point $\bq(\bx,t)$, which is the Lagrangian coordinate of the particle that is
located at the Eulerian position $\bx$ at time $t$ \cite{Burgersbook,Bec2007}.
Moreover, this particle has kept its initial velocity and we have
\beq
\bu(\bx,t) = \bu_0(\bq) = \frac{\bx-\bq(\bx,t)}{t} .
\label{vxv0q}
\eeq
If we have a shock at position $\bx$ there are several degenerate solutions
to (\ref{psixpsi0q}) and the velocity is discontinuous (as seen from 
Eq.(\ref{vxv0q}), as we move from one solution $\bq_-$ to another one $\bq_+$ 
when we go through $\bx$ from one side of the shock surface to the other side) 
while the density is infinite. The solution (\ref{psixpsi0q}) has a nice 
geometrical interpretation in terms of paraboloids \cite{Burgersbook,Bec2007}.
Thus, let us consider the family of upward paraboloids $\cP_{\bx,c}(\bq)$ 
centered at $\bx$ and of height $c$, with a curvature radius $t$, 
\beq
\cP_{\bx,c}(\bq)=\frac{|\bq-\bx|^2}{2t}+c .
\label{Paraboladef}
\eeq
Then, moving down $\cP_{\bx,c}(\bq)$ from $c=+\infty$, where the paraboloid is
everywhere well above the initial potential $\psi_0(\bq)$ (this is possible
for the initial conditions (\ref{ndef}) since we have $|\psi_0(\bq)| \sim
q^{(1-n)/2}$, which grows more slowly than $q^2$ at large distances),
until it touches the surface defined by $\psi_0(\bq)$, the abscissa $\bq$ of this
first-contact point is the Lagrangian coordinate $\bq(\bx,t)$. If first-contact
occurs simultaneously at several points there is a shock at the Eulerian 
location $\bx$. One can build in this manner the inverse Lagrangian map
$\bx\mapsto\bq(\bx,t)$.

For the initial conditions (\ref{ndef}) that we consider in this paper,
the rescaled initial velocity potential $\psi_0(\lambda\bq)$ has the same
probability distribution as $\lambda^{(1-n)/2}\psi_0(\bq)$ for any $\lambda>0$,
when we normalize by $\bu_0(0)=0$ and $\psi_0(0)=0$, as seen in 
Eq.(\ref{scalingu0psi0}).
Then, the explicit solution (\ref{psixpsi0q}) gives the scaling laws
\beqa
\psi(\bx,t) & \law & t^{\frac{1-n}{n+3}} \;
\psi\left(t^{\frac{-2}{n+3}}\bx,1\right) , \label{scale_psi} \\
\bu(\bx,t) & \law & t^{\frac{-n-1}{n+3}} \;
\bu\left(t^{\frac{-2}{n+3}}\bx,1\right) , \\
\bq(\bx,t) & \law & t^{\frac{2}{n+3}} \;
\bq\left(t^{\frac{-2}{n+3}}\bx,1\right) \label{scale_q} .
\eeqa
For the spherical overdensity $\eta_r$ and the spherical velocity increment
$\ctheta_r$ this yields
\beq
\cP(\eta_r;t) = \cPb\left(\eta;\frac{r}{t^{2/(n+3)}}\right) , \;\;
\cP(\ctheta_r;t) = \cPb\left(\ctheta;\frac{r}{t^{2/(n+3)}}\right) ,
\label{scaling_etha_theta}
\eeq
that is, the distributions $\cP(\eta_r;t)$ and $\cP(\ctheta_r;t)$ of the
overdensity and velocity divergence at scale $r$ and time $t$ only depend
on the ratio $r/t^{2/(n+3)}$.

These scalings mean that the dynamics is self-similar: a rescaling of time is
statistically equivalent to a rescaling of distances, as
\beq
\lambda>0: \;\; t\rightarrow \lambda t, \;\; \bx \rightarrow 
\lambda^{\frac{2}{n+3}} \bx .
\label{selfsimilar}
\eeq
Thus, the system displays a hierarchical evolution as
increasingly larger scales turn nonlinear. More precisely, since in the
inviscid limit there is no preferred scale for the power-law initial conditions
(\ref{ndef}), the only characteristic scale at a given time $t$ is the
so-called integral scale of turbulence, $L(t)$, which is generated by the
Burgers dynamics and grows with time as in (\ref{selfsimilar}),
\beq
L(t)\propto t^{2/(n+3)} .
\label{Lt}
\eeq
It measures the typical distance between shocks,
and it separates the large-scale quasi-linear regime, where the energy spectrum
and the density power spectrum keep their initial power-law forms (\ref{E0n})
and (\ref{PdeltaL}), $P_{\delta}(k,t) \propto t^2 k^{n+3-d}$, from the small-scale
nonlinear regime, which is governed by shocks and point-like masses, where
the density power spectrum reaches the universal white-noise behavior
(i.e. $P_{\delta}(k,t)$ has a finite limit for $k\gg 1/L(t)$).

This self-similar evolution only holds for $n<1$, so that $|\psi_0(\bq)|$ grows
at larger scales, see for instance Eq.(\ref{sigpsirdef}), and $n>-3$,
so that $|\psi_0(\bq)|$ grows more slowly than $q^2$ and the solution
(\ref{psixpsi0q}) is well-defined \cite{Gurbatov1997}. 
This is the range that we consider in this paper.
The persistence of the initial power law at low $k$ for the energy spectrum,
$E(k,t) \propto k^{n+1-d}$, that holds in such cases, is also called the 
principle of permanence of large eddies \cite{Gurbatov1997}.

\section{Perturbative expansion and Zeldovich dynamics}
\label{Perturbative-expansion}

Although the Burgers dynamics can be integrated through the Hopf-Cole
solution (\ref{psixpsi0q}), the computation of its statistical properties
for random initial conditions remains a difficult problem for general $n$
and $d$. Only in the peculiar one-dimensional cases $n=0$ 
\cite{She1992,AvellanedaE1995,Avellaneda1995,Frachebourg2000,Valageas2009} 
and $n=-2$ \cite{She1992,Sinai1992,Bertoin1998,Valageas2008}, with $d=1$,
where the initial velocity is a white noise or a Brownian motion,
one can derive explicit analytical results, taking advantage of the Markovian
character of these two specific stochastic processes.
For general $n$ and $d$ one must resort to approximation methods, such as
perturbative expansions, as with most nonlinear dynamics.
In particular, at early times we may look for the solution of the equations
of motion (\ref{Burgers})-(\ref{continuity}) as an expansion over powers
of time,
\beqa
\deltat(\bk,t) & = & \sum_{p=1}^{\infty} t^p \int\dd\bk_1..\dd\bk_p 
\, \delta_D(\bk_1+...+\bk_p-\bk) \nonumber \\
&& \times F_p(\bk_1,..,\bk_p) \, \thetat_0(\bk_1) ... \thetat_0(\bk_p) ,
\label{deltatp}
\eeqa
and
\beqa
\thetat(\bk,t) & = & \sum_{p=1}^{\infty} t^{p-1} \int\dd\bk_1..\dd\bk_p 
\, \delta_D(\bk_1+...+\bk_p-\bk) \nonumber \\
&& \times G_p(\bk_1,..,\bk_p) \, \thetat_0(\bk_1) ... \thetat_0(\bk_p) .
\label{thetatp}
\eeqa
The Dirac factors express the invariance through translations of the
equations of motion, $F_1=G_1=1$ from Eqs.(\ref{linear}), and the
higher-order kernels $F_p$ 
and $G_p$ obey a recursion relation that is obtained by substituting the
expansion (\ref{deltatp})-(\ref{thetatp}) into the equations of motion
(\ref{Burgers})-(\ref{continuity}). This yields in Fourier space
\beqa
\hspace{-0.5cm} p F_p(\bk_1,..,\bk_p) - G_p(\bk_1,..,\bk_p) & = & \nonumber \\
&& \hspace{-4.2cm} \sum_{\ell=1}^{p-1} 
\frac{\bk_{1,p}.\bk_{1,\ell}}{|\bk_{1,\ell}|^2} \,
G_{\ell}(\bk_1,..,\bk_{\ell}) \, F_{p-\ell}(\bk_{\ell+1},..,\bk_p) ,
\label{Fp}
\eeqa
and
\beqa
\hspace{-0.5cm} (p-1) G_p(\bk_1,..,\bk_p) & = & \sum_{\ell=1}^{p-1} 
\frac{|\bk_{1,p}|^2 \, (\bk_{1,\ell}.\bk_{\ell+1,p})}
{2 \, |\bk_{1,\ell}|^2 \, |\bk_{\ell+1,p}|^2} \nonumber \\
&& \hspace{-1cm} \times G_{\ell}(\bk_1,..,\bk_{\ell}) \,
G_{p-\ell}(\bk_{\ell+1},..,\bk_p) ,
\label{Gp}
\eeqa
where we note $\bk_{i,j}=\bk_i+\bk_{i+1}+..+\bk_{j}$ with $j\geq i$.
This gives for $p=2$ the kernels
\beq
F_2^s(\bk_1,\bk_2)= \frac{(\bk_{12}.\bk_1)(\bk_{12}.\bk_2)}{2k_1^2k_2^2} ,
\label{F2s}
\eeq
and
\beq
G_2^s(\bk_1,\bk_2)= \frac{k_{12}^2(\bk_1.\bk_2)}{2k_1^2k_2^2} ,
\label{G2s}
\eeq
where we defined $F_2^s(\bk_1,\bk_2)=[F_2(\bk_1,\bk_2)+F_2(\bk_2,\bk_1)]/2$
and $G_2^s$ the symmetrized kernels.
In Eqs.(\ref{deltatp})-(\ref{G2s}) we took the inviscid limit
$\nu=0^+$. Then, the effects of the infinitesimal viscosity (i.e. the formation
of shocks) completely disappear in these perturbative expansions.
This implies that taking shocks into account requires non-perturbative methods.

Note that the expansions (\ref{deltatp})-(\ref{G2s}) over powers of time
are also expansions over powers of the initial velocity fluctuations 
$\theta_0(\bx)$, or equivalently over powers of the linear density mode
$\delta_L(\bx,t)$ given in Eq.(\ref{linear}).
Since the amplitude of the linear density fluctuations decreases at large
scales, as seen in Eq.(\ref{sig2def}), the perturbative expansions apply to
both limits of early time or large scale.
In particular, in these limits the distributions $\cP(\eta_r)$ and
$\cP(\ctheta_r)$ converge to the Gaussian of variance $\sigr^2$.
Then, at early times, or when the system is smoothed over
large scales, the displacements of particles are small and one recovers at
leading order the linear theory of section~\ref{Linear-theory}, that is set
by the initial conditions. A simple example is provided by the case
$\{n=-2,d=1\}$ of one-dimensional initial Brownian velocity
\cite{Bertoin1998,Valageas2008}.
However, this only holds for $n<d-2$, where the linear density variance
$\sigr$ is well defined. For $n\geq d-2$ it is not possible to neglect shocks
as soon as $t\neq 0$, and the distributions $\cP(\eta_r)$ and $\cP(\ctheta_r)$ 
remain far from Gaussian at any time and scale.
This is for instance the behavior obtained in the case $\{n=0,d=1\}$ of
one-dimensional initial white-noise velocity 
\cite{AvellanedaE1995,Avellaneda1995,Frachebourg2000,Valageas2009}.
We shall recover these two different behaviors in the following sections.

The perturbative approach (\ref{deltatp})-(\ref{thetatp}) is the standard
method used in cosmology to study the gravitational dynamics at large
scales and early times \cite{Peebles1980,Goroff1986,Bernardeau2002} (in this case
the equation of motion (\ref{Burgers}) gets two new linear terms, associated
with the gravitational force and a friction term that comes from the expansion
of the Universe and the change to comoving coordinates, but the nonlinearity
is the same and the perturbative expansion is similar).
Indeed, in the standard cold dark matter model \cite{Peebles1982},
the amplitude of the linear density fluctuations decreases at larger
scales (i.e $-3<n<1$ as in the present paper, with the same definition
of the power-spectrum index $n$ for $d=3$), and the perturbative
approach allows to describe the large scale structure of the Universe
(e.g. beyond the scale associated with clusters of galaxies today),
that is, the cosmic web formed by voids, filaments and walls that join the
nonlinear high-density objects such as galaxies or clusters of galaxies.
In the hydrodynamical context, perturbative expansions over powers of time,
such as (\ref{deltatp})-(\ref{thetatp}), have been used for instance in 
\cite{Kraichnan1970,Kaneda1999}
to study Eulerian and Lagrangian two-point correlations. They can also serve
as a basis for Pad\'e approximants that attempt to improve the convergence
of the series \cite{Kaneda1999,Valageas2007}.

From the point of view of the perturbative expansions 
(\ref{deltatp})-(\ref{G2s}), the Burgers dynamics (\ref{Burgers}) becomes
equivalent in the inviscid limit to the Zeldovich dynamics 
\cite{Zeldovich1970}, obtained by
setting the right hand side in Eq.(\ref{Burgers}) to zero.
This describes the free motion of collisionless particles, that always
keep their initial velocity $\bu_0$ and can cross each other.
In a Lagrangian framework, the trajectory of the particle of initial
Lagrangian coordinate $\bq=0$ always reads as
\beq
\bx(\bq,t) = \bq+ t\,\bu_0(\bq) ,
\label{Zel1}
\eeq
as in Eq.(\ref{vxv0q}) that only held before shocks.
Before orbit-crossings the conservation of matter gives for the density
field
\beq
\rho(\bx) \dd\bx= \rho_0\dd\bq , \;\; \mbox{hence} \;\;  1+\delta(\bx) = 
\left|\det\left(\frac{\pl\bx}{\pl\bq}\right)\right|^{-1} .
\label{det}
\eeq
This gives
\beq
1+\delta(\bx) = \int\dd\bq \; \delta_D[\bx-\bq-t\bu_0(\bq)] ,
\label{deltaDirac}
\eeq
which remains valid after orbit crossing as we integrate over all streams
that pass through position $\bx$ at time $t$. In Fourier space this yields
\beq
\deltat(\bk) = \int\frac{\dd\bq}{(2\pi)^d} \, e^{-\ii\bk.\bq} 
\left(e^{-\ii \bk.t\bu_0(\bq)}-1\right) .
\label{deltakexp}
\eeq
Then, expanding the exponential over $\bu_0$ directly gives the symmetric
kernels $F_p^s$ associated with the expansion (\ref{deltatp})
\cite{Grinstein1987},
\beq
F_p^s(\bk_1,..,\bk_p) = \frac{1}{p!} \, \frac{\bk_{1,p}.\bk_1}{k_1^2}
... \frac{\bk_{1,p}.\bk_p}{k_p^2} ,
\label{FpZel}
\eeq
which agrees with (\ref{F2s}) for $p=2$.
From the perturbative expansions (\ref{deltatp})-(\ref{G2s}) we can compute
the cumulants of the smoothed density contrast $\eta_r$ and velocity
divergence $\ctheta_r$ (in the quasi-linear regime where shocks do not
contribute, that is, leading-order terms at early times and large scales for
$n\leq d-3$, as discussed in section~\ref{Quasi-linear} below). 
For instance, substituting the expression
(\ref{deltatp}), the density three-point correlation reads in Fourier space as
\cite{Fry1984,Bernardeau2002}
\beqa
\hspace{-0.8cm} \lag\delta(\bk_1)\delta(\bk_2)\delta(\bk_3)\rag_c & = &
\delta_D(\bk_1+\bk_2+\bk_3) \nonumber \\
&& \hspace{-3cm} \times \left[ 2 P_{\delta_L}(k_1,t) 
P_{\delta_L}(k_2,t) F_2^s(\bk_1,\bk_2) + {\rm cyc.} 
+ ... \right] ,
\label{Bispectrum}
\eeqa
where ``cyc.'' stands for two terms associated with cyclic permutations over
$\{\bk_1,\bk_2,\bk_3\}$ of the previous term, while the dots stand for 
higher-order terms. Then, from Eq.(\ref{deltarFkr})
the cumulant of order three of the overdensity within radius $r$ writes
\beqa
\lag\eta_r^3\rag_c & = & 6 \int\dd\bk_1\dd\bk_2 \, P_{\delta_L}(k_1,t) 
P_{\delta_L}(k_2,t) F_2^s(\bk_1,\bk_2) \nonumber \\
&& \times W(k_1r) W(k_2r) W(|\bk_1+\bk_2|r) + ... 
\label{etarp=3}
\eeqa
Using the properties of Bessel functions, such as their addition theorem,
one obtains for instance in dimension $d=3$ 
\cite{Bernardeau1994,BernardeauKof1995},
\beq
d=3 : \;\;\; \lag\eta_r^3\rag_c = (1-n) \, \sigr^4 + ...
\label{S3pert}
\eeq
One can use this method to derive the leading-order term of all cumulants
$\lag\eta_r^p\rag_c$ and $\lag\ctheta_r^p\rag_c$. Then, from the characteristic
function $\varphi(y)$, defined from the Taylor series (\ref{phidef}) below,
one can reconstruct the distributions $\cP(\eta_r)$ and $\cP(\ctheta_r)$
in the quasi-linear regime, $\sigr\ll 1$ \cite{BernardeauKof1995,Bernardeau2002}.
We shall describe in section~\ref{Quasi-linear} below another method
that directly gives the generating function $\varphi(y)$ without using the
expansions (\ref{deltatp})-(\ref{G2s}) and that allows to go beyond the
singularities associated with the Taylor series (\ref{phidef}).
Let us recall here that the previous results only hold for the case $n<d-2$, 
where the linear theory is meaningful (i.e. $\sigr$ is well-defined).

\section{Quasi-linear limit}
\label{Quasi-linear}

\subsection{Distribution of the density within spherical cells}
\label{density-within-spherical-cells}

We consider here the probability distribution, $\cP(\eta_r)$, of the overdensity
$\eta_r$ within spherical cells introduced in (\ref{deltardef}).
More precisely, we investigate its asymptotic form in the quasi-linear limit,
defined as $\sigr\rightarrow 0$. Therefore, we restrict ourselves to the range 
$n<d-2$ (in addition to $-3<n<1$) so that the linear variance $\sigr^2$
is well defined, see Eq.(\ref{sig2def}). Taking advantage of the statistical
isotropy of the system, we apply to the Burgers dynamics the steepest-descent
method (instanton technique) that was devised in \cite{Valageas2002a} for
the collisionless gravitational dynamics.

\subsubsection{Action $\cS[\delta_L]$}
\label{Action}

To obtain the quasi-linear limit of the probability distribution $\cP(\eta_r)$
it is convenient to first introduce the moment generating function $\Psi(y)$,
\beq
\Psi(y) = \lag e^{-y\eta_r}\rag = \int_0^{\infty} \dd\eta_r \, e^{-y\eta_r} \,
\cP(\eta_r) ,
\label{Psidef}
\eeq
from which we can recover $\cP(\eta_r)$ through the inverse Laplace transform
\beq
\cP(\eta_r) = \inta\frac{\dd y}{2\pi\ii} \, e^{y\eta_r} \Psi(y) .
\label{PetarPsi}
\eeq
Since the system is fully defined by the Gaussian linear density field at the
time of interest, $\delta_L(\bx)$ (we usually omit the time dependence
as $t$ can be seen as a mere parameter, since we only consider equal-time  
statistics), the average (\ref{Psidef}) can be written as the path-integral
\beq
\Psi(y) = (\det C_{\delta_L}^{-1})^{1/2} \int \cD\delta_L \, 
e^{-y\eta_r[\delta_L]-\frac{1}{2} \delta_L . C_{\delta_L}^{-1} . \delta_L} ,
\label{Psipathint}
\eeq
where $\eta_r[\delta_L]$ is the functional that affects to the initial condition
defined by the linear density field $\delta_L(\bx)$ the nonlinear overdensity 
$\eta_r$, built by the Burgers dynamics (\ref{Burgers})-(\ref{continuity})
at time $t$, within the spherical cell of radius $r$ centered (for instance)
on the origin $\bx=0$. 
Here and in the following we use the short-hand notation for scalar products
\beq
\delta_L . C_{\delta_L}^{-1} . \delta_L = \int\dd\bx_1\dd\bx_2 \, 
\delta_L(\bx_1) C_{\delta_L}^{-1}(\bx_1,\bx_2) \delta_L(\bx_2) ,
\eeq
where $C_{\delta_L}^{-1}$ is the inverse of the two-point
correlation (\ref{CdeltaL}).
Equation (\ref{Psipathint}) is exact but the difficulty
of the problem is hidden in the nonlinear functional $\eta_r[\delta_L]$.
In order to make some progress, we consider the quasi-linear limit,
$\sigr\rightarrow 0$, associated with large scales or early times.
Then, it is convenient to rescale the moment generating function as
\cite{Valageas2002a}
\beq
\Psi(y) = e^{-\varphi(y\sigr^2)/\sigr^2} ,
\label{psiphidef}
\eeq
where $\varphi(y)$ is the cumulant generating function, which has the
Taylor expansion
\beq
\varphi(y) = - \sum_{p=1}^{\infty} \frac{(-y)^p}{p!} 
\frac{\lag\eta_r^p\rag_c}{\sigr^{2(p-1)}} .
\label{phidef}
\eeq
Substituting Eq.(\ref{psiphidef}) into Eq.(\ref{Psipathint}) gives
\beq
e^{-\varphi(y)/\sigr^2} =  (\det C_{\delta_L}^{-1})^{1/2} \int \cD\delta_L \, 
e^{-\cS[\delta_L]/\sigr^2} ,
\label{path1}
\eeq
with the action $\cS[\delta_L]$ given by
\beq
\cS[\delta_L] = y \, \eta_r[\delta_L] + \frac{\sigr^2}{2} \,
\delta_L . C_{\delta_L}^{-1} . \delta_L
\label{SdLdef}
\eeq
The rescaling (\ref{psiphidef}) allows us to derive the quasi-linear limit
through a steepest-descent method, since the action $\cS[\delta_L]$ no longer
depends on the amplitude of the two-point correlation $C_{\delta_L}$
(as $\sigr^2\propto C_{\delta_L}$) and the path-integral (\ref{path1})
is clearly dominated by the minimum of the action $\cS$ in the limit 
$\sigr\rightarrow 0$.

Here we may note that the use of a path-integral formalism to analyze
dynamical systems such as Eqs.(\ref{Burgers})-(\ref{continuity}) is a standard
approach, following the operator formalism of Martin-Siggia-Rose
\cite{MSR1973} or the functional method of Phythian 
\cite{Phythian1977,Jensen1981,Dominicis1976}. In such a framework, the
path-integral (\ref{Psipathint}) is rewritten in terms of the nonlinear fields
$\delta(\bx,t)$ and $\theta(\bx,t)$, so that $\eta_r$ is a simple linear
functional of $\delta$ as in (\ref{deltardef}), by introducing a Dirac
functional such as $\delta_D[\pl_t \bu + (\bu.\nabla)\bu - \nu \Delta \bu
-\xi]$ (and similarly for $\thetat$) to enforce the equation of motion
(\ref{Burgers}), where, depending on the system, $\xi(\bx,t)$ can represent
both a stochastic external forcing and the random initial conditions.
Taking care of the Jacobian, which is usually equal to unity thanks to
causality \cite{Zinn}, one obtains a path integral such as (\ref{path1}),
but over the nonlinear density field $\delta(\bx,t)$ and its conjugate
$\lambda(\bx,t)$, rather than over $\delta_L(\bx)$, and over the velocity
pair $\{\theta(\bx,t),\mu(\bx,t)\}$. This procedure is described in details
in \cite{Valageas2007} for the Zeldovich dynamics recalled in 
section~\ref{Perturbative-expansion} above, that amounts to set the 
right hand side in the Burgers equation (\ref{Burgers}) to zero, 
see also \cite{Valageas2007a} for the collisionless gravitational dynamics.
For noisy dynamics, where one adds a stochastic external forcing, 
this method is presented for instance in \cite{Gurarie1996,Fogedby1998} 
for the forced Burgers dynamics and in \cite{Falkovich1996} 
for the forced Navier-Stokes dynamics.

Then, one obtains a cubic action $\cS[\delta,\theta;\lambda,\mu]$.
Expanding over the cubic term gives back the perturbative results discussed
in section~\ref{Perturbative-expansion}, as one recovers an expansion over
powers of the initial power spectrum $P_{\delta_L}$. On the other hand,
this path integral can serve as a basis for other expansion schemes,
such as large-$N$ methods \cite{Valageas2007}, that recover at leading order
Kraichnan's direct interaction approximation when applied to the Navier-Stokes
equations \cite{Mou1993,Kraichnan1959}.

Alternative expansion schemes, where one does not expand over powers of
some coupling constant or parameter, are provided by steepest-descent
methods (instanton techniques \cite{Zinn}) where one expands around a
saddle-point of the action $\cS$. If this saddle-point is non-perturbative
this approach can go beyond perturbative expansions such as those described
in section~\ref{Perturbative-expansion}, as we shall see more clearly
in section~\ref{Asymptotic-tails} below. This approach has been applied to
the forced Burgers dynamics in \cite{Gurarie1996,Fogedby1998} and to the forced
Navier-Stokes dynamics in \cite{Falkovich1996}. In particular, this allows to
obtain the right exponential tail of the probability distribution of the
velocity increment \cite{Gurarie1996,Balkovsky1997} and its left power-law tail
\cite{Moriconi2009}.

In the noiseless case, a problem that arises when one
tries to apply this method to the standard action
$\cS[\delta,\theta;\lambda,\mu]$, directly obtained from the equations of
motion as described above, is that this action is highly singular when there
is no external forcing. Indeed, in such cases the dynamics is fully
deterministic, so that the system is fully defined by the initial condition
$\theta_0(\bx)$ (or by $\delta_L(\bx)$ at a given reference time).
Then, one can check that this action $\cS$ is only finite for fields
that obey the equations of motion (\ref{Burgers})-(\ref{continuity}) and infinite
elsewhere, which simply means that the path integral only counts fields that
are solutions of the dynamics, as it should. Therefore, the action is only
finite over a lower dimensional subspace parameterized by $\theta_0(\bx)$,
that is, the time degree of freedom of the fields $\delta(\bx,t)$ and
$\theta(\bx,t)$ is not real. Then, the action has no finite second-derivative
and the steepest-descent approach is not very well defined. Moreover,
any expansion point must be an exact solution of the dynamics so that this
approach does not bring much progress.

By contrast, the path integral (\ref{path1}) only involves the true degrees of
freedom of the system, parameterized by the linear density field
$\delta_L(\bx,t)$ at the given time of interest (i.e. we do not integrate
over non-existent time degrees of freedom). Then, the action $\cS[\delta_L]$
is finite and has a well-defined second derivative, at least close to
$y=0$ and $\delta_L=0$, so that the steepest-descent approach rests on firm
grounds. Moreover, the difficulty associated with the nonlinear functional
$\eta_r[\delta_L]$ would not be overcome by using the standard action
$\cS[\delta,\theta;\lambda,\mu]$, since in this case too we would need to
study exact solutions of the dynamics. Thus, the action (\ref{SdLdef})
is well-suited to the application of the steepest-descent approach to
deterministic dynamics, as we shall see in the following. 
In practice, in order to handle the term $\eta_r[\delta_L]$, one must be able to
obtain saddle-points where the dynamics can be explicitly solved in simple terms.
In our case, it is natural to take advantage of the statistical isotropy of the
system to look for spherically symmetric solutions of the dynamics. Then, this
requires to focus on spherically symmetric observables, such as $\eta_r$ and
$\ctheta_r$ defined in Eqs.(\ref{deltardef})-(\ref{cthetadef}), so that
spherical initial conditions can also be saddle-points of the action
$\cS[\delta_L]$. In fact, in such a case, a minimum of the action with respect
to spherically symmetric initial conditions is automatically a saddle-point
with respect to non-spherically symmetric initial conditions (but not necessarily
a minimum). In our case, we shall see below in
section~\ref{Spherical-saddle-point} that we really obtain a local minimum in
the quasi-linear regime (i.e. for small $y$ and $\delta_L$). Then, even if there
exists another local minimum reached for some non-spherical initial conditions,
which requires a finite fluctuation $\delta_L$, such a contribution is
exponentially subdominant in the quasi-linear limit, $\sigr\rightarrow 0$,
so that we obtain exact results in this limit, without the need to explicitly
study the functional $\eta_r[\delta_L]$ over all possible non-spherical states.

Note that for non-spherically symmetric observables (for instance we could
choose cubic cells to define the mean density $\eta_r$ and velocity divergence
$\ctheta_r$) we could apply the same approach and look for local minima with
respect to spherically symmetric initial conditions. However, these would no
longer be saddle-points with respect to non-spherical fluctuations, and would not
be the true minima of the action. Then, one would only obtain lower bounds for
the asymptotic behavior of the distribution $\cP(X)$, where $X$ is any 
observable and is not necessarily spherically symmetric, in the quasi-linear
or rare-event limits.
One can expect that the exponents obtained in this fashion would remain correct,
but the numerical factors within exponential tails would only be approximate.
The asymptotic behaviors obtained within this approximation would clearly 
show the same qualitative properties as those obtained for
the spherical observables studied here (since one uses the same initial states).
Thus, it is straightforward to obtain lower bounds for any distribution
$\cP(X)$, in the quasi-linear or rare-event limits, from the method described 
in this article and we restrict ourselves to the spherically symmetric observables
(\ref{deltardef})-(\ref{cthetadef}) in the following.

\subsubsection{Spherical saddle-point}
\label{Spherical-saddle-point}

As explained above, and as for the gravitational dynamics \cite{Valageas2002a}, 
taking advantage of the spherical symmetry of the action (\ref{SdLdef}),
we can look for a spherical saddle-point. Indeed, since the first functional
derivative, $\cD\cS/\cD\delta_L(\bq)$, taken at a spherical linear density
field $\delta_L(\bq)$, is spherically symmetric, it only depends on $|\bq|$.
Then, the first variation $\Delta\cS$ due to a non-radial perturbation
$\Delta\delta_L(\bq)$ vanishes,
\beq
\Delta\cS = \int\dd\bq \, \frac{\cD\cS}{\cD\delta_L(\bq)} \, \Delta\delta_L(\bq)
= 0 ,
\eeq
when
\beq
\delta_L(\bq) = \delta_L(|\bq|) \;\;\; \mbox{and} \;\;\;
\int_{|\bq|=q} \dd\hat{\bq} \, \Delta\delta_L(\bq) = 0 ,
\eeq
where the second equality is the integral over angular variables at any
radius $q$. Therefore, a saddle-point with respect to spherically symmetric
states (i.e. radial degrees of freedom) is automatically a saddle-point
with respect to angular degrees of freedom, whence a true saddle-point
with respect to any infinitesimal perturbation $\Delta\delta_L(\bq)$.
Then, we can restrict the
action $\cS[\delta_L]$ to spherically symmetric initial conditions and
look for its minimum within this subspace. For such initial conditions,
the action can be expressed in terms of the one-dimensional field 
$\delta_{Lq'}$, defined as in Eq.(\ref{deltardef}) over $0<q'<\infty$,
(we note by the letter $q$ initial Lagrangian radii, to distinguish them from
the Eulerian radii $r$ reached at time $t$). This reads as
\beq
\cS[\delta_{Lq'}] = y \, \eta_r[\delta_{Lq'}] + \frac{\sigr^2}{2} \,
\delta_{Lq_1'} . C_{\delta_{Lr}}^{-1} . \delta_{Lq_2'} ,
\label{Sdr'def}
\eeq
where $q'$ is a dummy variable and $C_{\delta_{Lr}}(q_1',q_2')$ is the 
covariance introduced in Eq.(\ref{Cdr1dr2def}).
Then, saddle-points of the action (\ref{Sdr'def}) are given by the
condition $\cD \cS/\cD \delta_{Lq'}=0$ over $0<q'<\infty$, that is,
\beq
y\frac{\cD\eta_r}{\cD\delta_{Lq'}} + \sigr^2 \int_0^{\infty} \dd q'' 
C_{\delta_{Lr}}^{-1}(q',q'') \, \delta_{Lq''} = 0 .
\eeq
Multiplying by the operator $C_{\delta_{Lr}}$ this reads as
\beq
\delta_{Lq'} = \frac{-y}{\sigr^2} \int_0^{\infty} \dd q'' 
C_{\delta_{Lr}}(q',q'') \frac{\cD\eta_r}{\cD\delta_{Lq''}} .
\label{saddle}
\eeq
Next, we note that if there have been no collisions (i.e. no shocks) 
until time $t$, the spherical collapse or expansion has remained well ordered, 
and the mass $m$ within the radius $r$ comes from the matter that was initially
located within a Lagrangian radius $q$ at time $t=0$. Then, the overdensity 
$\eta_r=m/(\rho_0V)$
is also given by $\eta_r=(q/r)^d$ and it only depends on the initial Lagrangian 
coordinate $q$ of the shell that is located at radius $r$ at time $t$.
On the other hand, in the inviscid limit the Burgers dynamics (\ref{Burgers}) 
implies that particles that have not collided yet have kept their initial 
velocity $\bu_0$. Therefore, for a spherical state the initial Lagrangian radius
$q$ is related to the Eulerian radius $r$ by $r=q+t \, u_{0q}$, whence
\beq
\eta_r=(q/r)^d = \left(1+\frac{t \, u_{0q}}{q}\right)^{-d} = \cF(\delta_{Lq}) ,
\label{etarF}
\eeq
with
\beq
\cF(\delta_{Lq}) = \left(1-\frac{\delta_{Lq}}{d}\right)^{-d} ,
\label{FdLq}
\eeq
where $u_{0q}$ is the initial radial velocity at radius $q$ and we used
Eq.(\ref{u0rdef}) (for spherical initial conditions we have 
$\bu_0(\bx)=u_{0x} \bxh$).
Thus, the overdensity $\eta_r$ only depends on the initial velocity
at the Lagrangian coordinate $q$, whence on the linear density contrast
$\delta_{Lq}$ within the Lagrangian radius $q$. As a consequence, it is
independent of infinitesimal perturbations to the initial profile 
$\delta_{Lq'}$ over inner or outer shells ($q'<q$ or $q'>q$), that only 
redistribute matter at smaller or larger radii. On the other hand, under an
infinitesimal perturbation $\Delta\delta_{Lq'}$ the Lagrangian radius $q$ and 
the overdensity $\eta_r$ are modified as $q\rightarrow q+\Delta q$ and 
$\eta_r\rightarrow\eta_r+\Delta\eta_r$. From Eq.(\ref{etarF}) we obtain
at first order,
\beqa
\Delta\eta_r & = & \cF'(\delta_{Lq}) \left[ 
\left.\frac{\dd\delta_{Lq'}}{\dd q'}\right|_{q} \Delta q + \Delta\delta_{Lq} 
\right] , \\
\frac{\Delta\eta_r}{\eta_r} & = & d \, \frac{\Delta q}{q} .
\eeqa
This leads to $\Delta\eta_r \propto \Delta\delta_{Lq}$, which means that
the functional differential $\cD\eta_r/\cD\delta_{Lq''}$ in Eq.(\ref{saddle}) 
is a Dirac distribution centered on $q''=q$, in agreement with the previous
discussion, and we directly obtain the initial profile of the saddle-point as
\beq
\delta_{Lq'} \propto C_{\delta_{Lr}}(q',q) \;\;\; \mbox{whence} \;\;\;
\delta_{Lq'} = \delta_{Lq} \, \frac{C_{\delta_{Lr}}(q',q)}{\sigq^2} .
\label{profile}
\eeq
Using Eq.(\ref{u0rdef}), this also gives for the initial velocity profile
\beq
u_{0q'} = u_{0q} \, \frac{q'}{q} \, \frac{\delta_{Lq'}}{\delta_{Lq}} 
= u_{0q} \, \frac{C_{u_{0r}}(q',q)}{\sigma_{u_{0q}}^2} .
\label{profile_u0}
\eeq
Next, the amplitude $\delta_{Lq}$, or the Lagrangian coordinate $q$, can be 
determined by substituting the profile (\ref{profile}) into the action 
(\ref{Sdr'def}) and looking for its minimum with respect to $\delta_{Lq}$. 
This reads as
\beq
\cS= y \cF(\delta_{Lq}) + \frac{\delta_{Lq}^2\,\sigr^2}{2\sigq^2} .
\label{SdLq}
\eeq
Then, defining the variable $\tau$ and the function $\cG(\tau)$ by
\beq
\tau = -\delta_{Lq} \frac{\sigr}{\sigq} , \;\;\;
\cG(\tau) = \cF(\delta_{Lq}) = \eta_r ,
\label{taudef}
\eeq
the action (\ref{SdLq}) and its derivative read as
\beq
\cS= y \cG(\tau)+\frac{\tau^2}{2} , \;\;\;\;\; 
\frac{\pl \cS}{\pl\tau} = y \cG'+\tau .
\label{Stau}
\eeq
Therefore, since at leading order in the quasi-linear limit, the cumulant 
generating function $\varphi(y)$ is given by the minimum of the action 
$\cS[\delta_L]$ from Eq.(\ref{path1}), it is given by the implicit system
\beq
\varphi(y)= y \cG(\tau)+\frac{\tau^2}{2} \;\;\;\; \mbox{with} \;\;\;\;
\tau= - y \cG'(\tau) .
\label{varphiy}
\eeq
Thus, the generating function $\varphi(y)$ is also the Legendre transform of
the function $-\tau(\cG)^2/2$, as defined by
\beq
\varphi(y)= \min_{\tau}\left[y\cG(\tau)+\frac{\tau^2}{2}\right] 
= \min_{\cG}\left[y\cG+\frac{\tau(\cG)^2}{2}\right] .
\label{Legendre}
\eeq
To make sure that
the solution (\ref{varphiy}) is indeed relevant, we must check that it is
indeed a local minimum of the action (and not a maximum), in agreement with
(\ref{Legendre}) and the original path integral (\ref{path1}).
This directly follows from the expression (\ref{SdLdef}). Indeed, for 
$y=0$ the saddle-point obtained above is simply $\delta_L=0$, i.e. $\tau=0$,
and the Hessian of the action at this point is $\sigr^2 C_{\delta_L}^{-1}$
which is strictly positive. Then, by continuity, for small $y$ the  
Hessian around the saddle-point given by Eq.(\ref{varphiy}) is positive
which ensures that it is a local minimum. 
As we shall see below in section~\ref{Cumulant-generating-function},
for some cases it may only be a local minimum, but
the global minimum associated with finite density contrasts is irrelevant
in the quasi-linear limit: it corresponds to the tail of the distribution
$\cP(\eta_r)$ and it is exponentially suppressed in the limit
$\sigr\rightarrow 0$.

It is clear that the procedure described above must recover the results that
would be obtained for the leading-order term of the cumulants
$\lag\eta_r^p\rag_c$ (which is of order $\sigr^{2(p-1)}$ so that $\varphi(y)$
has indeed a finite quasi-linear limit in Eq.(\ref{phidef}))
from the perturbative expansion presented in 
section~\ref{Perturbative-expansion}. Indeed, in both cases we obtain
an expansion over powers of $\sigr^2$ (in the steepest-descent approach
subleading terms would be obtained from Eq.(\ref{path1}) by expanding the
action around its saddle-point and performing the Gaussian integrations),
as we actually start from the unperturbed solution $\delta_L=0$.
In fact, as shown for the case of the three-dimensional gravitational
dynamics \cite{Bernardeau1992},
it is possible to derive the quasi-linear generating function $\varphi(y)$
from the perturbative expansion (\ref{deltatp})-(\ref{thetatp}), using its
Taylor expansion (\ref{phidef}) and the leading-order term of each
cumulant $\lag\eta_r^p\rag_c$. This gives back $\varphi(y)$ as the solution
of the implicit system (\ref{varphiy}) for the unsmoothed case,
where one has $\cG(\tau)=\cF(\tau)$ (so that there is no dependence
on the initial conditions). 
Then, one can show that the same result is obtained in Lagrangian space,
where the perturbative expansions are built in terms of the Lagrangian
coordinates, and the shift from $\cF(\delta_L)$ to $\cG(\tau)$ as in
Eq.(\ref{taudef}) is obtained through a mapping from Lagrangian to
Eulerian space \cite{Bernardeau1994a}.

As shown in \cite{Valageas2002a} and described above, the steepest-descent
approach provides the quasi-linear generating function in a more direct
fashion. In particular, the integrations over angles, as in Eq.(\ref{etarp=3}),
are automatically included in the spherical dynamics, and the dependence
on $n$, associated with the mapping from Lagrangian to Eulerian coordinates
(through the ratio $\sigr/\sigq$ in Eq.(\ref{taudef})
that measures the ratio of initial power on scales $r$ and $q$)
is automatically provided by the form of the action $\cS[\delta_L]$.
In addition, the definition of $\varphi(y)$ as the Laplace transform
(\ref{path1}) allows to give a meaning to $\varphi(y)$ beyond the radius of
convergence of its Taylor series.
Finally, as discussed below, the profile (\ref{profile}) of the saddle-point
allows us to check the range of overdensity $\eta_r$ and Laplace conjugate $y$
to which these results apply. Indeed, they only hold as long as the
saddle-point has not formed shocks yet, which can only be checked from
the knowledge of Eq.(\ref{profile}).

\begin{figure}
\begin{center}
\epsfxsize=7.5 cm \epsfysize=5.8 cm {\epsfbox{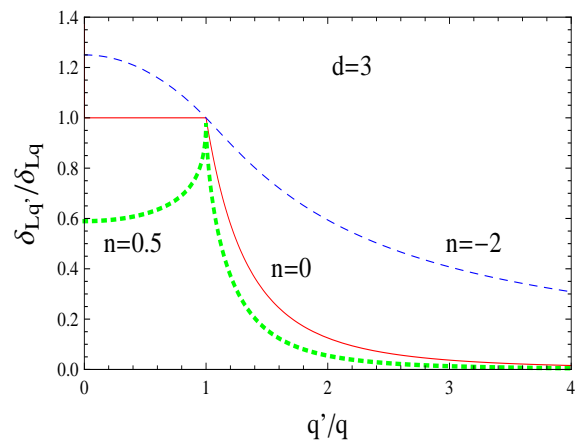}}
\end{center}
\caption{(Color online) 
The linear density profile of the spherical saddle-point for the 
cases $n=-2,0$ and $0.5$, in dimension $d=3$. This shows the integrated 
linear density contrast $\delta_{Lq'}$ within the sphere of radius $q'$,
from Eq.(\ref{profilen}), and $q$ is the initial Lagrangian radius of the
shell that is at the radius of interest $r$ at time $t$.}
\label{figprofile}
\end{figure}

\begin{figure}
\begin{center}
\epsfxsize=7.5 cm \epsfysize=5.8 cm {\epsfbox{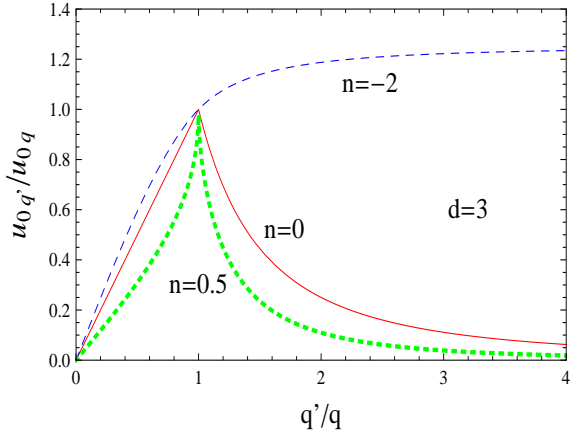}}
\end{center}
\caption{(Color online) 
The linear velocity profile of the spherical saddle-point for the 
cases $n=-2,0$ and $0.5$, in dimension $d=3$, as in Fig.~\ref{figprofile}. 
This shows the initial radial velocity $u_0(q')$ at Lagrangian radius $q'$.}
\label{figvel_profile}
\end{figure}

For the power-law power spectra (\ref{PdeltaL}) the radial linear profile 
(\ref{profile}) of the saddle-point reads as
\beq
\frac{\delta_{Lq'}}{\delta_{Lq}} = \left(\frac{2}{1+x}\right)^{n+3} 
\frac{_2F_1(\frac{n+3}{2},\frac{d+1}{2};d+1;\frac{4x}{(1+x)^2})}
{_2F_1(\frac{n+3}{2},\frac{d+1}{2};d+1;1)} ,
\label{profilen}
\eeq
with $x=q'/q$, see Eq.(\ref{Cdr1dr2}). 
This also gives the linear velocity profile through Eq.(\ref{u0rdef}).
We show in Figs.~\ref{figprofile}, \ref{figvel_profile}, the density and
velocity profiles obtained in the three-dimensional case for $n=-2,0$ and $0.5$.
For integer values of $n$ and odd $d+n$ the hypergeometric function in 
Eq.(\ref{profilen}) simplifies and we give in Table~\ref{Table_profile}
the explicit forms obtained for some low dimensional cases. Note that
the profile is singular at the Lagrangian radius $q$, associated with the
radius $r$ of the Eulerian cell. We also give in the last
row the simple profile obtained in the limit $d\rightarrow\infty$, where the 
singularity disappears. Note that in this infinite dimensional limit the profile
still depends on $n$, in agreement with the fact that the profile 
(\ref{profilen}) decays as $x^{-(n+3)}$ at large distance, independently of $d$.

\begin{table}
\begin{center}
\begin{tabular}{c|c||c|c||c|c}
& & \multicolumn{2}{c||}{$\frac{\delta_{Lq'}}{\delta_{Lq}}$} & \multicolumn{2}{c}{$\frac{u_{0q'}}{u_{0q}}$} \rule[-0.3cm]{0cm}{0.6cm} \\ \cline{3-6}
\; $n$ \; & \; $d$ \; & \,$x<1$\, & $x>1$ & $x<1$ & $x>1$ \rule[-0.15cm]{0cm}{0.5cm}\\ \hline\hline
0 & 3 & 1 & $\frac{1}{x^3}$ & $x$ & $\frac{1}{x^2}$ \rule[-0.25cm]{0cm}{0.6cm} \\ \hline
-1 & 2 & 1 & $\frac{1}{x^2}$ & $x$ & $\frac{1}{x}$ \rule[-0.25cm]{0cm}{0.6cm} \\ \hline
-2 & 1 & 1 & $\frac{1}{x}$ & $x$ & 1 \rule[-0.25cm]{0cm}{0.6cm} \\ \hline
-2 & 3 & $\frac{5-x^2}{4}$ & $\frac{5x^2-1}{4x^3}$ & \,$\frac{x(5-x^2)}{4}$\, & $\frac{5x^2-1}{4x^2}$ \rule[-0.25cm]{0cm}{0.7cm} \\  \hline
$n$ & $\infty$ &  \multicolumn{2}{c||}{\,$\left(\frac{1+x^2}{2}\right)^{-(n+3)/2}$\,} &  \multicolumn{2}{c}{\;$x\left(\frac{1+x^2}{2}\right)^{-(n+3)/2}$\,} \rule[-0.25cm]{0cm}{0.8cm} \\
\end{tabular}
\end{center}
\caption{The linear integrated-density and velocity profiles of the spherical 
saddle-point for some values of $n$ and $d$, where Eq.(\ref{profilen}) 
simplifies. Here $x=q'/q$, where $q$ is the Lagrangian radius associated with
the Eulerian radius of interest $r$. The profile is singular at $x=1$, except
in the limit of infinite dimension, $d\rightarrow\infty$, shown in the last row.}
\label{Table_profile}
\end{table}

As expected, for all values of $n$ in the range $-3<n<1$ that we consider in 
this paper, the density contrast vanishes at large distance. For $n\leq d-3$ it 
is monotonically decreasing but for $n>d-3$, which corresponds to significant
initial power at high wavenumbers, it shows a peak at radius $q$. 
On the other hand, the radial velocity vanishes at the center $q'=0$,
in agreement with spherical symmetry, but it only decays at large distance
for $n>-2$. For $n<-2$ it keeps growing at large distance (note that the
initial velocity field only shows homogeneous increments for $n<-1$, so that
this growth is not surprising).

\begin{table}
\begin{center}
\begin{tabular}{c||c|c}
$n$ & \,$\eta_r>1, \; \ctheta_r>0$\, & $\eta_r<1, \; \ctheta_r<0$  \rule[-0.15cm]{0cm}{0.5cm}\\ \hline\hline
$n>d-3$ & \multicolumn{2}{c}{shock as soon as $t\neq 0$} \rule[-0.22cm]{0cm}{0.6cm} \\ \hline
$-2<n\leq d-3$ & no shock & \,shock below a threshold \! \rule[-0.25cm]{0cm}{0.6cm} \\ \hline
$-3<n\leq -2$ & no shock & no shock \rule[-0.25cm]{0cm}{0.6cm} \\
\end{tabular}
\end{center}
\caption{This Table shows whether the saddle-point (\ref{profile}) forms a shock
after a finite time, which corresponds to a finite threshold for the density
$\eta_r$ or the velocity divergence $\ctheta_r$. We only consider the range
$-3<n<1$ and $d\geq 1$. If $n>d-3$ shocks form as soon as $t\neq 0$ so that
the saddle-point (\ref{profile}) is never valid (but it should give a reasonable
approximation if $n<d-2$).}
\label{Table_shocks}
\end{table}

To make sure that the saddle-point obtained above is relevant we must
check that no shocks have formed, so that Eq.(\ref{etarF}) is valid.
The naive Lagrangian map, $\bx=\bq+t\bu_0(\bq)$, shows that a shock
occurs when $\det(\pl \bx/\pl \bq)=0$, that is when $1+t\dd u_{0q'}/\dd q'=0$ for
the spherical saddle-point. The profiles show a singularity at radius $q$ 
of the form $|q'-q|^{d-n-2}$. Then, for $n>d-3$ the velocity has a spike at
radius $q$ with infinite left and right derivatives, so that shocks appear as 
soon as $t\neq 0$. For $n\leq d-3$, $|u_{0q'}|$ shows a sublinear (or linear) 
growth with $q'$ hence there will be no shock, except at the center, for 
overdense saddle-points (particles reach the center before 
$\dd u_{0q'}/\dd q'$ reaches $-1/t$).
For underdense saddle-points, a shock appears after a finite time for
$-2<n\leq d-3$, while no shocks form for $-3<n\leq-2$ since the radial
velocity grows with radius. 
We summarize in Table~\ref{Table_shocks} these behaviors associated with
different ranges of the index $n$ of the initial energy spectrum. 

Since in the quasi-linear limit we only probe small density fluctuations
we can use the saddle-point obtained above for $n\leq d-3$, as shocks
only appear after some finite time (or never). For $n>d-3$ we should modify the
saddle-point to take into account shocks. However, since for moderate
times and density fluctuations this should only change the profile close to
the Lagrangian radius $q$ and give small modifications to the quasi-linear 
generating function $\varphi(y)$ we shall keep Eq.(\ref{varphiy}) below for
$n<d-2$ (for $n\geq d-2$ where the linear variance $\sigr$ diverges it is not
possible to neglect shocks).
Note that for large $d$ this problem disappears, see also the last row
of Table~\ref{Table_profile}. 
For all cases shown in Table~\ref{Table_profile} the saddle-point obtained above
is relevant in the quasi-linear regime, as can be checked from the explicit forms
of the velocity profiles and in agreement with the previous discussion.

\subsubsection{Cumulant generating function $\varphi(y)$}
\label{Cumulant-generating-function}

Close to the origin $y=0$, the solution of the implicit system (\ref{varphiy})
always satisfies
\beq
\tau\rightarrow 0, \;\; y \rightarrow 0 : \;\; \cG\sim 1-\tau, \;\;\; 
y \sim \tau \;\;\; \mbox{and} \;\;\; \varphi \sim y -\frac{y^2}{2} .
\label{phiy0}
\eeq
Keeping only these low-order terms corresponds to the linear regime and gives 
back the Gaussian of variance $\sigr^2$ for the probability distribution 
$\cP(\eta_r)$ on very large scales and early times. 
This agrees with the fact that the initial conditions are Gaussian and it means
that over large scales or at early times, that is when $\sigr^2\ll 1$ and 
for $|\delta_r| \ll 1$, we recover linear theory and the probability 
distributions are still governed by the Gaussian initial conditions.
Of course, this breaks down for the cases $d-2<n<1$ where the linear variance
$\sigr^2$ itself is not well defined, so that even on large scales the 
probability distributions are strongly non-Gaussian, as seen for instance
in \cite{Avellaneda1995,AvellanedaE1995,Frachebourg2000,Valageas2009} 
for the case $\{n=0,d=1\}$.

For the case $n\leq d-3$, where the quasi-linear regime considered in this section
applies, the deviations from the Gaussian, associated with higher-order 
cumulants, appear for finite density contrast $\delta_r$, that is for finite $y$
and $\tau$. Note that this corresponds to rare events, as expected for a 
steepest-descent approach to be valid.
Thus, in the quasi-linear limit we are sensitive to small but finite
values of $\{\delta_r,y,\tau\}$ around zero. Moreover, at leading order
the moments and cumulants of the overdensity are given by the expansion
(\ref{phidef}) of $\varphi(y)$ around the origin. In particular, the solution 
(\ref{varphiy}) directly gives the leading order value of $\lag\eta_r^p\rag_c$,
that can also be derived from the perturbative expansion of the equations
of motion (\ref{Burgers})-(\ref{continuity}) described in
section~\ref{Perturbative-expansion}.

The previous results are valid for any initial energy spectrum, provided
$\sigr$ is well defined. For the power-law power spectra (\ref{PdeltaL}),
using Eq.(\ref{sig2def}), we obtain from the definition (\ref{taudef}),
\beq
\tau = - \delta_{Lq} \left(\frac{r}{q}\right)^{-\frac{n+3}{2}} = 
- \delta_{Lq} \, \eta_r^{\frac{n+3}{2d}} ,
\label{taudL}
\eeq
whence
\beq
\cG(\tau) = \cF\left(-\tau\cG^{-\frac{n+3}{2d}}\right) = 
\left(1+\frac{\tau}{d}\cG^{-\frac{n+3}{2d}}\right)^{-d} .
\label{Gtau}
\eeq
In terms of the inverse function $\tau(\cG)$ this reads as
\beq
\tau(\cG) = d \left(\cG^{\frac{n+1}{2d}} - \cG^{\frac{n+3}{2d}} \right) ,
\label{tauG}
\eeq
so that $\varphi(y)$ is given by the parametric representation
\beqa
\varphi & = & \frac{d}{2}\left[ (d\!-\!n\!-\!1)\cG^{\frac{n+1}{d}} 
- 2 (d\!-\!n\!-\!2)\cG^{\frac{n+2}{d}} \right.  \nonumber \\
&& \left. \hspace{0.5cm} + (d\!-\!n\!-\!3)\cG^{\frac{n+3}{d}} \right] ,
\label{phiG}
\eeqa
\beq
y= -\frac{d}{2} \cG^{-1} \left[ (n\!+\!1)\cG^{\frac{n+1}{d}} 
- 2 (n\!+\!2)\cG^{\frac{n+2}{d}} + (n\!+\!3)\cG^{\frac{n+3}{d}} \right] .
\label{yG}
\eeq
Expanding around $\tau=0$, $y=0$ and $\cG=1$, we obtain the series 
expansion of $\varphi(y)$. Comparing with Eq.(\ref{phidef}) we 
obtain for the third and fourth-order cumulants in the quasi-linear limit,
\beq
\sigr\rightarrow 0 : \;\;\;\; 
S_3=\frac{\lag\eta_r^3\rag_c}{\lag\eta_r^2\rag_c^2} = 3\frac{d-n-2}{d} ,
\label{S3}
\eeq
which agrees with (\ref{S3pert}) for $d=3$, and
\beq
S_4=\frac{\lag\eta_r^4\rag_c}{\lag\eta_r^2\rag_c^3} = 
\frac{83+16d^2+84n+21n^2-36d(n+2)}{d^2} .
\label{S4}
\eeq
Note that although $S_3$ and $S_4$ as defined above are called the
skewness and the kurtosis in the cosmological litterature, they are
not exactly the skewness and the kurtosis defined in standard probability
theory, the latter being defined as $\lag\eta_r^3\rag_c/\lag\eta_r^2\rag_c^{3/2}$
and $\lag\eta_r^4\rag_c/\lag\eta_r^2\rag_c^2$. The reason for the use of 
(\ref{S3})-(\ref{S4}) is that these quantities have a finite value in the
quasi-linear limit discussed above (as seen in \cite{Valageas2009} they also
have a finite value in the small-scale limit, associated with the highly nonlinear
regime).

\begin{table*}
\begin{center}
\begin{tabular}{c|c||c|c|c||c|c|c}
\; $n$ \; & \; $d$ \; & $\varphi(y)$ & $y$ & $\cG=\eta_r$ & $\varphi(y)$ & $y$ & $\cG=\eta_r$ \rule[-0.25cm]{0cm}{0.6cm} \\ \hline\hline
0 & 3 & $\frac{27-9\sqrt{9-6y}}{(6-\sqrt{9-6y})^2}$ & $-\frac{9}{2}<y<\frac{3}{2}$ & $\frac{1}{8}<\cG<\infty$ & $\frac{27+9\sqrt{9-6y}}{(6+\sqrt{9-6y})^2}$ & $-\infty<y<\frac{3}{2}$ & $0<\cG<\frac{1}{8}$ \rule[-0.3cm]{0cm}{0.7cm} \\ \hline
-1 & 2 & $\frac{y}{1+y/2}$ & $-2<y<\infty$ & $0<\cG<\infty$ & & & \rule[-0.25cm]{0cm}{0.6cm} \\ \hline
-2 & 1 & $\sqrt{1+2y}-1$ & $-\frac{1}{2}<y<\infty$ & $0<\cG<\infty$ & & & \rule[-0.25cm]{0cm}{0.7cm} \\ \hline
-2 & 3 & \,$\frac{6^{3/2}+3\sqrt{6+16y}}{\sqrt{3+\sqrt{9+24y}}} - 9$\, & \,$-\frac{3}{8}<y<\infty$\, & \,$0<\cG<2\sqrt{2}$\, & \,$\frac{6^{3/2}-3\sqrt{6+16y}}{\sqrt{3-\sqrt{9+24y}}} - 9$\, & \,$-\frac{3}{8}<y<0$\, & \,$2\sqrt{2}<\cG<\infty$ \rule[-0.38cm]{0cm}{0.9cm} \\ \hline\hline
$n$ & $\infty$ & \multicolumn{6}{c}{$\varphi=\tau+\frac{\tau^2}{2}, \;\; y=\tau e^{\tau} , \;\; \cG= e^{-\tau}   ; \;\;\;  -\frac{1}{e}<y<\infty , \;\;0 <\cG<e ; \;\;\; -\frac{1}{e}<y<0 , \;\; e<\cG<\infty$} \rule[-0.2cm]{0cm}{0.6cm} \\
\end{tabular}
\end{center}
\caption{The cumulant generating function $\varphi(y)$ of the overdensity
$\eta_r$, in the quasi-linear limit $\sigr\rightarrow 0$, for a few values
of $n$ and $d$, where explicit solutions of the system (\ref{phiG})-(\ref{yG})
can be obtained. The columns 3-5 show the quasi-linear branch, and the 
associated range of $\{y,\cG\}$ that contains the point $\{0,1\}$. The columns
6-8 show the second branch, associated with very rare events, that appears
in some cases. The last row shows the infinite-dimensional limit, 
$d\rightarrow \infty$, which no longer depends on $n$ but has no explicit form
and shows two branches.}
\label{Table_varphi}
\end{table*}

For small integer values of $n$ and $d$ we can obtain explicit expressions
for the solution of the implicit system (\ref{phiG})-(\ref{yG}), by solving 
for $\cG(y)$ and substituting into $\varphi(\cG)$.
We give in Table~\ref{Table_varphi} our results for a few such cases.
Note that from the meaning of the variable $\cG$ as the overdensity $\eta_r$
within the radius $r$ for the spherical saddle-point, see Eq.(\ref{taudef}),
$\varphi(y)$ is a priori determined by Eqs.(\ref{phiG})-(\ref{yG}) by letting
$\cG$ vary over the range $0<\cG<\infty$.
For the cases $\{n=-1,d=2\}$ and $\{n=-2,d=1\}$ we obtain a generating
function $\varphi(y)$ that shows a singularity $y_s$ on the negative real axis,
with $y_s=-2$ or $-1/2$, and the range $y>y_s$ corresponds to the full range 
$\cG>0$. In the complex plane, there is usually a branch cut for $y<y_s$
or a pole at $y_s$. For the cases $\{n=0,d=3\}$ and $\{n=-2,d=3\}$ it happens
that the function $y(\cG)$ is no longer monotonic over $0<\cG<\infty$
so that the inverse $\cG(y)$ is bivaluate and $\varphi(y)$ shows two branches.
We show in columns 3-5 of Table~\ref{Table_varphi} the quasi-linear branch,
that contains the point $\{y=0,\cG=1,\varphi=0\}$ and corresponds to moderate 
density fluctuations. Columns 6-8 show the second branch that corresponds to
large fluctuations (very low densities, $0<\cG<1/8$, for $\{n=0,d=3\}$;
very high densities, $2\sqrt{2}<\cG<\infty$, for $\{n=-2,d=3\}$).

\begin{figure}
\begin{center}
\epsfxsize=7.5 cm \epsfysize=5.8 cm {\epsfbox{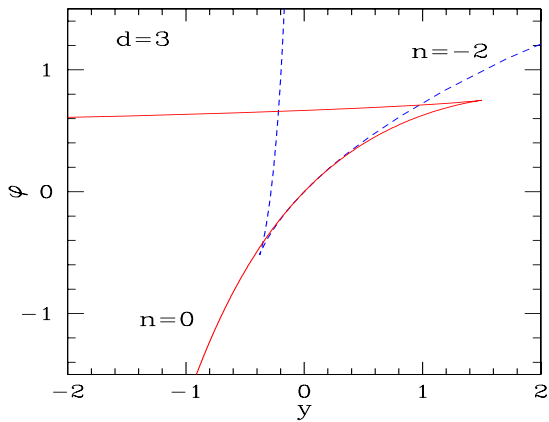}}
\end{center}
\caption{(Color online)
The density cumulant generating function $\varphi(y)$ in dimension $d=3$
for the power indices $n=0$ (solid line) and $n=-2$ (dashed line),
from rows 2 and 5 of Table~\ref{Table_varphi}.
In both cases there is a singularity on the real axis, with $y_s=3/2$ for
$n=0$ and $y_s=-3/8$ for $n=-2$. The branch that runs through the origin is
associated with moderate density fluctuations and is the relevant one
for the expansion (\ref{phidef}) in terms of cumulants at leading order in the
quasi-linear limit.}
\label{figphiy}
\end{figure}

For the general case, the behavior of the cumulant generating function 
$\varphi(y)$ defined by the implicit system (\ref{varphiy}) and the presence
of singularities can be obtained from the asymptotic behaviors at large and
small overdensities $\cG$, using Eqs.(\ref{tauG})-(\ref{yG}).
For large densities we obtain
\beqa
\cG\rightarrow+\infty & : & \tau \sim -d \cG^{\frac{n+3}{2d}} , \;\;\;
y \sim -\frac{d(n+3)}{2} \cG^{\frac{n+3-d}{d}} , \nonumber \\ 
&& \varphi \sim 
\frac{d(d-n-3)}{2}\left[\frac{-2y}{d(n+3)}\right]^{\frac{n+3}{n+3-d}} .
\label{largeG}
\eeqa
Thus, we have two possible behaviors for $\cG\rightarrow+\infty$,
\beq
n>d-3 : \;\;\; \tau \rightarrow -\infty, \;\;  y \rightarrow -\infty, \;\;
\varphi \rightarrow -\infty ,
\label{n+3-d>0}
\eeq
\beq
n<d-3 : \;\;\; \tau \rightarrow -\infty, \;\;  y \rightarrow 0^-, \;\;
\varphi \rightarrow +\infty .
\label{n+3-d<0}
\eeq
As explained in section~\ref{Spherical-saddle-point} and 
Table~\ref{Table_shocks}, the saddle-point approach studied here only exactly
applies to $n\leq d-3$ as shocks
form for $n>d-3$. However, we mention the case $n>d-3$ in (\ref{n+3-d>0})
because this method should still provide a reasonable approximation for
$d-3<n<d-2$. The case $n=d-3$ shows an intermediate behavior, as
in the limit $\cG\rightarrow+\infty$ we obtain $y\rightarrow-d^2/2$, and
$\varphi\rightarrow -1$ if $d=1$ or $\varphi\rightarrow -\infty$ if $d>1$.
Thus it is closer to the case $n>d-3$.
Then, we can see that for $n\geq d-3$ larger densities are associated with more 
negative $y$ and $\varphi$ and the function $\varphi(y)$ is regular and 
monotonically increasing over $]-\infty,0]$ (or $]-d^2/2,0]$).
This behavior is shown by the case $\{n=0,d=3\}$ in Fig.~\ref{figphiy}.
For $n<d-3$, since from (\ref{phiy0}) we have $y=0$ at $\cG=1$, 
the limit $y\rightarrow 0^-$ for large densities implies that
the function $y(\cG)$ is not monotonic over $\cG\in[1,+\infty[$ and shows 
a minimum $y_s<0$ at some value $\cG_s>1$. Around this point we have 
$y-y_s \propto (\cG-\cG_s)^2$.
This gives rise to a square-root singularity $\sqrt{y-y_s}$ for the function
$\varphi(y)$, which shows two branches going from this point.
A first branch goes through the point $\{y=0,\varphi=0\}$, it is the branch
associated with moderate fluctuations, below $\cG_s$, that is most relevant in 
the quasi-linear limit. The second branch is associated with large overdensities
above $\cG_s$. This behavior is shown by the case $\{n=-2,d=3\}$ 
in Fig.~\ref{figphiy}.

For low densities we obtain
\beqa
\cG\rightarrow 0 & : & \tau \sim d \cG^{\frac{n+1}{2d}} , \;\;\;
y \sim -\frac{d(n+1)}{2} \cG^{\frac{n+1-d}{d}} , \nonumber \\ 
&& \varphi \sim 
\frac{d(d-n-1)}{2}\left[\frac{-2y}{d(n+1)}\right]^{\frac{n+1}{n+1-d}} .
\label{smallG}
\eeqa
Since we assumed $n<d-2$, so that the linear variance $\sigr^2$ is well defined,
we have $n+1-d<-1$ and this gives rise to the two behaviors:
\beq
n<-1 : \;\;\; \tau \rightarrow +\infty, \;\;  y \rightarrow +\infty, \;\;
\varphi \rightarrow +\infty ,
\label{n<-1}
\eeq
\beq
n>-1 : \;\;\; \tau \rightarrow 0^+, \;\;  y \rightarrow -\infty, \;\;
\varphi \rightarrow 0^+ .
\label{n>-1}
\eeq
Thus, for $n<-1$ the function $\varphi(y)$ is regular and monotonically 
increasing over $[0,+\infty[$ (case $\{n=-2,d=3\}$ in Fig.~\ref{figphiy})
while for $n>-1$ it shows a square-root 
singularity at some finite value $y_s>0$, associated with an underdensity
$\cG_s<1$, from which two branches leave (case $\{n=0,d=3\}$ 
in Fig.~\ref{figphiy}).

In the large-dimension limit, $d\rightarrow\infty$, we obtain from 
Eqs.(\ref{FdLq}), (\ref{taudL}), $\cF(\delta_L)= e^{\delta_L}$ and
$\tau=-\delta_L$. This gives the parametric representation of $\varphi(y)$
shown in the last row of Table~\ref{Table_varphi}. Since $0<\cG<\infty$
corresponds to $-\infty<\tau<\infty$ and $y(\tau)$ has a minimum at $\tau_s=-1$
this generating function $\varphi(y)$ shows two branches. The comparison
with appendix A of \cite{Valageas2002a} shows that in the quasi-linear limit
this leads to a log-normal distribution $\cP(\eta_r)$ (contrary
to some statistical models, used to describe fully developed turbulence, this
is not related to some underlying multiplicative cascade process).
Note that the dependence of $\varphi(y)$ and $\cP(\eta_r)$ on $n$ disappears
in this limit $d\rightarrow\infty$, even though the density profile of the
associated saddle-point keeps a dependence on $n$, see last row of 
Table~\ref{Table_profile}.

As noticed in \cite{Valageas2008}, it happens that in the case $\{n=-2,d=1\}$
the quasi-linear result shown in the fourth row in Table~\ref{Table_varphi}
actually gives the exact cumulant generating function defined 
as $\overline{\varphi}(y)= -\sum S_p (-y)^p/p!$, with 
$S_p=\lag\eta_r^p\rag_c/\lag\eta_r^2\rag_c^{(p-1)}$.
Note that for the quasi-linear limit we defined $\varphi(y)$ in 
Eq.(\ref{phidef}) using $\sigr^2$ instead of $\lag\eta_r^2\rag_c$ in the 
coefficients $S_p$, which is only equivalent at leading order.
In this case $\{n=-2,d=1\}$ we actually have the exact equality 
$\lag\eta_r^2\rag_c=\sigr^2$, see \cite{Valageas2008}.
For generic cases we expect the exact generating function $\overline{\varphi}(y)$
and the variance $\lag\eta_r^2\rag_c$ to deviate at small scales or late times 
from their quasi-linear limits $\varphi(y)$ and $\sigr^2$.

\begin{figure}
\begin{center}
\epsfxsize=7.5 cm \epsfysize=5.8 cm {\epsfbox{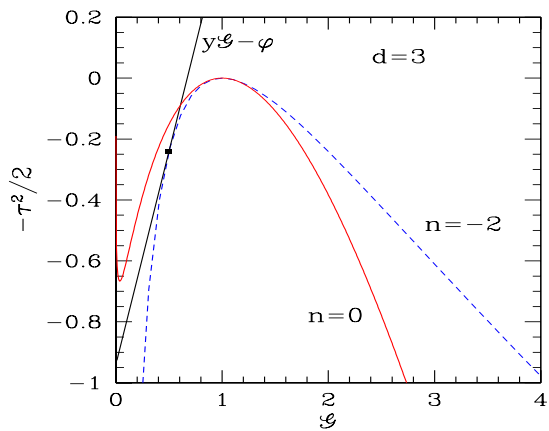}}
\end{center}
\caption{(Color online)
The Legendre transformation (\ref{Legendre}) of the curve 
$-\tau^2(\cG)/2$, which gives the generating function $\varphi(y)$.
The first-contact line $y\cG+c$, of fixed slope $y$ and height $c$ decreasing
from $+\infty$, with the curve $-\tau^2(\cG)/2$, intersects the vertical axis
at $(0,-\varphi)$ (i.e. for $c=-\varphi$). We show the cases $n=0$ and $n=-2$
in three dimensions.}
\label{figLeg}
\end{figure}

We display in Fig.~\ref{figphiy} the cumulant generating function $\varphi(y)$ 
that we obtain in dimension $d=3$ for the two indices $n=0$ and $n=-2$.
This provides an illustration of all the behaviors (\ref{n+3-d>0})-(\ref{n>-1}).
The appearance of these singular behaviors can also be seen from the geometrical
construction of the Legendre transform (\ref{Legendre}), that we show in 
Fig.~\ref{figLeg} for these three-dimensional cases, $n=0$ and $n=-2$.
For a given $y$, $-\varphi$ is obtained as the intercept on the vertical axis
of the first-contact straight line $y\cG+c$, of slope $y$, with the curve 
$-\tau^2(\cG)/2$, decreasing its height $c$ from $+\infty$.
Thus, the Legendre transform (\ref{Legendre}) follows the concave hull of
the function $-\tau^2(\cG)/2$ and it is regular if the latter is concave over 
$0<\cG<\infty$. 
Note that this is obviously the case in the linear regime where $\tau=1-\cG$. 

For $n=0$ we have $\tau(0)=0$ (the curve $-\tau^2(\cG)/2$ shows a steep up-turn
at very low $\cG$ in Fig.~\ref{figLeg}) so that for $y>0$ the global minimum is 
$\tau=0$: the point in the range $\cG_s<\cG<1$ with a tangent of slope $y>0$ is 
only a local minimum and there is a local maximum in the range $0<\cG<\cG_s$. 
The local minimum corresponds to the regular branch in Fig.~\ref{figphiy} and 
Table~\ref{Table_varphi}, that runs through $\varphi(0)=0$, while the local maximum
corresponds to the second branch in Fig.~\ref{figphiy} and 
Table~\ref{Table_varphi}.
For $n=-2$ we have $\tau(\cG)^2\propto \cG^{1/3}$ at large overdensities, from
Eq.(\ref{largeG}). Then, for $y_s<y<0$ we again have a local minimum, with
$1<\cG<\cG_s$, and a local maximum, with $\cG>\cG_s$.

In the quasi-linear limit, $\sigr\rightarrow 0$, such large density fluctuations
are exponentially suppressed by a term of order $e^{-\tau^2/(2\sigr^2)}$, as
seen in Eq.(\ref{Petatau}) below, so that
it is sufficient to define the generating function by the branch that runs 
through the origin.

\subsubsection{Probability distribution $\cP(\eta_r)$}
\label{Probability-distribution}

Finally, from the cumulant generating function $\varphi(y)$ we obtain 
through an inverse Laplace transform the
probability distribution $\cP(\eta_r)$ in the quasi-linear limit. Using
Eqs.(\ref{PetarPsi}),(\ref{psiphidef}), we have
\beq
\cP(\eta_r) = \inta\frac{\dd y}{2\pi\ii\sigr^2} \, 
e^{[y\eta_r-\varphi(y)]/\sigr^2} .
\label{Petaphi}
\eeq
It is  best to compute the integral (\ref{Petaphi}) exactly, using the
branch of $\varphi(y)$ that runs through $\varphi(y)=0$ in case this
function is multivalued (then it only applies to some range of overdensities
$\eta_r$ around unity). However, in the quasi-linear limit at fixed $\eta_r$,
it is again possible to evaluate the distribution $\cP(\eta_r)$ from a 
steepest-descent method. The saddle-point of the exponent in (\ref{Petaphi})
is given by $\eta_r=\varphi'(y)$. On the other hand, from (\ref{varphiy})
we have $\varphi'(y)=\cG$, whence $\cG=\eta_r$. Therefore, as expected
the probability distribution $\cP(\eta_r)$ at point $\eta_r$ is governed
by the saddle-point described in section~\ref{Spherical-saddle-point}
such that its overdensity $\cG$ is equal to $\eta_r$, and we obtain
from Eqs.(\ref{varphiy}), (\ref{taudef}),
\beq
\cP(\eta_r) \sim e^{-\tau(\eta_r)^2/(2\sigr^2)} 
= e^{-\delta_{Lq}^2/(2\sigq^2)} .
\label{Petatau}
\eeq
Thus, in the saddle-point approximation associated with the quasi-linear
limit there is a precise correspondence between the overdensity $\eta_r$
and its Laplace conjugate $y$, and the variable $\tau$ expresses the Gaussian 
weight of the initial velocity (or linear density contrast) as shown by
Eq.(\ref{Petatau}). The nontrivial relation $\tau(\eta_r)$ describes both
the evolution (\ref{etarF}) of the density of a Lagrangian region,
which only depends on the dimension $d$, and the effect (\ref{taudef}),
(\ref{taudL}), of the change of size from $q$ to $r$, which involves the
initial power-spectrum index $n$ through the dependence of initial power on
scale. In particular, for the power-law power spectra (\ref{PdeltaL}),
using Eq.(\ref{tauG}), Eq.(\ref{Petatau}) reads as
\beq
\sigr\rightarrow 0 : \;\; \ln\cP(\eta_r) \sim -\frac{d^2}{2\sigr^2}
\left( \eta_r^{\frac{n+1}{2d}}-\eta_r^{\frac{n+3}{2d}}\right)^2 .
\label{Petataun}
\eeq
In the large dimensional limit, $d\rightarrow\infty$, we have seen that
$\tau(\eta_r)=-\ln(\eta_r)$ (last row of Table~\ref{Table_varphi}),
whence
\beq
d\rightarrow\infty , \;\; \sigr\rightarrow 0 : \;\;
\ln\cP(\eta_r) \sim -\frac{\ln^2(\eta_r)}{2\sigr^2} .
\label{Petadinf}
\eeq
Note however that Eqs.(\ref{Petataun})-(\ref{Petadinf}) only hold for 
densities $\eta_r$ such that the saddle-point (\ref{profile}) has not formed
shocks yet.
As discussed in section~\ref{Spherical-saddle-point} and summarized in
Table~\ref{Table_shocks}, this implies $n\leq d-3$ and it gives a lower bound 
for $\eta_r$ if $n>-2$. These lower bounds are given by the last column in
Table~\ref{Table_Peta_ql} and they will be obtained in
section~\ref{Saddle-point-with-shocks} below.

\begin{table}
\begin{center}
\begin{tabular}{c|c||c|c}
\,$n$\, & \,$d$\, & $\ln\cP(\eta_r)$ & $\eta_r$ \rule[-0.2cm]{0cm}{0.5cm} \\ \hline\hline
0 & 3 & \,$-\frac{r^3}{2t^2}\,\left(\eta_r^{1/6}-\eta_r^{1/2}\right)^2$\, & \,$\eta_r>(2/3)^3$ \rule[-0.35cm]{0cm}{0.9cm} \\ \hline
-1 & 2 & $-\frac{r^2}{2t^2}\,\left(1-\eta_r^{1/2}\right)^2$ & $\eta_r>1/4$ \rule[-0.35cm]{0cm}{0.9cm} \\ \hline
-2 & 1 & $-\frac{r}{2t^2}\,\left(\eta_r^{-1/2}-\eta_r^{1/2}\right)^2$ & $\eta_r>0$ \rule[-0.35cm]{0cm}{0.9cm} \\ \hline
-2 & 3 & \;$-\frac{5r}{8t^2}\,\left(\eta_r^{-1/6}-\eta_r^{1/6}\right)^2$\; & $\eta_r>0$ \rule[-0.35cm]{0cm}{0.9cm} \\ \hline
n & \,$\infty$\, & $-\frac{(2r^2)^{\frac{n+3}{2}}}{2t^2}\ln^2(\eta_r)$ & \rule[-0.25cm]{0cm}{0.85cm} \\
\end{tabular}
\end{center}
\caption{Asymptotic behavior of the probability distribution $\cP(\eta_r)$
in the quasi-linear regime $\sigr\rightarrow 0$ (i.e. $t\rightarrow 0$ or
$r\rightarrow\infty$), for the initial conditions of Table~\ref{Table_correl}.
The last column shows the range of overdensities $\eta_r$ where these results
apply. If the lower threshold is not zero (i.e. $-2<n\leq d-3$, 
see Table~\ref{Table_shocks}), it means that the 
spherical saddle-point forms shocks for lower densities.}
\label{Table_Peta_ql}
\end{table}

As seen from the last expression (\ref{Petatau}), the tails of the probability
distribution $\cP(\eta_r)$ are simply governed at leading order by the
initial Gaussian weight $e^{-\delta_{Lq}^2/(2\sigq^2)}$ of the associated
initial fluctuation $\delta_{Lq}$ at the Lagrangian scale $q$. 
In fact, Eq.(\ref{Petatau}) could be directly obtained from a Lagrange multiplier
method, without introducing the generating function $\varphi(y)$. Indeed,
in the rare-event limit we may write
\beq
\mbox{rare events}: \;\;  \cP(\eta) \sim \max_{\{\delta_L[\bq]{\displaystyle |}
\eta_r[\delta_L]=\eta\}} e^{-\frac{1}{2} \delta_L.C_L^{-1}.\delta_L} .
\label{rareP}
\eeq
That is, $\cP(\eta)$ is governed by the maximum of the Gaussian weight
$e^{-(\delta_L.C_L^{-1}.\delta_L)/2}$ subject to the constraint 
$\eta_r[\delta_L]=\eta$ (assuming there are no degenerate maxima).
Then, we can obtain this maximum by minimizing the action 
$\cS[\delta_L]/\sigr^2$ of Eq.(\ref{SdLdef}), where $y$ plays the role of
a Lagrange multiplier. This gives the saddle-point (\ref{profile}), and
the amplitude $\delta_{Lq}$ and the radius $q$ are directly obtained from the
constraint $\eta=\cF(\delta_{Lq})$, as in Eq.(\ref{etarF}). Then, we do not
need the explicit expression of the Lagrange multiplier $y$, as this is
sufficient to obtain the last expression of the asymptotic tail (\ref{Petatau}).
Nevertheless, it is useful to introduce the generating function $\varphi(y)$,
which makes it clear that the Lagrange multiplier $y$ is also the Laplace
conjugate of the nonlinear overdensity $\eta$ as in Eq.(\ref{Psidef}),
since it is also of interest by itself, as it yields the density cumulants
through the expansion (\ref{phidef}). Moreover, it is easier to check
through the action $\cS$ and the generating function $\varphi(y)$ that the
path integral (\ref{path1}) is indeed dominated by a saddle-point. On the
other hand, as noticed above, in the quasi-linear regime it is best to compute
the distribution $\cP(\eta_r)$ from the integral (\ref{Petaphi}), expressed in
terms of $\varphi(y)$, as the property $\varphi(0)=0$ automatically ensures
that the probability distribution is properly normalized to unity (in the
case of the gravitational dynamics this has been seen to give a good match
with numerical results \cite{Valageas2002a,Bernardeau2002}).

We can note that at this
order the nonlinear distribution $\cP(\eta_r)$ could be described by a
spherical-dynamics model, where one makes the approximation 
$\cP(\eta_r)\dd\eta_r = \cP_L(\delta_{Lq})\dd\delta_{Lq}$ with
$\eta_r=\cF(\delta_{Lq})$ and $\cP_L$ is the initial distribution of the
linear density contrast, as developed for instance in \cite{Valageas1998}
for the collisionless gravitational dynamics. 
Note that such a phenomenological model can be readily extended to non-Gaussian
initial conditions. However, one needs the steepest-descent framework
described in the previous sections to justify the behavior (\ref{Petatau})
for the rare-event tails. Moreover, in cases where collisions (shocks)
take place, such a phenomenological model becomes ambiguous, while the
saddle-point approach remains valid and allows to derive exact results,
as we shall describe in section~\ref{Saddle-point-with-shocks} below.

We show in Table~\ref{Table_Peta_ql} the asymptotic behaviors (\ref{Petataun})
obtained for the initial conditions given in Table~\ref{Table_correl}, as well
as the lower bound $\eta_*$ below which the saddle-point (\ref{profile}) forms
shocks. The value of $\eta_*$ will be derived in section~\ref{Asymptotic-tails}
below, where we take into account shocks. Note that it is not related to
the value $\cG_s$ where the cumulant generating function is singular,
which was given in Table~\ref{Table_varphi}.

\begin{figure}
\begin{center}
\epsfxsize=7.5 cm \epsfysize=5.8 cm {\epsfbox{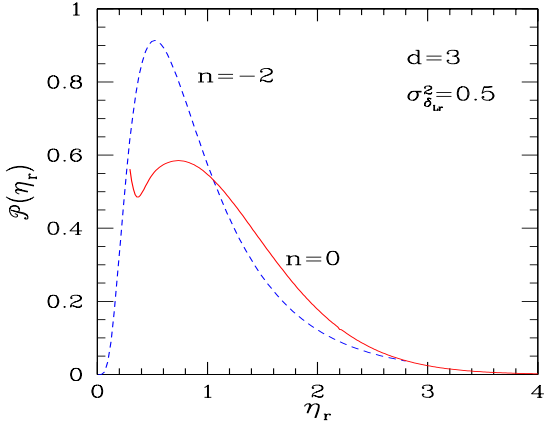}}
\end{center}
\caption{(Color online)
The probability distribution $\cP(\eta_r)$ in the quasi-linear
limit from Eq.(\ref{Petaphi}). We show the cases $n=0$ and $n=-2$
in three dimensions, for a linear variance $\sigr^2=0.5$.}
\label{figPeta}
\end{figure}

We show the distribution $\cP(\eta_r)$ in Fig.~\ref{figPeta} for a linear
variance $\sigr^2=0.5$ and the cases $n=0$ and $n=-2$ in $d=3$.
For finite $\sigr$ it is better to use the integral (\ref{Petaphi}),
rather than the asymptotic result (\ref{Petatau}), as
it ensures that the distribution is normalized to unity and captures the
asymmetry of the distribution with the shift of its peak.
Note that in the case $\{n=-2,d=1\}$ the inverse Laplace transform 
(\ref{Petaphi}) of the quasi-linear generating function given in the fourth row
of Table~\ref{Table_varphi} gives the explicit expression
\beq
n=-2, d=1 : \; \cP(\eta_r)= \frac{\eta_r^{-3/2}}{\sqrt{2\pi}\sigr}
\, e^{-(\sqrt{\eta_r}-\frac{1}{\sqrt{\eta_r}})^2/(2\sigr^2)} .
\label{PBrown}
\eeq
Again, as seen in \cite{Valageas2008}, it happens that in this case the result 
(\ref{PBrown}) is actually exact. 
In generic cases, deviations from the quasi-linear limiting distribution should 
appear at small scales and late times.

From the geometrical construction described in Fig.~\ref{figLeg} we can see
that singularities for the function $\varphi(y)$ occur at inflexion points
of the curve $-\tau^2(\cG)/2$. In particular, in agreement with the analysis
of Eqs.(\ref{largeG})-(\ref{n+3-d<0}), at large densities a singularity
appears as soon as $\tau^2/\cG\rightarrow 0$ for $\cG\rightarrow\infty$,
which implies that $-\tau^2(\cG)/2$ is no longer concave at large $\cG$.
From Eq.(\ref{Petatau}) this simply corresponds to a sub-exponential 
large-density tail of the form $\sim e^{-\eta_r^{\alpha}}$ with $\alpha<1$.
Then, it is clear that the integral (\ref{Psidef}) is divergent for $y<0$
so that the exact cumulant generating function has a branch cut on the
negative real axis, $y<0$, even though the cumulants of all orders may be 
finite. In this case, the singularity at $y_s<0$ and the two branches
observed on $y_s<y<0$ are related to this branch cut and to this
sub-exponential large-density tail.

For the low-density tail, this construction shows that when $\tau(\cG)$
and $\tau'(\cG)$ remain finite for $\cG\rightarrow 0$, a singularity appears
on the positive real axis, $y_s>0$, or $\varphi(y)$ is restricted to a finite
range $y<y_s$. Thus, singularities on the positive real axis are generically
associated with distributions that do not vanish in the limit 
$\eta_r\rightarrow 0$.

In any case, in the quasi-linear limit the large-fluctuation regime associated
with the second branch of $\varphi(y)$ is irrelevant, as it is exponentially
suppressed by a factor of the form $e^{-1/\sigr^2}$, see Eq.(\ref{Petatau}).
Therefore, we only plot
in Fig.~\ref{figPeta} the range associated with the quasi-linear branch
of $\varphi(y)$, that is, $\eta_r<2\sqrt{2}$ for $n=-2$, 
see Table~\ref{Table_varphi}. For $n=0$, this would require $\eta_r>1/8$,
but as shown in Table~\ref{Table_Peta_ql} and explained in 
section~\ref{Casen0d3} below, shocks appear before this threshold, as soon
as $\eta_r<(2/3)^3$, below which Eq.(\ref{Petataun}) is no longer valid. Therefore,
in the case $\{n=0,d=3\}$, we only plot the result (\ref{Petaphi}), obtained
from the quasi-linear generating function, above this lower-density bound,
$\eta_r>(2/3)^3$. In the limit $\sigr\rightarrow 0$ the weight associated
with such low-density regions decays exponentially as $e^{-1/\sigr^2}$
(disregarding the numerical factor), but at $\sigr^2=0.5$ this region is already
non-negligible. Note that the upturn at low density shown in Fig.~\ref{figPeta}
is not necessarily a signature of the breakdown of the quasi-linear limit
in this domain at $\sigr^2=0.5$. Indeed, as discussed in
section~\ref{Casen0d3}, the distribution $\cP(\eta_r)$ does not decay
exponentially at low density, see also Table~\ref{Table_tailshock}, and
it may even grow as a power law. As shown in \cite{Valageas2009}, this is
for instance what happens for the case $\{n=0,d=1\}$, where at low density
the probability distribution shows an inverse square root tail,
$\cP(\eta_r) \propto 1/\sqrt{\eta_r}$.

\subsection{Velocity divergence (i.e. spherical velocity increment)}
\label{Velocity-divergence}

We now consider the probability distribution, $\cP(\ctheta_r)$, of the mean
velocity divergence $\ctheta_r$, defined in Eq.(\ref{cthetadef}),
that is also the velocity increment over distance $2r$ averaged over
all directions (up to a normalization factor).
Following the method described in 
section~\ref{density-within-spherical-cells} for the spherical overdensity 
$\eta_r$, we can also obtain the quasi-linear limit of $\cP(\ctheta_r)$ by
a steepest-descent approach. Thus, we again define the moment and cumulant
generating functions $\Psi(y)$ and $\varphi(y)$ as in 
Eqs.(\ref{Psidef})-(\ref{phidef}), which can be expressed as a path integral
(\ref{path1}) with an action $\cS[\delta_L]$ as in (\ref{SdLdef}), where
the nonlinear functional $\eta_r[\delta_L]$ is replaced by $\ctheta_r[\delta_L]$.
The action still being spherically symmetric, we can also look for a spherical
saddle-point. If particles on the sphere $S$ have not been shocked yet
they have kept their initial velocity $\bu_0$, so that $\ctheta_r$ only
depends on the initial radial velocity $\bu_{0q}$ of the particles at the 
Lagrangian radius $q$ that have moved to radius $r$ at time $t$. Then, by the 
same reasoning as for the overdensity, we obtain as expected the same 
profile (\ref{profile}) for the saddle-point, but a different function
$\cF(\delta_{Lq})$. Indeed, since the radii $q$ and $r$ are related by 
$r=q+t\,u_{0q}$ we have from Eq.(\ref{cthetadef})
\beq
\ctheta_r=-\frac{d}{r} \, t \, u_r(t) = -\frac{d}{r} \, t \, u_{0q} 
= d\left(\frac{q}{r}-1\right) .
\label{thetaqr}
\eeq
Note that Eq.(\ref{thetaqr}) shows that for spherical dynamics the quantity
$\ctheta_r$ obeys
\beq
\ctheta_r \geq -d .
\label{thetamin}
\eeq
This lower bound corresponds to very fast expansion, so that the particles 
observed at radius $r$ come from the center $\bq=0$. Note that this requires
the initial velocity field to be singular at the origin, whence $n>-1$,
in agreement with the exponential falloff (\ref{Petainf}), (\ref{Petainfexp}),
obtained at low densities for $n<-1$, as shown in section~\ref{Asymptotic-tails}
below. In such cases, where rarefaction regions can appear (i.e. truly empty
regions), it may happen that the cell of radius $r$ is within a larger
empty domain, so that there are no particles on the sphere $S$. However,
in the quasi-linear limit, where we consider small values of $|\ctheta_r|$,
we do not consider this case. 
Next, from Eq.(\ref{u0rdef}) the linear density contrast is given by
\beq
\delta_{Lq} = -d \, \frac{t\,u_{0q}}{q} = d\left(1-\frac{r}{q}\right) ,
\eeq
whence
\beq
\ctheta_r=\cF(\delta_{Lq}) \;\; \mbox{with} \;\; 
\cF(\delta_{Lq})= -d+\frac{d}{1-\frac{\delta_{Lq}}{d}} .
\label{thetaF}
\eeq
Again, since the function $\cF$ simply describes the spherical Burgers dynamics
it only depends on the dimension $d$ and not on the initial power spectrum.
Then, we obtain the expression (\ref{SdLq}) with this new function $\cF$,
and we can define the associated variables $\tau$ and $\cG$ as in (\ref{taudef}),
so that the cumulant generating function $\varphi(y)$ is given by the relations
(\ref{varphiy}) and (\ref{Legendre}).
The spherical saddle-point being identical to the one obtained in
section~\ref{Spherical-saddle-point}, the profile (\ref{profilen}) still applies
for power-law power spectra, as well as Figs.~\ref{figprofile}, 
\ref{figvel_profile}. In particular,
if $n>d-3$ shocks appear as soon as $t\neq 0$ so that this saddle-point is no
longer exact in such cases. As in section~\ref{density-within-spherical-cells}
we shall not consider the modifications that appear in such cases, as they
should remain small in the quasi-linear regime (with $d-3<n<d-2$),
and we focus on cases such that $n\leq d-3$.

Close to the origin, since by symmetry we have $\lag\bu\rag=0$, whence
$\lag\ctheta_r\rag=0$, we always have (compare with (\ref{phiy0}))
\beq
\tau\rightarrow 0, \;\; y \rightarrow 0 : \;\; \cG\sim -\tau, \;\;\; 
y \sim \tau \;\;\; \mbox{and} \;\;\; \varphi \sim -\frac{y^2}{2} ,
\label{theta_phiy0}
\eeq
and keeping only these low-order terms gives back the linear Gaussian of
variance $\sigr^2$. For the power-law initial power spectra (\ref{ndef})
we obtain from Eqs.(\ref{taudef}), (\ref{thetaqr}),
\beq
\tau= -\delta_{Lq} \left(\frac{r}{q}\right)^{-\frac{n+3}{2}} 
= -\delta_{Lq} \left(1+\frac{\ctheta_r}{d}\right)^{\frac{n+3}{2}} ,
\label{theta_taudL}
\eeq
which leads to
\beq
\tau(\cG)= d \left(1+\frac{\cG}{d}\right)^{\frac{n+1}{2}} 
- d \left(1+\frac{\cG}{d}\right)^{\frac{n+3}{2}} ,
\label{thetatauG}
\eeq
and to the parametric representation of $\varphi(y)$,
\beqa
\varphi & = & -d y -\frac{d^2}{2}\left[ n \left(1+\frac{\cG}{d}\right)^{n+1} 
- 2 (n+1)\left(1+\frac{\cG}{d}\right)^{n+2} \right.  \nonumber \\
&& \left. \hspace{1.9cm} + (n+2) \left(1+\frac{\cG}{d}\right)^{n+3} \right] ,
\label{thetaphiG}
\eeqa
\beqa
\hspace{-0.9cm} y & = & -\frac{d}{2} \left(1+\frac{\cG}{d}\right)^{-1} 
\left[ (n+1) \left(1+\frac{\cG}{d}\right)^{n+1}  \right. \nonumber \\
&& \hspace{-0.9cm} \left. - 2 (n+2)\left(1+\frac{\cG}{d}\right)^{n+2} 
+ (n+3) \left(1+\frac{\cG}{d}\right)^{n+3} \right] 
\label{thetayG}
\eeqa
We can see that, contrary to the Eqs.(\ref{phiG})-(\ref{yG}) associated with
the density contrast, the dependence on $d$ of the system 
(\ref{thetaphiG})-(\ref{thetayG}) simplifies as $d^{-2}\varphi(d y)$
no longer depends on the dimension $d$. This implies for the probability
distribution the scaling
\beq
\cP_d(\ctheta_r;\sigr^2) = \frac{1}{d} 
\cP_1\left(\frac{\ctheta_r}{d};\frac{\sigr^2}{d^2}\right) ,
\label{Pthetascalingd}
\eeq
where we noted $\cP_d(\ctheta_r;\sigr^2)$ the quasi-linear probability density 
of $\ctheta_r$ in dimension $d$ when the linear variance is $\sigr^2$.
Thus, in the quasi-linear limit the change of dimension is fully absorbed by
a rescaling of the velocity divergence $\ctheta_r$ and of the linear variance
$\sigr^2$. Therefore, contrary to the case of the overdensity studied in
section~\ref{density-within-spherical-cells}, the properties of $\varphi(y)$
and $\cP(\ctheta_r)$, such as the presence of singularities and sub-exponential
tails, only depend on $n$ and not on the dimension $d$.

\begin{table*}
\begin{center}
\begin{tabular}{c||c|c|c||c|c|c}
\; $n$ \; & $\varphi(y)$ & $y$ & $\cG=\ctheta_r$ & $\varphi(y)$ & $y$ & $\cG=\ctheta_r$ \rule[-0.25cm]{0cm}{0.6cm} \\ \hline\hline
0 & \,$\frac{d^2}{27}-\frac{dy}{3}-\frac{d^2}{27}\left(1-\frac{6y}{d}\right)^{3/2}$\, & \,$-\infty<y<\frac{d}{6}$\, & \,$-\frac{d}{3}<\cG<\infty$\, & \,$\frac{d^2}{27}-\frac{dy}{3}+\frac{d^2}{27}\left(1-\frac{6y}{d}\right)^{3/2}$\, & \,$-\frac{d}{2}<y<\frac{d}{6}$\, & \,$-d<\cG<-\frac{d}{3}$ \rule[-0.3cm]{0cm}{0.8cm} \\ \hline
-1 & $-\frac{y^2}{2}$ & $-\infty<y<d$ & $-d<\cG<\infty$ & & & \rule[-0.27cm]{0cm}{0.7cm} \\ \hline
-2 & $-d^2-d y + d^2\sqrt{1+\frac{2y}{d}}$ & $-\frac{d}{2}<y<\infty$ & $-d<\cG<\infty$ & & & \rule[-0.25cm]{0cm}{0.8cm} \\ \hline\hline
$d=\infty$ & $-\frac{y^2}{2}$ & $-\infty<y<\infty$ & $-\infty<\cG<\infty$ & & & \rule[-0.27cm]{0cm}{0.7cm} \\
\end{tabular}
\end{center}
\caption{The cumulant generating function $\varphi(y)$ of the velocity divergence
$\ctheta_r$, in the quasi-linear limit $\sigr^2\rightarrow 0$, for integer values
of $n$ and arbitrary dimension $d$, where explicit solutions of the system 
(\ref{thetaphiG})-(\ref{thetayG}) can be obtained. The columns 2-4 show the 
quasi-linear branch, and the associated range of $\{y,\cG\}$ that contains the 
point $\{0,0\}$. The columns 5-7 show the second branch, associated with very 
rare events, that appears in some cases (here only for $n=0$). The last row 
shows the infinite-dimensional limit, $d\rightarrow \infty$, which no longer 
depends on $n$ and corresponds to the Gaussian.}
\label{Table_theta_varphi}
\end{table*}

\begin{figure}
\begin{center}
\epsfxsize=7.5 cm \epsfysize=5.8 cm {\epsfbox{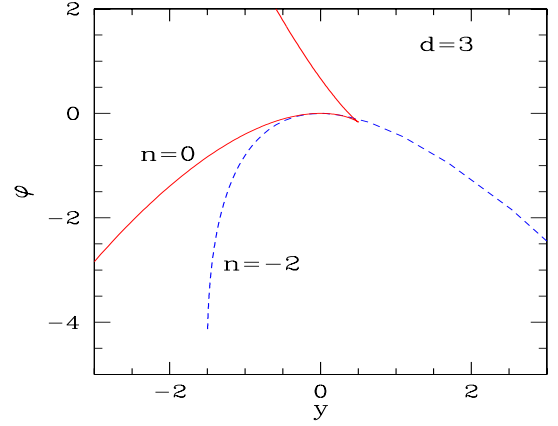}}
\end{center}
\caption{(Color online)
The cumulant generating function $\varphi(y)$ for the velocity
divergence $\ctheta_r$, in dimension $d=3$ for the power indices $n=0$ 
(solid line) and $n=-2$ (dashed line), from rows 2 and 4 of 
Table~\ref{Table_theta_varphi}.
In both cases there is a singularity on the real axis, with $y_s=1/2$ for
$n=0$ and $y_s=-3/2$ for $n=-2$, but only one branch for $n=-2$. 
The branch that runs through the origin is associated with moderate velocity 
fluctuations and is the relevant one for the expansion in terms of cumulants 
in the quasi-linear limit.}
\label{figtheta_phiy}
\end{figure}

Expanding near the origin, we obtain for the third and fourth-order cumulants
in the quasi-linear limit:
\beq
\sigr\rightarrow 0 : \;\;\; 
S_3=\frac{\lag\ctheta_r^3\rag_c}{\lag\ctheta_r^2\rag_c^2} = -3\frac{n+1}{d} ,
\label{thetaS3}
\eeq
\beq
S_4=\frac{\lag\ctheta_r^4\rag_c}{\lag\ctheta_r^2\rag_c^3} = 
\frac{3(n+1)(7n+9)}{d^2} .
\label{thetaS4}
\eeq
As discussed below (\ref{cthetadef}) this also gives the cumulants associated with
the third and fourth-order spherical velocity structure functions in the 
quasi-linear limit. As for Eqs.(\ref{S3})-(\ref{S4}), these quantities are not
the standard skewness and kurtosis, and the powers in the denominators are such
that they have a finite non-zero quasi-linear limit.

For integer values of $n$ we can also derive explicit solutions to 
Eqs.(\ref{thetaphiG})-(\ref{thetayG}), which we show in 
Table~\ref{Table_theta_varphi}. As for the density studied in 
section~\ref{Cumulant-generating-function}, the quasi-linear generating function
$\varphi(y)$ can show two branches when the function $y(\cG)$ is not monotonic
over $\cG\in]-d,+\infty[$ (the variable $\cG$ now covers the range $]-d,+\infty[$,
as seen from Eq.(\ref{thetamin})). This occurs for $n=0$, shown in the second
row in Table~\ref{Table_theta_varphi}, while for $n=-1$ and $n=-2$ there is only
one branch.

In dimension $d=1$ we have from Eqs.(\ref{thetaqr}), (\ref{etarF}),
\beq 
d=1 : \;\;\; \ctheta_r= \frac{q}{r}-1 = \eta_r -1 ,
\label{d1_etar_cthetar}
\eeq
so that the distributions $\cP_1(\eta_r)$ and $\cP_1(\ctheta_r)$ are identical
up to a shift of unity. Then, we can check that the results shown in
Tables~\ref{Table_varphi} and \ref{Table_theta_varphi} for the case 
$\{n=-2,d=1\}$ are consistent. In particular, the generating function given 
in the fourth row of Table~\ref{Table_theta_varphi} yields the probability 
distribution
\beqa
\hspace{-1.1cm} n=-2 & : & \cP(\ctheta_r) = \frac{1}{\sqrt{2\pi}\sigr} 
\left(\frac{\ctheta_r}{d}+1\right)^{-3/2} \nonumber \\
&& \hspace{-1.1cm} \times \exp\left[ -\frac{d^2}{2\sigr^2}
\left(\sqrt{\frac{\ctheta_r}{d}+1}
-\frac{1}{\sqrt{\frac{\ctheta_r}{d}+1}}\right)^2 \right] .
\label{thetaPBrown}
\eeqa
We can check that this agrees with relation (\ref{d1_etar_cthetar}) and
Eq.(\ref{PBrown}) for $d=1$. Moreover, in dimension $d=1$ this result
(\ref{thetaPBrown}) again happens to be exact \cite{Valageas2008}.

For $n=-1$ we simply obtain the Gaussian (third row in 
Table~\ref{Table_theta_varphi})
\beq
n=-1: \; \varphi(y) = -\frac{y^2}{2} , \; 
\cP(\ctheta_r) = \frac{1}{\sqrt{2\pi}\sigr} e^{-\ctheta_r^2/(2\sigr^2)} .
\label{thetaPn-1}
\eeq
Thus, the effects of the nonlinear evolution (\ref{thetaF}) and of the
change of scale $q\rightarrow r$ (encoded in the change from $\cF$ to $\cG$)
compensate in such a way that at leading order in the quasi-linear limit
the cumulants $\lag\ctheta_r^p\rag_c$ vanish, whence 
$\lag\ctheta_r^p\rag_c \ll \sigr^{2(p-1)}$ for $\sigr\rightarrow 0$ and 
$p\geq 3$, in agreement with Eqs.(\ref{thetaS3})-(\ref{thetaS4}).
However, note that the distribution (\ref{thetaPn-1}) differs from the linear
Gaussian in the sense that $\ctheta_r$ is restricted to $\ctheta_r\geq -d$,
from Eq.(\ref{thetamin}). Of course, the weight associated with 
this lower bound becomes exponentially small in the
quasi-linear limit, so that it cannot be seen in the leading-order value of the
cumulants $\lag\ctheta_r^p\rag_c$, whence in the quasi-linear generating 
function $\varphi(y)$.

On the other hand, in the limit of large dimension, $d\rightarrow \infty$,
we obtain $\cG(\tau)=-\tau$ which gives back the linear Gaussian of 
Eq.(\ref{thetaPn-1}), again in agreement with 
Eqs.(\ref{thetaS3})-(\ref{thetaS4}). 
However, contrary to the distribution (\ref{thetaPn-1})
associated with $n=-1$ at finite $d$, in the limit $d\rightarrow \infty$
the Gaussian extends down to $-\infty$, since the lower bound (\ref{thetamin})
is repelled to $-\infty$. Note that this was not the case for the overdensity
$\eta_r$, where the probability distribution did not tend to the Gaussian but
to a lognormal distribution for $d\rightarrow \infty$, see the last row in
Table~\ref{Table_varphi} and Eqs.(\ref{S3})-(\ref{S4}).

\begin{table}
\begin{center}
\begin{tabular}{c|c||c|c}
\,$n$\, & \,$d$\, & $\ln\cP(\ctheta_r)$ & $\ctheta_r$ \rule[-0.2cm]{0cm}{0.5cm} \\ \hline\hline
0 & 3 & \,$-\frac{r^3}{2t^2}\,\left[(1\!+\!\frac{\ctheta_r}{3})^{1/2}-(1\!+\!\frac{\ctheta_r}{3})^{3/2}\right]^2$\, & \,$\ctheta_r>-1$ \rule[-0.35cm]{0cm}{0.9cm} \\ \hline
-1 & 2 & $-\frac{r^2}{8t^2}\,\ctheta_r^2$ & $\ctheta_r>-1$ \rule[-0.3cm]{0cm}{0.8cm} \\ \hline
-2 & 1 & $-\frac{r}{2t^2}\,\left[(1\!+\!\ctheta_r)^{-1/2}-(1\!+\!\ctheta_r)^{1/2}\right]^2$ & $\ctheta_r>-1$ \rule[-0.35cm]{0cm}{0.9cm} \\ \hline
-2 & 3 & \;$-\frac{5r}{8t^2}\,\left[(1\!+\!\frac{\ctheta_r}{3})^{-1/2}-(1\!+\!\frac{\ctheta_r}{3})^{1/2}\right]^2$\; & $\ctheta_r>-3$ \rule[-0.35cm]{0cm}{0.9cm} \\ \hline
n & $\infty$ & $-\frac{(2r^2)^{\frac{n+3}{2}}}{2t^2}\,\ctheta_r^2$ &  \rule[-0.25cm]{0cm}{0.85cm} \\
\end{tabular}
\end{center}
\caption{Asymptotic behavior of the probability distribution $\cP(\ctheta_r)$
in the quasi-linear regime, for a few integer values of $n$ and $d$, using the
normalization of Table~\ref{Table_correl} for the initial conditions. 
The last column shows the range of spherical velocity increment $\ctheta_r$
where these results apply. If the lower threshold is not equal to $-d$ 
(i.e. $-2<n<d-3$), it means that the spherical saddle-point forms shocks for 
lower $\ctheta_r$.}
\label{Table_Ptheta_ql}
\end{table}

We show in Fig.~\ref{figtheta_phiy} the quasi-linear cumulant generating
function $\varphi(y)$, obtained in three dimensions for $n=0$ and $n=-2$.
Although there is a singularity on the real axis for both cases, there is only 
one branch for $n=-2$.

Next, the quasi-linear probability distribution $\cP(\ctheta_r)$ is obtained
from the cumulant generating function $\varphi(y)$ by an inverse Laplace
transform, as in Eq.(\ref{Petaphi}). 
In the quasi-linear limit, as for the overdensity $\eta_r$, it obeys the
asymptotic behavior (\ref{Petatau}). Then, using Eq.(\ref{thetatauG}) we obtain
\beq
\sigr\!\!\rightarrow\!0 \! : \;\! \ln\cP(\ctheta_r) \sim \frac{-d^2}{2\sigr^2}\!
\left[ \!\left(\!1\!+\!\frac{\ctheta_r}{d}\!\right)^{\!\!\!\frac{n+1}{2}} 
\!\!\!\!-\! \left(\!1\!+\!\frac{\ctheta_r}{d}\!\right)^{\!\!\!\frac{n+3}{2}}
\right]^2 
\label{Pthetataun}
\eeq
and in the large dimensional limit, where $\tau=-\ctheta_r$,
\beq
d\rightarrow\infty , \;\; \sigr\rightarrow 0 : \;\;
\ln\cP(\ctheta_r) \sim -\frac{\ctheta_r^2}{2\sigr^2} ,
\label{Pthetadinf}
\eeq
which only hold for $n<d-3$ and above a low-$\ctheta_r$ threshold if $n>-2$,
see Table~\ref{Table_shocks}. Note that Eq.(\ref{Pthetataun}) may be directly
obtained from the asymptotic tail (\ref{Petataun}) derived for the density
distribution, $\cP(\eta_r)$, by substituting
\beq
\ctheta_r=d\,(\eta_r^{1/d}-1) , \;\;\;\; 
\eta_r=\left(1+\frac{\ctheta_r}{d}\right)^d .
\label{theta_eta}
\eeq
This relation follows from Eqs.(\ref{etarF}), (\ref{thetaqr}), that
express both $\eta_r$ and $\ctheta_r$ in terms of the Lagrangian radius
$q$ of the saddle-point, so that there is a unique correspondence between
$\eta_r$ and $\ctheta_r$ for these spherical saddle-points.

\begin{figure}
\begin{center}
\epsfxsize=7.5 cm \epsfysize=5.8 cm {\epsfbox{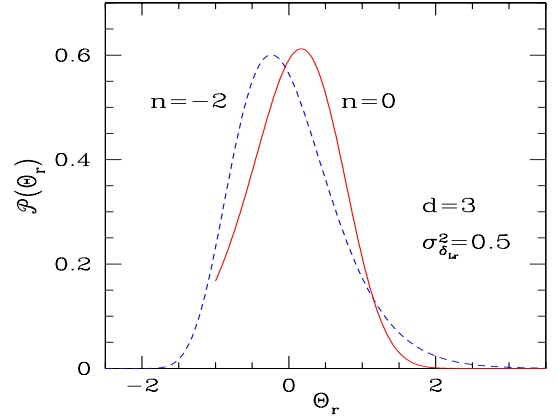}}
\end{center}
\caption{(Color online)
The probability distribution $\cP(\ctheta_r)$ of the velocity divergence
$\ctheta_r$ (spherical velocity increment) in the quasi-linear limit. 
We show the cases $n=0$ and $n=-2$
in three dimensions, for a linear variance $\sigr^2=0.5$.}
\label{figPtheta}
\end{figure}

We show in Table~\ref{Table_Ptheta_ql} the asymptotic behaviors 
(\ref{Pthetataun}) obtained for the initial conditions given in 
Table~\ref{Table_correl}, as well as the lower bound $\ctheta_*$ below which
the saddle-point (\ref{profile}) forms shocks, which will be derived in 
section~\ref{Asymptotic-tails} below where we take into account shocks.
Again, the value (and the existence) of $\ctheta_*$ is not related to the
value (and the existence) of $\cG_s$ where the cumulant generating function 
is singular, which was given in Table~\ref{Table_theta_varphi}.

We show in Fig.~\ref{figPtheta} our results for $\cP(\ctheta_r)$ for the cases 
$n=0$ and $n=-2$ in three dimensions, for a linear variance $\sigr^2=0.5$,
as in Fig.~\ref{figPeta}. Again, we only plot the distribution over the range
associated with the quasi-linear branch of $\varphi(y)$. For $n=-2$ this
actually covers the whole range $\ctheta_r>-d$, but for $n=0$ this corresponds 
to $\ctheta_r>-d/3$, see Table~\ref{Table_theta_varphi}.
Note that in the latter case this lower bound happens to coincide with the
lower bound $\ctheta_*$ where shocks appear (for $d=3$).
The comparison between Figs.~\ref{figPtheta} and \ref{figPeta} shows that the 
distribution of the velocity divergence $\ctheta_r$ remains closer to the 
Gaussian than the distribution of the overdensity $\eta_r$. This can also be 
seen from the fact that the singularities $y_s$ are farther from the origin 
$y=0$ (compare Tables~\ref{Table_theta_varphi} and \ref{Table_varphi} and 
Figs.~\ref{figtheta_phiy} and \ref{figphiy}). On the other hand, in agreement
with (\ref{thetaS3}), we can see that the skewness has opposite signs for $n=0$ 
and $n=-2$, as the peak of the distribution shifts to either side of 
$\ctheta_r=0$, while the skewness of the density was always positive for
$n<d-2$, see Eq.(\ref{S3}), which covers the range where $\sigr^2$ is well 
defined.

\section{Asymptotic tails}
\label{Asymptotic-tails}

The results obtained in section~\ref{Asymptotic-tails} applied to the
quasi-linear limit, $\sigr\rightarrow 0$, for the case $n\leq d-3$, so that
$\sigr$ is well defined and shocks only appear after a finite time
(if $d-3<n<d-2$ shocks appear as soon as $t\neq 0$ but $\sigr$
is still well defined and these results should provide a reasonable 
approximation).
We now consider the limit of rare events, that is very large density and
velocity fluctuations
at fixed linear variance $\sigr^2$, or at fixed $\sigpsir^2$ if $\sigr^2$
is divergent.
Thus we study the tails of the probability distributions $\cP(\eta_r)$ and 
$\cP(\ctheta_r)$, for any value of $\sigr$ or $\sigpsir$. As in 
section~\ref{Quasi-linear} we can use a steepest-descent approach and look
for the minimum of the action $\cS$, defined as in (\ref{SdLdef}). This will
give the tails of the probability distribution through Eq.(\ref{Petatau}).

\subsection{Saddle-point without shocks}
\label{Saddle-point-without-shocks}

\subsubsection{Rare density fluctuations}
\label{Rare-density}

For $n\leq d-3$ we can use the same action $\cS[\delta_L]$ as in (\ref{SdLdef})
and we obtain the same saddle-point defined by Eqs.(\ref{profile}) and 
(\ref{Legendre}) provided no shocks have formed.
As discussed in section~\ref{Spherical-saddle-point}, this constraint is
satisfied for overdensities if $n\leq d-3$ and for underdensities if
$-3<n\leq-2$. Therefore, in such cases Eq.(\ref{Petatau}) remains valid,
where $\tau(\eta_r)$ is still given by Eq.(\ref{tauG}), provided the
saddle-point approximation is justified. Thus, we must show that as we consider
very large density fluctuations the Laplace transform $\Psi(y)$ and the
distribution $\cP(\eta_r)$ are dominated by an increasingly narrow region
around this saddle-point. We no longer have a fixed action $\cS[\delta_L]$
multiplied by a prefactor that goes to infinity, as was the case in the 
quasi-linear regime for (\ref{path1}), with $1/\sigr^2\rightarrow\infty$.
Therefore, the analysis is more complicated as we should study the Hessian
of the action at the saddle-point, which requires second-order perturbation
theory around this saddle-point, taking into account angular degrees of freedom.
Here we shall simply show that the steepest-descent approximation is well
justified for $\Psi(y)$ (i.e. $\varphi(y)$) with respect to the family of
initial states (\ref{profile})-(\ref{profile_u0}), parameterized by the
overdensity $\cG$ within radius $r$.

In this subspace, the action $\cS[\delta_L]$ is an ordinary function
$\cS(\cG)$, given by Eq.(\ref{Stau}) as $\cS(\cG)=y\cG+\tau(\cG)^2/2$.
Thus, at the saddle-point the first and second derivative read as
\beq
\frac{\dd\cS}{\dd\cG} = 0 , \;\;\;\;
\frac{\dd^2\cS}{\dd\cG^2} = \frac{\dd^2}{\dd\cG^2} 
\left(\frac{\tau(\cG)^2}{2}\right) .
\eeq
Let us first consider the case of large underdensities, $\cG\rightarrow 0$,
with $-3<n\leq -2$. Then, from Eq.(\ref{tauG}) we have
\beq
\cG\rightarrow 0: \;\; \frac{\dd^2\cS}{\dd\cG^2} \sim 
\frac{(n+1)(n+1-d)}{2} \, \cG^{\frac{n+1}{d}-2} .
\eeq
Then, over this subspace, disregarding prefactors associated with changes of
variables, we write the analog of (\ref{path1}) as
\beq
e^{-\varphi(y)/\sigr^2} \sim \int\dd\cG \, e^{-\cS(\cG)/\sigr^2} ,
\label{intG}
\eeq
and expanding the action around the saddle-point $\cG_c$, defined by the 
condition $\dd\cS/\dd\cG=0$, we find that only values sufficiently close
to $\cG_c$ contribute to the integral, with
\beq
|\cG-\cG_c| \sim \frac{1}{\sqrt{\cS''(\cG)}} \sim \cG_c^{1-\frac{n+1}{2d}} ,
\eeq
whence (since $-3<n\leq -2$)
\beq
\cG_c \rightarrow 0 : \;\;\; \frac{|\cG-\cG_c|}{\cG_c} 
\sim \cG_c^{-(n+1)/(2d)} \rightarrow 0 .
\label{DeltaGund}
\eeq
Since $\cS(\cG)$ behaves as a power law, Eq.(\ref{DeltaGund}) implies
that higher-order terms beyond the Gaussian around $\cG_c$ are subdominant.
Thus, at large positive $y$, which corresponds to 
$\cG_c\rightarrow 0$, see (\ref{n<-1}), the Laplace transform $\varphi(y)$
is dominated by its saddle-point (if we only consider the subspace
described by the profile (\ref{profile})). Then, assuming that this remains
true when we take into account other degrees of freedom, and that there is
no deeper minimum associated with strong deviations from spherical symmetry,
we recover at large $y$ the behavior (\ref{smallG}) obtained in 
section~\ref{Cumulant-generating-function} where we studied the quasi-linear
regime. Next, the distribution $\cP(\eta_r)$ is still given by the inverse
Laplace transform (\ref{Petaphi}), and it only remains to show that this
integral is again dominated by its saddle-point, in the limit 
$\eta_r\rightarrow 0$. As noticed in section~\ref{Probability-distribution},
the saddle-point $y_c$ of the exponential satisfies $\eta_r=\varphi'(y_c)$,
whence $\eta_r=\cG_c$ where $\cG_c$ is related to $y_c$ through the Legendre
transformation (\ref{varphiy}). Thus, in the limit $\eta_r\rightarrow 0$
we obtain $y_c\rightarrow +\infty$ and from the previous results the
behavior (\ref{smallG}). Next, expanding the exponential around
$y_c$, we find that contributions to $\cP(\eta_r)$ come from the range
\beq
|y-y_c| \sim \frac{1}{\sqrt{-\varphi''(y_c)}} 
\sim y_c^{1-\frac{n+1}{2(n+1-d)}} ,
\eeq
where we used the large-$y$ behavior (\ref{smallG}). This yields
\beq
y_c\rightarrow \infty: \;\;\; \frac{|y-y_c|}{y_c} 
\sim y_c^{-\frac{n+1}{2(n+1-d)}} \rightarrow 0 ,
\label{Deltayund}
\eeq
which again ensures that the integral (\ref{Petaphi}) is dominated by
the Gaussian integration around
its saddle-point in the limit $\eta_r\rightarrow 0$. Therefore, the 
steepest-descent approximation is legitimate at very low densities, whatever
the value of $\sigr$, and we recover the rare-event tail (\ref{Petatau}).
As in (\ref{Petataun}), this yields the low-density tail
\beq
-3<n\leq -2, \;\; \eta_r \rightarrow 0 :  \;\;
\cP(\eta_r) \sim e^{-d^2\eta_r^{\frac{n+1}{d}}/(2\sigr^2)} . 
\label{Petainf}
\eeq

We can apply the same procedure for large overdensities, $\eta_r\rightarrow
+\infty$. Here, a subtlety arises from the fact that for $n<d-3$ large
densities are associated with the second branch (\ref{n+3-d<0}) of $\varphi(y)$ 
discussed in section~\ref{Cumulant-generating-function}.
As recalled in section~\ref{Probability-distribution}, this simply corresponds
to cases where the high-density tail of the distribution $\cP(\eta_r)$
shows a sub-exponential decay, such as $e^{-\eta_r^{\alpha}}$ with 
$\alpha< 1$. From the definition (\ref{Psidef}) we can see at once that
in such cases the Laplace transform $\Psi(y)$, whence $\varphi(y)$, diverge
for negative $y$: they generically have a branch cut along the negative real
axis. Then, the saddle-point in the path integral (\ref{path1})
becomes a local maximum but one can still apply the steepest-descent method,
using an appropriate deformation of the integration contours in the complex
plane. This is discussed in details in section~3.6 and appendices A and B
of \cite{Valageas2002a} hence we do not give further comments on this point
here, see also the chapter on instanton contributions in \cite{Zinn}.
Then, as for underdensities, we can check that for large overdensities
the contributions to $\varphi(y)$ come from an increasingly narrow range
of $\cG$, when we restrict to the subspace parameterized by the profile
(\ref{profile}), as in (\ref{intG}). Thus, using again Eq.(\ref{tauG}),
we now find that the range of overdensities that contribute to $\varphi(y)$
behaves as
\beq
\cG_c \rightarrow +\infty : \;\;\; \frac{|\cG-\cG_c|}{\cG_c} 
\sim \cG_c^{-(n+3)/(2d)} \rightarrow 0 ,
\label{DeltaGover}
\eeq
so that $\varphi(y)$ is still governed by the saddle-point (assuming
there are no larger contributions associated with strong deviations from
spherical symmetry), and we recover the behavior (\ref{n+3-d<0}).
Then, the range of $y$ that contributes to the large-density tail is
\beq
y_c\rightarrow 0^-: \;\;\; \frac{|y-y_c|}{|y_c|} 
\sim |y_c|^{\frac{n+3}{2(d-n-3)}} \rightarrow 0 .
\label{Deltayover}
\eeq
Therefore, the steepest-descent approximation is again legitimate at very
high densities, whatever the value of $\sigr$, and we recover the rare-event
tail (\ref{Petatau}). As in (\ref{Petataun}), this yields the high-density tail
\beq
n\leq d-3, \;\; \eta_r \rightarrow \infty : \;\;
\cP(\eta_r) \sim e^{-d^2\eta_r^{\frac{n+3}{d}}/(2\sigr^2)} .
\label{Petasup}
\eeq
We can check that in the case $\{n=-2,d=1\}$ both asymptotics 
(\ref{Petainf}), (\ref{Petasup}), agree with the exact distribution
(\ref{PBrown}).

In the large dimensional limit, $d\rightarrow\infty$, we still obtain
Eq.(\ref{Petadinf}), which applies to both rare overdensities and 
underdensities, but only if $n\leq -2$ in the latter case.

In fact, the saddle-point approximation is valid as long as we consider the
limit of rare events, which corresponds either to the quasi-linear limit,
$\sigr\rightarrow 0$ (i.e. $r^{n+3}/t^2\rightarrow\infty$) at fixed density
$\eta_r$ and velocity increment $\ctheta_r$, or to the limit of extreme densities,
$\eta_r\rightarrow 0$ or $\eta_r\rightarrow\infty$, and extreme velocities,
$\ctheta_r\rightarrow -d$ or $\ctheta_r\rightarrow\infty$, at fixed time and
scale (i.e. at finite $\sigr$). Of course, the range of density fluctuations
to which these results apply is a priori repelled to increasingly large
fluctuations, $\eta_r \rightarrow \infty$ or $\eta_r \rightarrow 0$, 
as $\sigr$ grows (since they correspond to rare events, $\cP \ll 1$).
In such regimes, integrals such as (\ref{path1})
are governed at leading order by the minimum of the action $\cS$, which shows
a steep dependence on the initial conditions. This legitimates the 
steepest-descent approximation in these cases.

We can note that an exception to
this behavior occurs when the action happens to be singular at its minimum,
that is the initial conditions where $\cS$ is close to its minimum form a
subspace of measure zero. For ordinary integrals such as (\ref{intG}), this
corresponds to cases where the function $\cS(\cG)$ is discontinuous at
the point $\cG_c$, with $\cS(\cG_c)$ being strictly smaller than both left
and right limits, $\cS(\cG_c^-)$ and  $\cS(\cG_c^+)$. This possibility actually
appears for the case of the collisionless gravitational dynamics 
\cite{Valageas2002b}, where at large positive densities a strong radial orbit
instability appears (associated with the extreme sensitivity of the trajectories,
that actually diverges, as particles move through the center of the object).
This problem does not appear in the Burgers dynamics, since in any case 
particles that would reach the center would stick there. Thus, the infinitesimal
viscosity regularizes the dynamics and makes the sensitivity to non-spherical
perturbations finite. This well-behaved dependence on the initial conditions
clearly appears through the Hopf-Cole solution recalled in 
section~\ref{self-similarity}.

If we are only interested in the exponents that appear in the expressions
(\ref{Petainf}), (\ref{Petasup}), disregarding the numerical factors
and using Eq.(\ref{sig2def}) they read at leading order as
\beqa
\hspace{-0.7cm} n\leq d-3, \;\; \eta_r \rightarrow \infty \! & : & 
\ln\cP(\eta_r) \propto - r^{n+3} \eta_r^{\frac{n+3}{d}} \!/t^2 
\label{Petasupexp} \\
\hspace{-0.7cm} -3<n\leq -2, \;\; \eta_r \rightarrow 0 \! & : & 
\ln\cP(\eta_r) \propto - r^{n+3} \eta_r^{\frac{n+1}{d}} \!/t^2 
\label{Petainfexp}
\eeqa
We give in Table~\ref{Table_tailnoshock} the explicit expressions of the
tails (\ref{Petainf}), (\ref{Petasup}), for the initial conditions normalized
as in Table~\ref{Table_correl}, and we mark as ``shock'' the cases where
the saddle-point discussed above gives rise to shocks.
This agrees with Table~\ref{Table_Peta_ql} where we only keep the leading
term for $\eta_r\rightarrow \infty$ or $\eta_r\rightarrow 0$.

\begin{table*}
\begin{center}
\begin{tabular}{c|c||c|c||c|c}
\,$n$\, & \,$d$\, & \,$\ln\cP(\eta_r)$ for $\eta_r\rightarrow\infty$\, & $\ln\cP(\eta_r)$ for $\eta_r\rightarrow 0$ & \,$\ln\cP(\ctheta_r)$ for $\ctheta_r\rightarrow\infty$\, & $\ln\cP(\ctheta_r)$ for $\ctheta_r\rightarrow -d$ \rule[-0.25cm]{0cm}{0.7cm} \\ \hline\hline
0 & 3 & $-r^3\eta_r/(2t^2)$ & shock & $-r^3\ctheta_r^3/(54t^2)$ & shock \rule[-0.2cm]{0cm}{0.6cm} \\ \hline
-1 & 2 & $-r^2\eta_r/(2t^2)$ & shock & $-r^2\ctheta_r^2/(8t^2)$ & shock \rule[-0.2cm]{0cm}{0.6cm} \\ \hline
-2 & 1 & $-r\,\eta_r/(2t^2)$ & $-r/(2t^2\eta_r)$ & $-r\,\ctheta_r/(2t^2)$ & $-r/[2t^2(1+\ctheta_r)]$ \rule[-0.2cm]{0cm}{0.6cm} \\ \hline
-2 & 3 & $-5\,r\,\eta_r^{1/3}/(8t^2)$ & $-5\,r/(8t^2\eta_r^{1/3})$ & $-5\,r\,\ctheta_r/(24t^2)$ & $-15\,r/[8t^2(3+\ctheta_r)]$ \rule[-0.25cm]{0cm}{0.7cm} \\ \hline
n & $\infty$ & $-\frac{(2r^2)^{\frac{n+3}{2}}}{2t^2}\ln^2(\eta_r)$ & \,$-\frac{(2r^2)^{\frac{n+3}{2}}}{2t^2}\ln^2(\eta_r)$ \, if $n\leq -2$\, & $-\frac{(2r^2)^{\frac{n+3}{2}}}{2t^2}\,\ctheta_r^2$ & \,$-\frac{(2r^2)^{\frac{n+3}{2}}}{2t^2}\,\ctheta_r^2$ \, if $n\leq -2$ \rule[-0.25cm]{0cm}{0.85cm} \\
\end{tabular}
\end{center}
\caption{The tails of the distributions $\cP(\eta_r)$ and $\cP(\ctheta_r)$
for a few integer values of
$n$ and $d$ where the relevant saddle-point is regular (i.e. does not induce
shocks), from Eqs.(\ref{Petainf}), (\ref{Petasup}). The initial conditions are
those given in Table~\ref{Table_correl} (i.e. with the same normalization).
Cases marked as ``shock'' correspond to saddle-points that give rise to shocks,
so that the normalization factors in Eqs.(\ref{Petainf}), (\ref{Petasup}), and 
(\ref{Pthetasup})-(\ref{Pthetainf1}), are no longer valid. They are analyzed
in section~\ref{Saddle-point-with-shocks} (this includes the case $\{n=0,d=1\}$
for both large and small densities).
For $d\rightarrow\infty$ the expressions given at low densities and velocity
divergences only hold for $n\leq -2$ ($n>-2$ gives rise to shocks).}
\label{Table_tailnoshock}
\end{table*}

\subsubsection{Rare velocity fluctuations}
\label{Rare-velocity}

The discussion of $\cP(\eta_r)$ in the previous section directly extends to
the tails of the distribution $\cP(\ctheta_r)$ of the mean divergence
$\ctheta_r$ (spherical velocity increment), defined in (\ref{cthetadef}). 
Thus, for the same cases as in
section~\ref{Rare-density} where the saddle-point does not form shocks, the
rare-event tails are still given at leading order by the first exponential
in Eq.(\ref{Petatau}), where $\tau(\ctheta_r)$ is given by 
Eq.(\ref{thetatauG}), whence by Eq.(\ref{Pthetataun}). This yields
\beqa
\hspace{-1.1cm} n\leq d-3, \; \ctheta_r \rightarrow \infty & : & \nonumber \\
&& \hspace{-3.2cm} \ln\cP(\ctheta_r) \sim -\frac{d^2}{2\sigr^2} 
\left(\frac{\ctheta_r}{d}\right)^{n+3} 
\propto -\frac{r^{n+3}\ctheta_r^{n+3}}{t^2}  , 
\label{Pthetasup}
\eeqa
and
\beqa
\hspace{-2cm} -3<n\leq -2, \; \ctheta_r \rightarrow -d : && \nonumber \\
&& \hspace{-3cm} \ln\cP(\ctheta_r) \sim \frac{-d^2}{2\sigr^2} 
\left(1+\frac{\ctheta_r}{d}\right)^{n+1} 
\label{Pthetainf1} \\
&& \hspace{-1.5cm} \propto -\frac{r^{n+3}}{t^2} 
\left(1+\frac{\ctheta_r}{d}\right)^{n+1} .
\label{Pthetainf}
\eeqa
In the large dimension limit we still have Eq.(\ref{Pthetadinf}),
which holds for $\ctheta_r\rightarrow \infty$ for any $n$, but for 
$\ctheta_r\rightarrow -\infty$ only if $n\leq -2$.
We also give these results in Table~\ref{Table_tailnoshock}.

Note that the appearance of a second branch for $\varphi(y)$ is not
simultaneous for $\eta_r$ and $\ctheta_r$, as noticed in 
section~\ref{Velocity-divergence} and as can be seen from the 
comparison of Tables~\ref{Table_varphi} and \ref{Table_theta_varphi}.
Since for spherical initial conditions $\eta_r$ and $\ctheta_r$
are related by (\ref{theta_eta}), it is clear that the subtlety associated
with the second branch of $\varphi(y)$, that is, the branch cut of the
exact Laplace transform, is only a mathematical difficulty due to a
sub-exponential tail but has no physical effect. Thus, in cases where
only the density quasi-linear generating function $\varphi(y)$ shows
a second branch, we can first compute the high-$\ctheta_r$ tail of
$\cP(\ctheta_r)$, which only involves local minima of the action and does
not require a deformation of integration contours, and next use
the relation (\ref{theta_eta}) to derive the high-density tail of 
$\cP(\eta_r)$. More generally, as noticed in \cite{Valageas2002a},
in order to avoid the complications associated with sub-exponential tails,
we can simply compute the rare-event tail of the quantity $X=\eta_r^{\beta}$,
with a small enough $\beta$ so that $\cP(X)$ shows a super-exponential
decay, and next derive $\cP(\eta_r)$ from $\cP(X)$ through a simple change of
variable.
On the other hand, we could directly obtain 
Eqs.(\ref{Pthetasup})-(\ref{Pthetainf}) from Eqs.(\ref{Petainf}), 
(\ref{Petasup}), by using the relation (\ref{theta_eta}).

\subsection{Saddle-point with shocks}
\label{Saddle-point-with-shocks}

\subsubsection{Paraboloid construction and action $\cS[\psi_0]$}
\label{Paraboloid}

For cases where the constraints in (\ref{Petainf}), (\ref{Petasup}), are not
satisfied, that is when shocks cannot be ignored, we can no longer rely on
the regular saddle-point of the previous 
section~\ref{Saddle-point-without-shocks}. However, we can use the exact
solution (\ref{psixpsi0q}), and the geometrical
construction (\ref{Paraboladef}), to study solutions of the equations of motion
that contain shocks. In particular, following the approach used in the previous
sections, we can look for saddle-points of an appropriate action, that include
shocks, to obtain the tails of the distribution $\cP(\eta_r)$. However, to
use (\ref{psixpsi0q}) it is more convenient to work with the velocity potential
$\psi_0$, rather than with its Laplacian, $\theta_0=\delta_L/t$, that we used in 
section~\ref{Quasi-linear}. Thus, we now write the Laplace transform 
(\ref{Psidef}) as
\beq
\Psi(y) = (\det C_{\psi_0}^{-1})^{1/2} \int \cD\psi_0 \, 
e^{-y\eta_r[\psi_0]-\frac{1}{2} \psi_0 . C_{\psi_0}^{-1} . \psi_0} ,
\label{Psipathpsi0}
\eeq
where $\eta_r[\psi_0]$ is now the functional that affects to the initial 
condition $\psi_0$ the nonlinear overdensity $\eta_r$, built at time $t$ within
the cell of radius $r$ centered on the origin, $\bx=0$; and 
$C_{\psi_0}(\bx_1,\bx_2)$ is the two-point correlation of the initial potential.
Although this is no longer essential (since we do not consider here the limit 
$\sigpsir\rightarrow 0$) we rescale the transform $\Psi(y)$ in a fashion similar
to Eq.(\ref{psiphidef}),
\beq
\Psi(y) = e^{-\varphi(y\sigpsir^2)/\sigpsir^2} ,
\label{Psisigpsi}
\eeq
so as to obtain an action $\cS[\psi_0]$ that is similar to Eq.(\ref{SdLdef}),
\beq
\cS[\psi_0] = y \, \eta_r[\psi_0] + \frac{\sigpsir^2}{2} \,
\psi_0 . C_{\psi_0}^{-1} . \psi_0 ,
\label{Spsi0def}
\eeq
where $\sigpsir$ is the variance of the initial radial potential $\psi_{0r}$
at radius $r$, defined in Eq.(\ref{sigpsirdef}).

As in section~\ref{Quasi-linear} we can look for
spherical saddle-points. Then, we can look for the minimum of the action 
$\cS[\psi_0]$ within the subspace of spherically symmetric initial conditions,
where $\psi_0(\bq)=\psi_{0q}$ with $q=|\bq|$, and the restriction of the action
to this subspace reads as
\beq
\cS[\psi_{0q'}] = y \, \eta_r[\psi_{0q'}] + \frac{\sigpsir^2}{2} \,
\psi_{0q_1'} . C_{\psi_{0r}}^{-1} . \psi_{0q_2'} ,
\label{Spsi0q1q2}
\eeq
where $C_{\psi_{0r}}(q_1',q_2')$ is the radial covariance introduced in 
Eq.(\ref{Cpsi0r}).

\subsubsection{Regular saddle-point without shocks}
\label{Ordinary-saddle-point}

Let us first check that in the case where there is no shock, we recover from
(\ref{Spsi0q1q2}) the results obtained in section~\ref{Quasi-linear}.
As in Eq.(\ref{saddle}), a saddle-point of the action (\ref{Spsi0q1q2})
is characterized by
\beq
\psi_{0q'} = \frac{-y}{\sigpsir^2} \int_0^{\infty} \dd q'' 
C_{\psi_{0r}}(q',q'') \frac{\cD\eta_r}{\cD\psi_{0q''}} .
\label{saddlepsi0}
\eeq
Following the discussion below Eq.(\ref{FdLq}), since $\eta_r$ only depends on
the initial velocity at the Lagrangian radius $q$, and 
$u_{0q}=-\dd\psi_{0q}/\dd q$, the functional differential 
$\cD\eta_r/\cD\psi_{0q''}$ is zero for $q''\neq q$. However, it is no
longer a Dirac, $\delta_D(q''-q)$, but its first derivative, $\delta_D'(q''-q)$.
Indeed, from the geometrical construction
(\ref{Paraboladef}), the Lagrangian radius $q$ is obtained as the first-contact
point of the paraboloid $\cP_{\bx,c}(\bq)$ with the initial potential 
$\psi_0(\bq)$. Using the spherical symmetry this corresponds to the first-contact
point of the parabola $\cP_{r,c}(q)$ with the curve $\psi_{0q}$, which is 
characterized by the two equations (for contact and tangent slopes)
\beq
\psi_{0q} = \frac{(q-r)^2}{2t}+c , \hspace{1cm} \psi_{0q}' = \frac{q-r}{t} .
\label{contact}
\eeq
Then, as we change the initial potential by an infinitesimal perturbation
$\Delta\psi_{0q'}$, both $c$ and $q$ are modified by amounts 
$\Delta c$ and $\Delta q$, and the second Eq.(\ref{contact}) gives
\beq
\Delta q = \left(\frac{1}{t}-\psi_{0q}''\right)^{-1} \Delta\psi_{0q}'  
\eeq
Then, since $\eta_r=(q/r)^d$ as in Eq.(\ref{etarF}), we have 
$\Delta\eta_r \propto \Delta q \propto \Delta\psi_{0q}'$, whence
\beq
\frac{\cD\eta_r}{\cD\psi_{0q''}} \propto \delta_D'(q''-q) .
\label{derivpsi0}
\eeq
Thus, in agreement with the previous discussion, the derivative 
(\ref{derivpsi0}) vanishes for $q''\neq q$, but it is now the first
derivative of the Dirac distribution. Substituting into Eq.(\ref{saddlepsi0})
gives the linear profiles of the saddle-point as
\beqa
\psi_{0q'} & \propto & \frac{\pl}{\pl q} C_{\psi_{0r}}(q',q) , 
\label{psi0profile} \\
u_{0q'} & \propto & \frac{\pl^2}{\pl q'\pl q} C_{\psi_{0r}}(q',q) 
\propto C_{u_{0r}}(q',q) ,
\label{u0profile}
\eeqa
where we used Eq.(\ref{Cu0r}). Then, the comparison with Eq.(\ref{profile_u0})
shows that we have obtained the same spherical saddle-point as in 
section~\ref{Spherical-saddle-point}. This means that we recover the results
of section~\ref{Quasi-linear} in the quasi-linear limit 
(here $\sigpsir\rightarrow 0$), and of section~\ref{Saddle-point-without-shocks}
in the appropriate rare-event limits 
($\eta_r\rightarrow 0$ or $\eta_r\rightarrow\infty$). In particular, the tail
of the probability distribution reads at leading order as
\beq
\cP(\eta_r) \sim e^{-\frac{1}{2} \psi_0(\bq_1) . C_{\psi_0}^{-1} . \psi_0(\bq_2)}
= e^{-\frac{1}{2} \psi_{0q_1} . C_{\psi_{0r}}^{-1} . \psi_{0q_2}} ,
\label{Ptail}
\eeq
where the exponent is evaluated at the saddle-point, as in Eq.(\ref{Petatau})
to which it is equivalent.

\subsubsection{Taking shocks into account}
\label{shocks}

The advantage of the formulation (\ref{Spsi0q1q2}) in terms of the potential
$\psi_0$ is that we can now handle cases where shocks must be taken into 
account. Note that this applies in particular to the cases $n\geq d-2$ where
the variance $\sigr^2$ of the linear density contrast is divergent.
To this order, we generalize the previous configuration, with a unique
first-contact point between $\cP_{r,c}(q)$ and $\psi_{0q}$, to states where
the initial potential follows the parabola over a finite range $[q_-,q_+]$,
and remains below it elsewhere. This corresponds in particular to a shock
at radius $r$ that contains all the matter that was initially located within
Lagrangian radii $q_-$ and $q_+$ (all this matter merging at position $r$
at time $t$). 
Then, the nonlinear overdensity $\eta_r$ is not modified by infinitesimal
perturbations $\Delta\psi_{0q}$ over $q\notin [q_-,q_+]$ (since they do not
affect the first-contact parabola) and Eq.(\ref{saddlepsi0}) implies
\beq
q\geq 0 : \;\;\; \psi_{0q}= \int_{q_-}^{q_+} \dd q' \, C_{\psi_{0r}}(q,q') f(q') ,
\label{fdef}
\eeq
with some kernel $f(q')$ to be determined (in the case of a unique contact
point, i.e. no shock, we have seen above in Eq.(\ref{psi0profile}) that we
have $f(q')\propto\delta_D'(q'-q_+)$ and $q_-=q_+$). 
Let us note $\bCpsi$ the restriction
of the kernel $C_{\psi_{0r}}(q_1,q_2)$ to the interval $[q_-,q_+]$. Then,
since $\psi_{0q}=\cP_{r,c}(q)$ over the range $[q_-,q_+]$, Eq.(\ref{fdef}) 
implies over this interval:
\beq
q_-\leq q \leq q_+ : \;\; \cP_{r,c} = \bCpsi . f , \;\; \mbox{whence} \;\; 
f= \bCpsi^{\;-1} . \cP_{r,c}
\label{bCpsi}
\eeq
Moreover, substituting Eq.(\ref{fdef}) into the action (\ref{Spsi0q1q2})
yields
\beqa
\lefteqn{\hspace{-0.8cm} \cS = y\eta_r + \frac{\sigpsir^2}{2} \! \int_{q_-}^{q_+}
\!\! \dd q_1 \dd q_2 \, f(q_1) C_{\psi_{0r}}(q_1,q_2) f(q_2) }
\label{Sff} \\
&& \hspace{-1cm} = \! y\eta_r \!+\! \frac{\sigpsir^2\!}{2} \!\! \int_{q_-}^{q_+}
\!\!\! \dd q_1 \dd q_2 \, \cP_{r,c}(q_1) \bCpsi^{\;-1}(q_1,q_2) \cP_{r,c}(q_2)  
\label{SPP} \\
&& \hspace{-1cm} = y\eta_r + \frac{\sigpsir^2}{2} \int_{q_-}^{q_+} \dd q \, 
f(q)\cP_{r,c}(q) . 
\label{SfP}
\eeqa
Next, since $\eta_r$ and $\bCpsi^{\;-1}$ only depend on $q_-$ and $q_+$,
minimizing the action (\ref{SPP}) with respect to the height $c$ of the 
parabola $\cP_{r,c}$ gives
\beq
\frac{\pl\cS}{\pl c} = \sigpsir^2 \int_{q_-}^{q_+} \!\!\!
\dd q_1 \dd q_2 \, \bCpsi^{\;-1}(q_1,q_2) \cP_{r,c}(q_2) = 0 ,
\eeq
whence, using Eq.(\ref{bCpsi}),
\beq
\int_{q_-}^{q_+} \dd q \, f(q) = 0 .
\label{f=0}
\eeq
Thus, in order to minimize the action $\cS[\psi_0]$ over the spherically 
symmetric initial conditions that show a shock at radius $r$ we proceed in 
two steps. First, the minimum of $\cS$ over the class of profiles $\psi_0$ that
follow their first-contact parabola over a finite range $[q_-,q_+]$, to be 
determined afterwards, is obtained by solving Eqs.(\ref{bCpsi}) and (\ref{f=0}). 
This gives the kernel $f$ and the parabola height $c$ as a function of the 
parameters $q_-$ and $q_+$. Second, substituting into the action (\ref{SfP}) 
we minimize $\cS$ over $q_-$ and $q_+$. This provides the minimum of the action
(\ref{Spsi0q1q2}) over all spherical states such that the Eulerian radius $r$
maps to a continuous Lagrangian range $[q_-,q_+]$, provided the saddle-point 
obtained in this fashion remains strictly below the parabola $\cP_{r,c}$ 
outside of the interval $[q_-,q_+]$, which we must check afterwards since 
we have not imposed the constraint $\psi_0\leq\cP_{r,c}$ in the previous 
derivation. 
A priori it could happen that the minimum of the action is reached
for initial configurations that touch the first-contact parabola over disjoint
regions. Then, this would be seen by noticing that the minimum obtained through
the previous procedure touches or crosses the parabola $\cP_{r,c}$ somewhere
outside of the range $[q_-,q_+]$. In such a case, one would need to generalize 
the approach described above to initial states that follow their first-contact 
parabola over $k$ several disjoint intervals $[q_-^{(i)},q_+^{(i)}]$, 
$i=1,..,k$. Starting with the case $k=1$ discussed above, one could add a new
contact interval in a series of steps, until the minimizing profile remains 
below the first-contact parabola everywhere outside of the $k$ contact intervals.
Note that this method also includes the case where some intervals are reduced to
a point, which corresponds to the limit $q_+^{(i)}-q_-^{(i)}\rightarrow 0$.
Thus, this covers the case where the parabola would only have two (or a few)
isolated first-contact points.

Finally, we must specify the relation between the overdensity $\eta_r$ and the
Lagrangian coordinates $q_-$ and $q_+$. We must separate the cases of large 
overdensities and underdensities as
\beq
\eta_r>1: \;\;\; \eta_r=\left(\frac{q_+}{r}\right)^d , \;\;\;
\eta_r<1: \;\;\; \eta_r=\left(\frac{q_-}{r}\right)^d .
\label{etarq+q-}
\eeq
Indeed, since we have a shock at radius $r$, with a finite mass 
$m^{\rm shock}\propto (q_+^d-q_-^d)$, the density within radius $r$ and the 
enclosed mass are ambiguous. It is actually discontinuous at $r$, going from 
$m_-$ to $m_+$ with $m_+-m_-=m^{\rm shock}$. This also means that the minimum 
discussed above is unstable, in the sense that an infinitesimal perturbation will
move this mass inward or outward, so that $m(<r)$ goes to $m_-$ or $m_+$. Then, 
if we consider rare and large overdensities, the probability $\cP(\eta_r)$ will 
be governed by the initial conditions close to the previous minimum such that 
$m(<r)\simeq m_+$, which leads to the first Eq.(\ref{etarq+q-}).
Similarly, for extreme underdensities we obtain the second Eq.(\ref{etarq+q-}).
Note that this also shows that the action $\cS$ is not regular at the minimum 
obtained above and going beyond the leading term given by the exponential as in
Eq.(\ref{Ptail}) would require a careful analysis. However, this discontinuity
is not of the same kind as the one encountered for collisionless gravitational
collapse, associated with radial orbit instability and recalled in 
section~\ref{Rare-density} above, since by using (\ref{etarq+q-}) we simply
consider the left or right limit of $\cS$ (with respect to any degree of freedom)
and not an isolated point of zero measure.

\subsubsection{Computation of the saddle-point}
\label{Computation}

In practice, it can be difficult to solve Eq.(\ref{bCpsi}) for the kernel $f$,
and we did not obtain a general solution. However, for the power-law power spectra
(\ref{ndef}) with low integer values of $n$ and $d$, where the radial potential 
correlation $C_{\psi_{0r}}(q_1,q_2)$ takes the simple forms given in 
Table~\ref{Table_correl}, it is possible to derive explicit expressions for
$f(q)$ from Eq.(\ref{bCpsi}). It is convenient to first write $f(q)$ as a 
derivative,
\beq
f(q)= \frac{\dd g}{\dd q} \;\;\;\; \mbox{and} \;\;\;\;
g(q_+)=g(q_-)= 0 .
\label{fgdef}
\eeq
In the second equality we used Eq.(\ref{f=0}), which yields $g(q_+)=g(q_-)$, and
the fact that $g(q)$ being defined up to an additive constant we can choose
$g(q_-)=0$.
Next, substituting into the first Eq.(\ref{bCpsi}), integrating by parts
and derivating once, we obtain
\beq
q_- \leq q \leq q_+ : \;\;\; \frac{r-q}{t} = \int_{q_-}^{q_+} \dd q' \, 
C_{u_{0r}}(q,q') g(q') ,
\label{intgq}
\eeq
where we used the first relation (\ref{Cu0r}). Then, we can devise a systematic
procedure to solve Eq.(\ref{intgq}) when $d-n$ is an odd integer, that is for
\beq
d=n+1+2\ell  \;\;\; \mbox{with} \;\;\; \ell \in \mathbb{N} .
\label{elldef}
\eeq
Indeed, from Eqs.(\ref{Cu0r}) and (\ref{Cdr1dr2_3}) we have, with a normalization
factor $D$,
\beq
C_{u_{0r}}(q_1,q_2) = D \int_0^{\infty} \dd k \, k^n \, \Phi_k(q_1)
\Phi_k(q_2) ,  
\label{Cu0Phik}
\eeq
where we introduced the eigenfunctions on $[0,\infty[$ of the linear operator
$\cL$,
\beq
q \geq 0 : \;\; \Phi_k(q) = \frac{J_{d/2}(kq)}{(kq)^{d/2-1}} , \;\;\;\;\;
\cL . \Phi_k = k^2 \, \Phi_k ,
\label{Phikdef}
\eeq
with
\beq
\cL = - \frac{\dd^2}{\dd q^2} + \frac{1-d}{q} \frac{\dd}{\dd q} 
+ \frac{d-1}{q^2} .
\label{Ldef}
\eeq
On the other hand, we note from standard properties of Bessel functions 
(i.e. Hankel transforms) that
\beq
\int_0^{\infty} \dd k \, k^{d-1} \, \Phi_k(q_1) \Phi_k(q_2) = 
q_1^{1-d} \, \delta_D(q_1-q_2) .
\label{PhikDirac}
\eeq
Therefore, noting $\cL^{\dagger}$ the adjoint of $\cL$,
\beq
\cL^{\dagger} = - \frac{\dd^2}{\dd q^2} + \frac{d-1}{q} \frac{\dd}{\dd q} ,
\label{Ladjoint}
\eeq
we have when condition (\ref{elldef}) is satisfied, for $q_- \leq q \leq q_+$,
\beqa
\lefteqn{\int_{q-}^{q_+} \dd q' \, C_{u_{0r}}(q,q') 
\left( \cL^{\dagger\ell} q'^{d-1} \frac{r-q'}{Dt} \right)} \nonumber \\
&& =  \int_{q-}^{q_+} \dd q' \int_0^{\infty} \dd k k^n \Phi_k(q) 
\left( \cL^{\ell} \Phi_k(q') \right) q'^{d-1} \frac{r-q'}{t} + {\rm b.t.} 
\nonumber \\
&& = \frac{r-q}{t} + {\rm b.t.} \label{gsol2}
\eeqa
where we used Eqs.(\ref{Phikdef}), (\ref{PhikDirac}), and ``b.t.'' stands for
boundary terms at $q'=q_{\pm}$ generated by the integrations by parts over 
$q'$. Thus, we obtain the solution of Eq.(\ref{intgq}) as
\beq
g(q)= \cL^{\dagger\ell} q^{d-1} \frac{r-q}{Dt} + {\rm b.t.} 
\label{gqsol}
\eeq
where the boundary terms are of the Dirac type, such as $\delta_D(q-q_{\pm})$
and its derivatives, localized on the boundaries $q_{\pm}$.
Using the relation (\ref{elldef}) this yields
\beq
\ell=0 : \; g(q)\propto q^n\frac{r-q}{t} , \;\;\;\;
\ell \geq 1 : \; g(q)\propto \frac{q^n r}{t} + {\rm b.t.}
\label{gqsol2}
\eeq
Then, the kernel $f(q)$ can be obtained from Eq.(\ref{fgdef}). However, from
Eq.(\ref{fdef}) we can see that $g(q)$ directly gives the velocity profile
of the saddle-point as
\beq
u_{0q} = \int_{q-}^{q_+} \dd q' \, C_{u_{0r}}(q,q') g(q') ,
\label{u0qgq}
\eeq
which follows the linear slope (\ref{intgq}) in the interval $[q_-,q_+]$,
while the action (\ref{SfP}) writes
\beq
\cS =  y\eta_r + \frac{\sigpsir^2}{2} \int_{q_-}^{q_+} \dd q \, g(q) 
\frac{r-q}{t} .
\label{SgPp}
\eeq
Since the second equality (\ref{fgdef}) has already fulfilled the constraint 
(\ref{f=0}), associated with the minimization with respect to the parabola
height $c$, we only need to minimize $\cS$ over $q_-$ and $q_+$ to complete
the derivation of the saddle-point and of the rare-event tails of the
distribution $\cP(\eta_r)$ (after we check that this minimum does not give
rise to other shocks). In fact, from Eq.(\ref{SgPp}), in the present case the
analog of Eqs.(\ref{Petatau}) and (\ref{Ptail}) reads as
\beq
\cP(\eta_r) \sim e^{-\frac{1}{2}\int_{q_-}^{q_+} \dd q \, g(q) 
\frac{r-q}{t}} .
\label{Ptailgq}
\eeq
Thus, for overdensities $q_+$ is defined as a function of $\eta_r$ from
the first Eq.(\ref{etarq+q-}), and we only need to minimize $\cS$ over
$q_-$ to determine the saddle-point and the high-density tail (\ref{Ptailgq}).
For underdensities $q_-$ is set by the second Eq.(\ref{etarq+q-}) and we
must minimize $\cS$ over $q_+$ to obtain the low-density tail.

Again, this approach only holds for the rare-event tails, in both limits
of large scale/early time at fixed density, and of extreme density at fixed
scale and time, where the probability (\ref{Ptailgq}) is much smaller than
unity. In such regimes, the expression (\ref{Ptailgq}) gives the
asymptotic tail of the probability distribution at leading order.

\subsubsection{Tails of the velocity divergence distribution}
\label{tails-theta}

The method described in the previous section can also be applied to the
distribution of the velocity divergence, $\cP(\ctheta_r)$.
As for the quasi-linear regime studied in section~\ref{Velocity-divergence}
the only difference as we go from the overdensity to the velocity divergence 
is to replace $\eta_r$ by $\ctheta_r$ in the action $\cS$. In particular,
we recover the same saddle-point and the same rare-event tail (\ref{Ptailgq}),
and we only need to specify the relation between $\ctheta_r$ and the Lagrangian
radii $q_-$ and $q_+$.
Applying the discussion below Eq.(\ref{etarq+q-}) to $\ctheta_r$, which is
given by Eq.(\ref{thetaqr}) for regular points, we now write
\beqa
\ctheta_r>0 & : & \;\;\; \ctheta_r=d\left(\frac{q_+}{r}-1\right) , \\
\ctheta_r<0 & : & \;\;\; \ctheta_r=d\left(\frac{q_-}{r}-1\right) .
\label{thetarq+q-}
\eeqa
This gives back the relation (\ref{theta_eta}), so that 
the tails of the distribution $\cP(\ctheta_r)$ can again be obtained (at leading
order) from the tails of $\cP(\eta_r)$ by substituting the second 
Eq.(\ref{theta_eta}).

\subsubsection{Case $n=0,\;d=1$: white-noise initial velocity}
\label{Casen0d1}

Let us describe how this procedure works for the case $\{n=0,d=1\}$
(whence $\ell=0$ in Eq.(\ref{elldef})), where
the variance $\sigr^2$ of the linear density contrast is actually divergent,
so that shocks must always be taken into account. Using the initial velocity
correlation given in Table~\ref{Table_correl}, we immediately obtain
the solution of Eq.(\ref{intgq}),
\beq
q_-<q<q_+ : \;\; g(q)= \frac{r-q}{t} , 
\;\;\;\; g(q_{\pm}) = 0 ,
\label{gn0d1}
\eeq
which gives the linear velocity profile
\beq
q\in [q_-,q_+] :  u_{0q}= \frac{r-q}{t} , \;\;\;\; 
q\notin [q_-,q_+] :  u_{0q}= 0 .
\label{u0n0d1}
\eeq
Note that $g(q)$ is singular (discontinuous) at $q_{\pm}$, since $g(q_{\pm})=0$,
which gives Dirac terms $\delta_D(q-q_{\pm})$ for the kernel $f(q)$.
Then, the action (\ref{SgPp}) writes
\beq
\cS=y\eta_r+\frac{\sigpsir^2}{6 t^2} \left[(q_+-r)^3+(r-q_-)^3\right] .
\label{Sn0d1}
\eeq
For overdensities, $\eta_r>1$, the upper boundary $q_+$ is given by 
(\ref{etarq+q-}), $q_+=r \eta_r$, whereas $q_-$ is determined by minimizing
$\cS$. This gives $q_-=r$, since $q_->r$ is excluded as it would give further
shocks over the range $[r,q_-]$: we must check that the profile $\psi_{0r}$ 
does not cross the parabola $\cP_{r,c}$ outside of $[q_-,q_+]$, that is, that 
the velocity profile does not create a larger shock. For the simple profile 
(\ref{u0n0d1}) we do not need to consider $\psi_{0r}$ to check that no shocks
appear beyond $[q_-,q_+]$. Thus, the system is motionless over $[0,r[$ and 
$]q_+,+\infty[$, and particles in the range $]r,q_+[$ have the linear initial 
profile (\ref{u0n0d1}) and simultaneously merge at radius $r$ at time $t$.
Note that there appears a rarefaction interval (empty region) over $]r,q_+[$
as the initial velocity is discontinuous at $q_+$. This is due to the large
power at high $k$ of the initial white-noise energy spectrum (\ref{E0n}).
We can see that there are no other shocks over disjoint regions that modify
the density within radius $r$ at time $t$,
so that we have obtained the true minimum over symmetric initial conditions.
Then, Eq.(\ref{Ptailgq}) gives
\beq
\eta_r>1 : \;\; \cP(\eta_r) \sim e^{-r^3(\eta_r-1)^3/(6t^2)} .
\label{Psupn0d1}
\eeq
For underdensities, we obtain by a similar reasoning $q_-=r \eta_r$, $q_+=r$,
and
\beq
\eta_r<1 : \;\; \cP(\eta_r) \sim e^{-r^3(1-\eta_r)^3/(6t^2)} .
\label{Pinfn0d1}
\eeq
From Eq.(\ref{theta_eta}) we obtain at once for the velocity divergence
$\ctheta_r$, which is also the dimensionless velocity increment 
(\ref{theta1d}), the tails
\beq
\cP(\ctheta_r) \sim e^{-r^3|\ctheta_r|^3/(6t^2)} .
\label{Pthetan0d1}
\eeq
Since the case $\{n=0,d=1\}$ of white-noise one-dimensional initial velocity
can actually be solved \cite{Frachebourg2000,Valageas2009},
we can compare the results (\ref{Psupn0d1})-(\ref{Pinfn0d1}) with the exact
distribution $\cP(\eta_r)$. Using the notations of \cite{Valageas2009},
it is known to display the asymptotic behaviors at large scales,
\beqa
\lefteqn{\hspace{-1cm}  X\gg 1,\;\; |\eta_r-1|\gg X^{-1}, \;\; 
\eta_r \gg X^{-3} :} \nonumber \\
&& \cP(\eta_r) \sim e^{-\omega_1 X|\eta_r-1|-X^3|\eta_r-1|^3/12} ,
\label{PXsup}
\eeqa
and at small scales,
\beq
X\ll 1, \;\; \eta_r\gg X^{-1} : \;\; \cP(\eta_r) \sim 
e^{-\omega_1 X\eta_r-X^3\eta_r^3/12} ,
\label{PXinf}
\eeq
where we did not write power-law prefactors, and $-\omega_1$ is the zero
of the Airy function $\Ai(x)$ closest to the origin ($\omega_1\simeq 2.338$).
Here $X$ is the dimensionless length of the interval $[-r,r]$ of radius $r$,
whence of size $x=2r$,
\beq
X=\frac{2r}{(2Dt^2)^{1/3}} = \frac{2r}{(4t^2)^{1/3}} , \;\; \mbox{hence} 
\;\; \frac{X^3}{12}=\frac{r^3}{6t^2} ,
\eeq
since the normalization used in the present paper corresponds to $D=2$
\footnote{In \cite{Valageas2009}
the initial velocity correlation was normalized as 
$\lag u_0(q_1) u_0(q_2)\rag=D \, \delta_D(q_1-q_2)$,
so that the normalization used in the present paper, given in 
Table~\ref{Table_correl}, corresponds to $D=2$. Indeed, going from the velocity
$u_0(q)$ to its radial component $u_{0r}$, which in this one-dimensional case
writes $u_{0r}=[u_0(r)+u_0(-r)]/2$ (symmetric component), 
multiplies the two-point correlation by a factor $1/2$.}.
Thus, we can check that for large overdensities our saddle-point result 
(\ref{Psupn0d1}) agrees with the exact results (\ref{PXsup})-(\ref{PXinf})
at leading order, at both large and small scales. Of course, this only applies
to the rare-event tail, which is repelled to larger densities, $\eta_r\gg 1/X$,
at small scales in the highly nonlinear regime.
For large underdensities, we also recover the exact result (\ref{PXsup})
at leading order, that applies to large scales. It cannot give the low-density
tail in the highly nonlinear regime because this no longer corresponds to
rare events. Indeed, as described in \cite{Valageas2009}, at low densities
the distribution $\cP(\eta_r)$ shows an inverse square root tail, 
$\propto 1/\sqrt{\eta_r}$, and a Dirac contribution, $\delta_D(\eta_r)$, that
both have a weight, of order $e^{-\omega_1 X-X^3/12}$ at large scales, that
becomes of order unity at small scales. 
In fact, on small scales most cells of radius $r$ are actually empty, so that 
there no longer exists a rare-underdensities tail.
Note that this feature can actually be seen on the saddle-point result 
(\ref{Pinfn0d1}), as we can see that for $r < t^{2/3}$ the exponential becomes
of order unity for $\eta_r=0$. This shows that empty or almost empty regions
are no longer rare, and that the distribution $\cP(\eta_r)$ over this range
cannot be obtained by a saddle-point approach of the type described in this
article.

Finally, it is interesting to note that Eq.(\ref{Psupn0d1})
agrees with the behavior that would be obtained by a naive extension 
of Eq.(\ref{Petasupexp}) to $\{n=0,d=1\}$, even though the derivation of 
Eq.(\ref{Petasup}) does not apply to this case (the variance $\sigr^2$ is even
divergent). On the other hand, for underdensities Eq.(\ref{Petainfexp}) would
give $\ln\cP(\eta_r) \sim - t^{-2} r^3 \eta_r \rightarrow 0$ for 
$\eta_r \rightarrow 0$.
This shows at once that this cannot give the low-density part of the 
probability distribution, since we do not find a rare-event tail 
($\ln\cP(\eta_r)\nrightarrow-\infty$) but a probability of order unity,
which a saddle-point approach cannot describe. As discussed above, this
is not a failure of Eq.(\ref{Petainfexp}), since there is no rare
low-density tail as empty regions occur with a finite probability, that goes
to unity at small scales.

Of course, the discussion above also applies to the distribution 
$\cP(\ctheta_r)$. Thus, the tail (\ref{Pthetan0d1}) agrees with the exact
result for large positive $\ctheta_r$ in all regimes, and for negative
$\ctheta_r$ in the quasi-linear regime, while on small scales, in the highly
nonlinear regime, cells with $\ctheta_r\simeq -1$ (associated with almost
empty domains) are no longer rare and cannot be described by the method
used here. Again, the scalings obtained in the exponential (\ref{Pthetan0d1})
agree with a naive extension of Eq.(\ref{Pthetasup}) while the extension
of Eq.(\ref{Pthetainf}) correctly signals the absence of rare low-$\ctheta_r$
tail.

\subsubsection{Case $n=0,\;d=3$}
\label{Casen0d3}

We now consider the case $\{n=0,d=3\}$, which gives $\ell=1$ in 
Eq.(\ref{elldef}). From Eq.(\ref{gqsol2}) the regular part of $g(q)$ is 
proportional to $r/t$, and we find for the solution of
Eq.(\ref{intgq}) with the normalization of $C_{u_{0r}}$ given in 
Table~\ref{Table_correl},
\beq
g(q) = \frac{2r}{3t}+\frac{rq_-}{3t} \, \delta_D(q-q_-) 
+\frac{2rq_+-3q_+^2}{3t} \, \delta_D(q-q_+) .
\label{gn0d3}
\eeq
Here the Dirac terms should be understood as $\delta_D[q-(q_{\pm}\mp\epsilon)]$
with $\epsilon\rightarrow 0^+$ (i.e. they have an integral weight of unity 
within $[q_-,q_+]$). This gives the linear velocity profile
\beqa
\lefteqn{\hspace{-3.9cm} q<q_- \!: \; u_{0q} = \frac{q(r-q_-)}{tq_-} ,  \;\;\;\; 
q\in[q_-,q_+] \!: \; u_{0q} = \frac{r-q}{t} ,} \nonumber \\
&& \hspace{-4.1cm} q>q_+ \! : \; u_{0q} = \frac{q_+^2(r-q_+)}{tq^2} ,
\label{u0n0d3}
\eeqa
and the action
\beq
\cS=y\eta_r+\frac{\sigpsir^2}{6 t^2} 
\left( 4r^2q_+-6rq_+^2+3q_+^3-r^2q_- \right) .
\label{Sn0d3}
\eeq
For overdensities, $\eta_r>1$, the minimization over $q_-$ gives $q_-=q_+$.
Indeed, contrary to the previous case, $\{n=0,d=1\}$, it is now possible to
have $q_->r$ without building a larger shock, as already seen from 
Fig.~\ref{figvel_profile}, since we actually recover the
saddle-point of section~\ref{Saddle-point-without-shocks} without shocks, 
as $q_-=q_+$ (i.e. an isolated contact point between the parabola 
$\cP_{r,c}(q)$ and $\psi_{0q}$). Then, Eq.(\ref{Ptailgq}) writes as
$\cP(\eta_r) \sim e^{-r^3(\eta_r^{1/6}-\eta_r^{1/2})^2/(2t^2)}$, in agreement
with Eq.(\ref{Petataun}) and Table~\ref{Table_Peta_ql}, and we recover the tail
(\ref{Petasup}) and Table~\ref{Table_tailnoshock}.
Indeed, the constraint in (\ref{Petasup}) is satisfied, so that we already
knew that we had to recover that regular saddle-point.

For underdensities, $\eta_r<1$, the minimization over $q_+$ gives $q_+=2r/3$,
so that we only have a shock (i.e. $q_-<q_+$) for $q_-<2r/3$, that is for low
densities below $\eta_*=(2/3)^3$. This agrees with the discussion in 
section~\ref{Spherical-saddle-point} and Table~\ref{Table_shocks}, 
where we found that the regular saddle-point
(\ref{profile}) only develops a shock after a finite time, that is below
a nonzero density contrast threshold. 
In the quasi-linear limit, $r^3/t^2\gg 1$, where the range $\eta_*<\eta_r<1$
already corresponds to large density fluctuations, we can also use the method 
described in section~\ref{Quasi-linear} and we obtain the result of
Table~\ref{Table_Peta_ql}. The analysis described above from the action
(\ref{Sn0d3}) provides the density threshold $\eta_*=(2/3)^3$ written in
that Table. From the relation (\ref{theta_eta}), this also gives the velocity
divergence threshold $\ctheta_*=-1$, above which the quasi-linear distribution 
$\cP(\ctheta_r)$ is given by Table~\ref{Table_Ptheta_ql}.

For larger underdensities, $0<\eta_r<\eta_*$, we have $q_-<q_+$ and we
must use the action (\ref{Sn0d3}) that takes shocks into account, since
we can check that the profile (\ref{u0n0d3}) is
valid (there are no other shocks that modify the density within radius $r$).
Then, Eq.(\ref{Ptailgq}) gives
\beq
0< \eta_r< (2/3)^3 : \;\; \ln\cP(\eta_r) \sim -\frac{r^3}{6t^2} 
\left(\frac{8}{9}-\eta_r^{1/3}\right) .
\label{Pn0d3}
\eeq
Of course, we can check that at point $\eta_r=\eta_*$ Eq.(\ref{Pn0d3}) is
equal to the result (\ref{Petataun}), shown in Table~\ref{Table_Peta_ql},
which is provided by the regular saddle-point. We can see that below this
threshold the dependence on $\eta_r$ of $\cP(\eta_r)$ is modified by
shocks. Thus, Eq.(\ref{Pn0d3}) provides the very low density tail of
$\cP(\eta_r)$ in the quasi-linear limit, $r^3/t^2 \gg 1$.

In the nonlinear regime, $r^3/t^2\ll 1$, the result (\ref{Pn0d3}) becomes
of order unity over the range $0\leq \eta_r<\eta_*$. Note that this agrees
with the  naive extension of Eq.(\ref{Petainfexp}), which also yields the
correct exponents of $r,t$ and $\eta_r$.
Then, as for the case $\{n=0,d=1\}$ studied in the 
previous section~\ref{Casen0d1}, there is no rare underdensities tail,
and empty or almost empty regions are not rare. More precisely, there is
no exponential decay of $\cP(\eta_r)$ at low $\eta_r$, but power-law prefactors
associated with subleading order terms may give either a power-law growth
or falloff at $\eta_r\rightarrow -d$. However, the precise behavior
of the distribution $\cP(\eta_r)$ for $\eta_r\simeq 0$ cannot be derived through
a saddle-point method since there is no rare tail and one should take into account
many possible initial configurations.

The previous results directly extend to the distribution $\cP(\ctheta_r)$.
Thus, for large $\ctheta_r$ we recover the saddle-point of 
section~\ref{Saddle-point-without-shocks} and the tail (\ref{Pthetasup}),
which applies to both quasi-linear and highly nonlinear regimes.
For low  $\ctheta_r$, Eq.(\ref{Pn0d3}) gives
\beq
-3< \ctheta_r< -1 : \;\; \ln\cP(\ctheta_r) \sim \frac{r^3}{18t^2} 
\left(\frac{1}{3}+\ctheta_r\right) .
\label{Pthetan0d3}
\eeq
Again, this provides the very low-$\ctheta_r$ tail in the quasi-linear regime,
which disappears in the highly nonlinear regime where there is no longer
a rare-event low-$\ctheta_r$ tail, and this behavior can also be seen
in the naive extension of Eq.(\ref{Pthetainf}).

\subsubsection{Case $n=-1,\;d=2$}
\label{Casen1d2}

\begin{table*}
\begin{center}
\begin{tabular}{c|c||c|c||c|c}
\,$n$\, & \,$d$\, & \,$\ln\cP(\eta_r)$ for $\eta_r>1$\, & \,$\ln\cP(\eta_r)$ for $\eta_r<1$\, & \,$\ln\cP(\ctheta_r)$ for $\ctheta_r>0$\, & $\ln\cP(\ctheta_r)$ for $\ctheta_r<0$ \rule[-0.25cm]{0cm}{0.7cm} \\ \hline\hline
0 & 1 & $-\frac{r^3}{6t^2}\,(\eta_r-1)^3$ & $-\frac{r^3}{6t^2}\,(1-\eta_r)^3$ & $-\frac{r^3}{6t^2}\,\ctheta_r^3$ & $-\frac{r^3}{6t^2}\,(-\ctheta_r)^3$ \rule[-0.27cm]{0cm}{0.8cm} \\ \hline
0 & 3 & no shock & $-\frac{r^3}{6t^2}\,\left(\frac{8}{9}-\eta_r^{1/3}\right)$ \, for $\eta_r<\frac{8}{27}$ & no shock & $\frac{r^3}{18t^2}\,\left(\frac{1}{3}+\ctheta_r\right)$ \, for $\ctheta_r<-1$ \rule[-0.27cm]{0cm}{0.8cm} \\ \hline
-1 & 2 & no shock & \,$-\frac{r^2}{8t^2}\,\left[1-\ln(4\eta_r)\right]$ \, for $\eta_r<\frac{1}{4}$\, & no shock & \,$-\frac{r^2}{8t^2}\,[1-2\ln(2+\ctheta_r)]$ \, for $\ctheta_r<-1$ \rule[-0.27cm]{0cm}{0.8cm} \\
\end{tabular}
\end{center}
\caption{The tails of the distributions $\cP(\eta_r)$ and $\cP(\ctheta_r)$
for the initial conditions of Table~\ref{Table_correl}, in cases where shocks
appear. 
These results also hold for the quasi-linear limit, $t\rightarrow 0$ or 
$r\rightarrow\infty$, at fixed $\eta_r$ or $\ctheta_r$, but only below the 
thresholds given in this Table for the last two rows (higher densities
and velocity divergences are described by Tables~\ref{Table_Peta_ql}, 
\ref{Table_Ptheta_ql}). For $\{n=0,d=1\}$, these results hold in the quasi-linear
limit for any $\eta_r\neq 1$ and $\ctheta_r\neq 0$. In the highly nonlinear 
limit, $r\rightarrow 0$ or $t\rightarrow\infty$, the rare-event tails at 
low $\eta_r$ and $\ctheta_r$ disappear as low densities and velocity divergences
are no longer rare ($\ln\cP$ in this Table becomes of order unity and these
formulae are no longer valid).
Cases marked as ``no shock'' correspond to saddle-points that do not give rise
to shocks, so that the results of Tables~\ref{Table_Peta_ql}, 
\ref{Table_Ptheta_ql} and \ref{Table_tailnoshock} are valid.}
\label{Table_tailshock}
\end{table*}

We now turn to the case $\{n=-1,d=2\}$. As seen from 
Table~\ref{Table_tailnoshock}, shocks should only appear for underdensities,
as in the previous case $\{n=0,d=3\}$. The extension of Eq.(\ref{Petainfexp})
gives a vanishing power of $\eta_r$, so we can expect a logarithmic dependence
on $\eta_r$ (or a finite asymptote) for $\ln\cP(\eta_r)$ at low densities.
We again have $\ell=1$ in Eq.(\ref{elldef}), so that the regular part of
$g(q)$ is obtained from Eq.(\ref{gqsol2}) as $\propto r/(tq)$, and we find
\beq
g(q) = \frac{r}{2tq}+\frac{r}{2t} \, \delta_D(q-q_-) 
+\frac{r-2q_+}{2t} \, \delta_D(q-q_+) ,
\label{gn1d2}
\eeq
and
\beqa
\lefteqn{\hspace{-3.9cm} q<q_- \!: \; u_{0q} = \frac{q(r-q_-)}{tq_-} ,  \;\;\;\; 
q\in[q_-,q_+] \!: \; u_{0q} = \frac{r-q}{t} ,} \nonumber \\
&& \hspace{-4.1cm} q>q_+ \! : \; u_{0q} = \frac{q_+(r-q_+)}{tq} ,
\label{u0n1d2}
\eeqa
while the action writes
\beq
\cS= y\eta_r+\frac{\sigpsir^2}{4 t^2} 
\left( r^2 \ln\frac{q_+}{q_-} + 2r^2 - 4rq_+ + 2q_+^2 \right) .
\label{Sn1d2}
\eeq
As expected, for overdensities we recover $q_-=q_+$ (i.e. the regular 
saddle-point without shock) and 
$\cP(\eta_r) \sim e^{-r^2(\sqrt{\eta_r}-1)^2/(2t^2)}$, in agreement
with Eq.(\ref{Petataun}) and Table~\ref{Table_Peta_ql}, and we also recover
the tail (\ref{Petasup}) and Table~\ref{Table_tailnoshock}.

For underdensities we obtain $q_+=r/2$, so that we only have a shock below
$\eta_*=1/4$, which provides the density threshold written in 
Table~\ref{Table_Peta_ql}. Thus, as for the case $\{n=0,d=3\}$ and in agreement
with section~\ref{Spherical-saddle-point}, in the quasi-linear regime for
rare underdensities in the range $\eta_*<\eta_r<1$ the distribution $\cP(\eta_r)$
is obtained from the method described in section~\ref{Quasi-linear}, which
gives the result of Table~\ref{Table_Peta_ql}. In terms of the velocity
divergence $\ctheta_r$, this also provides the threshold $\ctheta_*=-1$
of Table~\ref{Table_Ptheta_ql}, and above this threshold the distribution
$\cP(\ctheta_r)$ is given by Eq.(\ref{Pthetataun}) and 
Table~\ref{Table_Ptheta_ql} in the quasi-linear limit.

For lower densities, in the range $0<\eta_r<\eta_*$, we obtain from 
Eq.(\ref{Sn1d2})
\beqa
\hspace{-1cm} 0<\eta_r<1/4 & : &\;\; \ln\cP(\eta_r) \sim -\frac{r^2}{8t^2} 
[1-\ln(4\eta_r)] , \label{Pinfn1d2a} \\ 
\mbox{whence} & & \cP(\eta_r) \sim (4\eta_r)^{r^2/(8t^2)} \, e^{-r^2/(8t^2)} . 
\label{Pinfn1d2}
\eeqa
Again, at the transition $\eta_r=\eta_*$ Eq.(\ref{Pinfn1d2a}) is equal to
Eq.(\ref{Petataun}) shown in Table~\ref{Table_Peta_ql}.
Thus, we obtain as expected a logarithmic dependence over $\eta_r$ for
$\ln\cP(\eta_r)$,
in agreement with (\ref{Petainfexp}). In the quasi-linear regime, $r^2/t^2\gg 1$,
Eq.(\ref{Pinfn1d2}) means that the low-density tail has a power-law behavior
$\cP(\eta_r) \sim \eta_r^{\alpha}$, with an exponent $\alpha \sim r^2/(8t^2)$
that grows at large scales and early times, so that the low-density falloff
is increasingly sharp. However, because there could be
a power-law prefactor in Eq.(\ref{Pinfn1d2}) due to sub-leading corrections
to the steepest-descent approximations, this is unlikely to give the exact
exponent $\alpha$ but only its behavior at large $r$ and small $t$. 
Then, in the nonlinear regime, $r^2/t^2\ll 1$, Eq.(\ref{Pinfn1d2}) is not 
sufficient to give the behavior of $\cP(\eta_r)$ for $\eta_r\rightarrow 0$, 
as these prefactors may either give a positive or negative exponent. 
This limiting configuration between the cases $n>-1$, where empty or almost 
empty regions have a finite probability at small scales, and $n<-1$, where 
low densities exhibit an exponential tail of the form (\ref{Petainfexp}), 
requires a finer analysis in the nonlinear regime.

For large velocity divergence $\ctheta_r$ we  recover the tail (\ref{Pthetasup})
associated with the regular saddle-point while for low $\ctheta_r$ 
Eqs.(\ref{Pinfn1d2a}) and (\ref{theta_eta}) yield
\beqa
\hspace{-0.6cm} -2\!<\!\ctheta_r\!<\!-1 & \!\!: & \ln\cP(\ctheta_r) \sim 
\frac{-r^2}{8t^2} [1\!-\!2\ln(2\!+\!\ctheta_r)] , \label{Pthetainfn1d2a} \\
\mbox{whence} & & \!\! \cP(\ctheta_r) \sim  (2\!+\!\ctheta_r)^{r^2/(4t^2)} 
\, e^{-r^2/(8t^2)} .
\label{Pthetainfn1d2}
\eeqa
This only gives the behavior at low $\ctheta_r$ in the quasi-linear regime,
$r^2/t^2\gg 1$, as in the nonlinear regime power-law prefactors may lead
either to a growth or decay of $\cP(\ctheta_r)$, but in both cases there is
no rare-event tail (i.e. no exponential falloff).

\subsubsection{Summary for low integer $n$ and $d$}
\label{Casend}

We summarize in Table~\ref{Table_tailshock} the results obtained from
the approach developed in the previous sections for the tails of the 
distributions $\cP(\eta_r)$ and $\cP(\ctheta_r)$, for the initial conditions
of Table~\ref{Table_correl} where shocks cannot be neglected. This
complements the Table~\ref{Table_tailnoshock} that applies to cases
where the saddle-point does not form shocks.

These rare-event tails apply to the large fluctuation limits, 
$\eta_r\rightarrow \infty$ and $\eta_r\rightarrow 0$, or
$\ctheta_r\rightarrow \infty$ and $\ctheta_r\rightarrow -d$, at fixed
time and scale, whatever the value of the variance $\sigr^2$ or $\sigpsir^2$.

They also apply to the quasi-linear limit, $\sigr\rightarrow 0$ or 
$\sigpsir\rightarrow 0$, that is at early times or large scales,
at fixed $\eta_r$ and $\ctheta_r$, below some finite thresholds $\eta_*$
and $\ctheta_*$ in the two cases $\{n=0,d=3\}$ and $\{n=-1,d=2\}$, and for
any $\eta_r\neq 1$ and $\ctheta_r\neq 0$ in the case $\{n=0,d=1\}$.  
 
For these three cases, in the highly nonlinear regime, $r\rightarrow 0$
or $t\rightarrow\infty$, the rare-event tail at low $\eta_r$ and $\ctheta_r$
disappears as these results predict that $\ln\cP$ becomes of order unity.
Then, low densities and velocity divergences are no longer rare (but the
probability distribution might still decay as a power law) and cannot be 
described by a saddle-point approach.

\subsection{Mass function of point-like singularities}
\label{Mass function}

\begin{table}
\begin{center}
\begin{tabular}{c|c||c}
\,$n$\, & \,$d$\, & \;$\ln[n(m)]$ \; for \; $m\rightarrow\infty$ \rule[-0.2cm]{0cm}{0.5cm} \\ \hline\hline
0 & 1 & $-m^3/(48 t^2\rho_0^3)$ \rule[-0.2cm]{0cm}{0.6cm} \\ \hline
0 & 3 & $-3m/(8\pi t^2\rho_0)$ \rule[-0.2cm]{0cm}{0.6cm} \\ \hline
-1 & 2 & $-m/(2\pi t^2\rho_0)$ \rule[-0.2cm]{0cm}{0.6cm} \\ \hline
-2 & 1 & $-m/(4 t^2\rho_0)$ \rule[-0.2cm]{0cm}{0.6cm} \\ \hline
-2 & 3 & $-\frac{5}{8 t^2}\,\left[3m/(4\pi\rho_0)\right]^{1/3}$ \rule[-0.25cm]{0cm}{0.7cm} \\
\end{tabular}
\end{center}
\caption{Large-mass tail of the mass function $n(m)$ of point-like
objects, for the initial conditions of Table~\ref{Table_correl}, from
Eq.(\ref{Petan0m}).}
\label{Table_n0m}
\end{table}

As the density and velocity fields evolve through the nonlinear Burgers
dynamics, starting from the scale-free initial conditions (\ref{ndef}),
the system displays an intricate self-similar progression from smaller
to larger scales. In particular, collisions between particles create
discontinuities (shocks) of dimension $d-1$, $d-2$, .., down to $0$,
lower dimensional objects arising at the intersection of higher-dimension
structures. For instance, if $d=3$, once particles have formed a two-dimensional
sheet of finite surface density, orthogonal to the direction of the largest
eigenvalue of the initial tidal tensor, they keep moving within this surface
and form critical lines and nodes.
Then, the typical distance between such objects increases as
$L(t)$, as in (\ref{Lt}), and their mass grows accordingly.
The mass and the overdensity within radius $r$ about a point $\bx$, 
contained in such a $D$-dimensional structure, scale as
\beqa
\hspace{-1cm} r\rightarrow 0 & : & \;\; m(<r) \sim \mu \, r^{D} , 
\label{mDmu} \\
\hspace{-1cm} && \;\; \eta_r = \frac{m(<r)}{\rho_0 V} \sim  r^{D-d} .
\label{etamu}
\eeqa
Thus, at small scales we can see that very large densities are associated
with the lowest-dimension objects, $D=0$, and the contribution of these
point-like masses to the probability distribution $\cP(\eta_r)$ reads as
\beq
r\rightarrow 0, \; \eta_r \rightarrow\infty : \;\; \cP(\eta_r) \dd\eta_r 
\sim V n(m) \dd m ,
\label{Petan0m}
\eeq
where $V$ is the volume of the sphere of radius $r$ and $n(m)$ is the mass
function of point-like masses, that is, $n(m)\dd m\dd \bx$ is the mean number
of such objects of $0-$dimension, with a mass in the range $[m,m+\dd m]$,
within the volume element $\dd\bx$. 
Then, from Eq.(\ref{Petasupexp}) we obtain for the high-mass tail
\beq
m\rightarrow\infty : \;\; \ln[n(m)] \propto - \frac{m^{(n+3)/d}}{t^2} .
\label{masstail}
\eeq
Indeed, we have seen in section~\ref{Saddle-point-with-shocks} that the scaling
(\ref{Petasupexp}), that was derived in section~\ref{Rare-density} for
$n\leq d-3$, actually extends to the full range $-3<n<1$, but the proportionality
factor is no longer set by Eq.(\ref{Petasup}).
However, in the range $n\leq d-3$, this numerical factor is given by
Eq.(\ref{Petasup}), while for $d-3<n<1$ it can be obtained from the analysis
described in section~\ref{Saddle-point-with-shocks}, and from
Table~\ref{Table_tailshock} for the associated integer values $n=0$ and $d=1$.
We show our results in Table~\ref{Table_n0m} for the high-mass tail of the
mass function $n(m)$ of point-like objects, for the initial
conditions of Table~\ref{Table_correl}.

Of course, as for the density and velocity distributions, these results agree
with the exact expressions that can be obtained in the two cases 
$\{n=0,d=1\}$ and $\{n=-2,d=1\}$ \cite{Avellaneda1995,Frachebourg2000,She1992,Sinai1992,Bertoin1998,Valageas2008,Valageas2009}. For more general cases, the 
scaling (\ref{masstail}) was already conjectured in 
\cite{She1992,Vergassola1994}, from numerical simulations and heuristic arguments,
and proved in \cite{Molchan1997} for $-1<n<1$ with $d=1$ (with upper and
lower bounds for the proportionality factor).

\subsection{Pre-shock singularities}
\label{Pre-shock}

Before we conclude, we should add a few comments on the comparison of this
work with studies of pre-shock singularities \cite{Bec2007}.
As shown in \cite{FrischBV2001}, for smooth initial conditions large densities
are localized near ``kurtoparabolic''
singularities residing on space-time manifolds of codimension two. They lead to
universal density tails $\cP(\eta) \sim \eta^{-7/2}$ in any dimension.
In one dimension, this corresponds to pre-shocks \cite{Bec2000,E1997}, that is,
when a shock forms the Lagrangian potential changes from a single extremum
to three extrema and at the
transition, where its second derivative vanishes, one can see through a Taylor
expansion that the Eulerian density field behaves as $x^{-2/3}$ close to the
singularity. Then, the contribution from the neighborhood of such events (both in
space and time) yields a power-law tail $\cP(\eta) \sim \eta^{-7/2}$.
This can also be extended to higher dimensions \cite{FrischBV2001,Bec2007}.
These results apply to the unsmoothed density field for smooth initial conditions.
By contrast, in the present article we study the smoothed density and velocity
fields, that is we always consider the mean density and velocity increment over
a finite radius $r$, for non-smooth initial conditions described by the
power-law power spectra (\ref{PdeltaL}).
Thus, these are two very different regimes. In particular, this explains why
we obtain probability distributions that depend on both the dimension $d$ and
the slope $n$ of the initial power spectrum (over the range $-3<n<1$), rather
than universal tails.
We can note from Eq.(\ref{Petasupexp}) that in the regime studied here the
probability distribution $\cP(\eta_r)$ shows a large-density exponential tail
with a characteristic cutoff $\eta_r \sim r^{-d}$ that goes to infinity as
$r\rightarrow 0$. Then, at very small scales (i.e. in the highly nonlinear
regime) an intermediate power-law regime can develop below this upper cutoff.
However, this power law is not universal either, since the exact results
obtained in \cite{Valageas2008,Valageas2009} show that for $d=1$
we have $\cP(\eta_r) \sim \eta_r^{-3/2}$ if $n=-2$, see also Eq.(\ref{PBrown}),
and $\cP(\eta_r) \sim \eta_r^{-1/2}$ if $n=0$.

We can note that for the forced Burgers equation similar universal power-law
tails can be obtained using instanton methods (i.e. looking for relevant
saddle-points, that correspond to shocks) \cite{Moriconi2009}, although
there is some debate on the exact value of the exponent, which might depend
on the influence of the boundary conditions \cite{E1997,Boldyrev2004}. Again,
these results consider a smooth forcing so that the exponent is set by the
dynamics of a single shock and is universal. For singular forcing (i.e.
with significant power at high wavenumbers) one might obtain non-universal
results for the density and velocity increments over finite radius $r$, in a
fashion similar to the free case studied here. However, this goes beyond the
scope of this article.

\section{Conclusion}
\label{conclusion}

We have studied in this article some asymptotic properties of decaying
Burgers turbulence in $d$ dimensions. Focussing on the case of random Gaussian
initial velocities and a uniform initial density, we considered power-law
initial energy spectra such that the evolution is self-similar. Thus, the
system displays a hierarchical evolution and the integral scale of turbulence,
$L(t)$, that is generated by the Burgers dynamics and separates the large-scale
quasi-linear regime from the small-scale highly nonlinear regime, grows with 
time as a power law, $L(t)\propto t^{2/(n+3)}$. 
Then, in order to take advantage of the statistical
homogeneity and isotropy of the system (once we have taken care of the
infrared divergence if $n\leq-1$), we have defined the spherical quantities,
$\eta_r$ and $\ctheta_r$, that are the overdensity and the velocity increment
over a sphere of radius $r$. This allows to preserve the statistical 
isotropy of the problem and to handle the case of large dimensions $d>1$.

We have first recalled how such a nonlinear dynamics can be studied through
standard perturbative expansions. Here this corresponds to expansions over
powers of time, or equivalently over powers of the initial velocity
fluctuations. This approach is quite flexible, as it does not require
any symmetry, but it becomes very heavy at high orders.
It can be somewhat simplified when one focusses on spherically symmetric
quantities such as $\eta_r$ and $\ctheta_r$, defined through a real-space
top-hat filter, but it remains cumbersome for arbitrary dimensions.
We have pointed out that from a perturbative point of view the Burgers
dynamics in the inviscid limit is equivalent to the Zeldovich dynamics.
This means that shocks are not taken into account and require non-perturbative
methods.

Next, we have described how to derive the asymptotic probability distributions
$\cP(\eta_r)$ and $\cP(\ctheta_r)$ reached in the quasi-linear regime from
a saddle-point approximation. This method allows to obtain at once the
asymptotic cumulant generating function $\varphi(y)$, the Taylor expansion
of which provides the leading-order term for each cumulant 
$\lag\eta_r^p\rag_c$, that would be obtained from the previous perturbative
expansion truncated at order $p-1$. 
In addition, the generating function $\varphi(y)$ being obtained
directly trough a steepest-descent computation, one can go beyond its
apparent singularities (associated with large high-order cumulants
and slowly decaying tails for the probability distributions) and make
sense of possible secondary branches, that are found when this function
appears to be multivalued. This approach takes advantage of the spherical
symmetry of the observables $\eta_r$ and $\ctheta_r$ to reduce the problem
to a one-dimensional system, as the relevant saddle-point (instanton) is also
spherically symmetric. This allows to derive simple results for arbitrary
dimension $d$, provided this instanton has not formed shocks yet.

Then, from the radial profile of this saddle-point, we have found that
these results only apply to the range of initial energy spectrum index
$-3<n\leq d-3$ (within $-3<n<1$), and only above a nonzero underdensity
if $-2<n\leq d-3$. For $n\geq d-2$ the quasi-linear regime does not really
exist. More precisely, the overdensity and velocity divergence $\eta_r$ and
$\ctheta_r$ are already divergent at linear order and their distributions
do not converge towards a Gaussian at early time or large scale, in spite
of the Gaussianity of the initial conditions. Thus, the system is always
dominated by shocks. For $d-3<n<d-2$ the linear theory is well defined,
so that one recovers Gaussian distributions at very early times or large 
scales, but the saddle-point forms shocks as soon as $t>0$. Then, the 
qualitative results obtained from this steepest-descent approach should remain
valid and still provide a reasonable quantitative approximation, as shocks appear
over a limited range of radii, but they are expected to be modified by 
prefactors of order unity.

Thus, in order to describe the cases $d-3<n<1$, as well as very large
underdensities for $-2<n\leq d-3$, it is necessary to take into account shocks.
We have shown how to modify this saddle-point method, taking advantage of the
geometrical interpretation of the Hopf-Cole solution in terms of first-contact
paraboloids, to handle these cases. This allows us to find out the instantons,
which contain shocks, that provide the leading-order behavior of the 
rare-event tails of $\cP(\eta_r)$ and $\cP(\ctheta_r)$.
Focussing on some low integer values of $n$ and $d$, where simple explicit
results can be derived, we have obtained the asymptotic tails of these
probability distributions, at any finite time and scale, for the cases
$\{n,d\}=\{0,1\}, \{0,3\}$ and $\{-1,2\}$. We note that the scalings actually
agree with a naive extension of those obtained from the regular saddle-point
computation. This also gives the high-mass tail of the mass function
of point-like singularities (i.e. Dirac peaks in the density field,
which correspond to shock strengths in the one-dimensional case).

Then, we find that for $n<-1$ the very low density tail shows an exponential
cutoff of the form $e^{-\eta_r^{(n+1)/2}}$, whereas for $n>-1$ there is no
exponential falloff (but there could be a power-law decline).
For the latter cases, in the quasi-linear regime, this part of the probability
distribution $\cP(\eta_r)$ corresponds to extremely rare underdensities and has
a negligible weight, and around moderate fluctuations, $|\eta_r-1| \ll 1$, the
distribution shows a falloff on both sides of the mean $\lag\eta_r\rag=1$. 
In the highly nonlinear regime, this intermediate low-density regime disappears
and low-density (and empty) regions are no longer rare. Then, one needs another
method to describe the low-density part of the probability distribution at small
scales.

Throughout this article, we have checked that our results agree with the
two one-dimensional cases of white-noise initial velocity ($n=0$) and
Brownian initial velocity ($n=-2$), where many exact results are known,
thanks to the Markovian properties shared by both cases, which allow
a derivation of explicit formulae through specific methods.
Note that these two cases are representative of the two classes of initial
conditions, $-1<n<1$ and $-3<n<-1$, where the initial velocity is dominated
by small/long wavelengths and which exhibit the different behaviors
discussed above. Hence they provide a good check of the general methods
presented in this article. In addition to the interest of the asymptotic
behaviors obtained here, we can hope that they could serve as a benchmark
to test other approximation schemes, devised to study additional quantities
such as typical events. Moreover, since the approach developed in this paper
is rather general - for instance it was already applied to the gravitational
dynamics - it may also prove useful for other systems.

\bibliography{ref}

\end{document}